\documentclass[twocolumn]{aastex631}

\usepackage{lipsum}
\usepackage{comment}
\usepackage{graphicx}	
\usepackage{amsmath}	
\usepackage{amssymb}	
\usepackage{xspace}
\usepackage{subfigure}
\usepackage{relsize} 
\usepackage{longtable}

\usepackage{array,etoolbox}
\preto\tabular{\setcounter{magicrownumbers}{0}}
\newcounter{magicrownumbers}

\usepackage{ragged2e}

\shorttitle{}
\shortauthors{}
\graphicspath{{./}{figures/}}

\newcommand{\BAYMAX}{\texttt{BAYMAX}\xspace}
\newcommand{\BF}{\mathcal{BF}\xspace}
\newcommand{\logBF}{$\log \mathcal{BF}$\xspace}

\begin{document}




\title{Searching for the Highest-z Dual AGN in the Deepest \emph{Chandra} Surveys}

\author{Brandon Sandoval}
\affiliation{Cahill Center for Astronomy and Astrophysics, California Institute of Technology, Pasadena, CA 91125, USA}
\affiliation{Kavli Institute of Particle Astrophysics and Cosmology, Stanford University, 452 Lomita Mall, Stanford, CA 94305, USA}

\author{Adi Foord}
\affiliation{Department of Physics, University of Maryland Baltimore County, 1000 Hilltop Cir, Baltimore, MD 21250, USA}
\affiliation{Kavli Institute of Particle Astrophysics and Cosmology, Stanford University, 452 Lomita Mall, Stanford, CA 94305, USA}

\author{Steven W. Allen}
\affiliation{Kavli Institute of Particle Astrophysics and Cosmology, Stanford University, 452 Lomita Mall, Stanford, CA 94305, USA}
\affiliation{Department of Physics, Stanford University, 382 Via Pueblo Mall, Stanford, CA 94305, USA}
\affiliation{SLAC National Accelerator Laboratory, 2575 Sand Hill Road, Menlo Park, CA 94025, USA}

\author{Marta Volonteri}
\affiliation{Institut d’Astrophysique de Paris, Sorbonne Universit{\'e}, CNRS, UMR 7095, 98 bis bd Arago, F-75014 Paris, France}

\author{Aaron Stemo}
\affiliation{Department of Physics and Astronomy, Vanderbilt University, 2301 Vanderbilt Place, Nashville, TN 37235, USA}

\author{Nianyi Chen}
\affiliation{McWilliams Center for Cosmology, Department of Physics, Carnegie Mellon University, Pittsburgh, PA 15213, USA}

\author{Tiziana Di Matteo}
\affiliation{McWilliams Center for Cosmology, Department of Physics, Carnegie Mellon University, Pittsburgh, PA 15213, USA}

\author{Kayhan G{\"u}ltekin}
\affiliation{Department of Astronomy and Astrophysics, University of Michigan, 1085 South University Ave, Ann Arbor, MI 48109, USA}

\author{Melanie Habouzit}
\affiliation{Zentrum f{\"u}r Astronomie der Universit{\"a}t Heidelberg, ITA, Albert-Ueberle-Str. 2, D-69120 Heidelberg, Germany}
\affiliation{Max-Planck-Institut f{\"u}r Astronomie, K{\"o}nigstuhl 17, D-69117 Heidelberg, Germany }

\author{Clara Puerto-S{\'a}nchez}
\affiliation{Zentrum f{\"u}r Astronomie der Universit{\"a}t Heidelberg, ITA, Albert-Ueberle-Str. 2, D-69120 Heidelberg, Germany}

\author{Edmund Hodges-Kluck}
\affiliation{NASA Goddard Space Flight Center, 8800 Greenbelt Rd, Greenbelt, MD 20771, USA}

\author{Yohan Dubois}
\affiliation{Institut d’Astrophysique de Paris, Sorbonne Universit{\'e}, CNRS, UMR 7095, 98 bis bd Arago, F-75014 Paris, France}

\begin{abstract}
\noindent We present an analysis searching for dual AGN among 62 high-redshift ($2.5 < z < 3.5$) X-ray sources selected from publicly available deep \emph{Chandra} fields. We aim to quantify the frequency of dual AGN in the high-redshift Universe, which holds implications for black hole merger timescales and low-frequency gravitational wave detection rates. We analyze each X-ray source using \BAYMAX, an analysis tool that calculates the Bayes factor for whether a given archival \emph{Chandra} AGN is more likely a single or dual point source. We find no strong evidence for dual AGN in any individual source in our sample. We then increase our sensitivity to search for dual AGN across the sample by comparing our measured distribution of Bayes factors to that expected from a sample composed entirely of single point sources, and again find no evidence for dual AGN in the observed sample distribution. Although our analysis utilizes one of the largest \emph{Chandra} catalogs of high-$z$ X-ray point sources available to study, the findings remain limited by the modest number of sources observed at the highest spatial resolution with \emph{Chandra} and the  typical count rates of the detected sources. Our non-detection allows us to place an upper-limit on the X-ray dual AGN fraction between $2.5<z<3.5$ of 4.8\%.  Expanding substantially on these results at X-ray wavelengths will require future surveys spanning larger sky areas and extending to fainter fluxes than has been possible with \emph{Chandra}. We illustrate the potential of the AXIS mission concept in this regard.
\end{abstract}



\section{Introduction} \label{sec:intro}
There is now broad consensus that supermassive black holes (SMBH) exist at the center of most massive ($M_* > 10^{10} M_{\odot})$ galaxies \citep{smbhcenter, SMBHGalaxyMass}. Thus, during galaxy mergers, we may expect systems of two interacting SMBHs. During such a merger, gas may be funneled down to the SMBHs causing them both to accrete and become active galactic nuclei (AGN; \citealt{agnfuel, DeboraMergerTriggerAGN, HopkinsMergerTriggerAGN}). Such systems can be classified as ``dual AGN'' at the earliest phase of merger evolution, where the two AGN are at kiloparsec scale separations and not yet gravitationally bound.

Given the importance of galaxy mergers in the hierarchical model of galaxy evolution, measurements of the prevalence of dual AGN as a function of redshift can help us better understand how SMBHs and galaxies evolve together over cosmic time \citep{hierarchical}. Such measurements can better constrain the timescales associated with SMBH mergers (i.e., \citealt{BegelmanDualAGNTimescale}), develop a lower-limit to the dual SMBH occupation fractions at various redshifts, and gain insight on SMBH merger rates to be detected with current and future detectors.

SMBH binaries are progenitors of gravitational radiation. Massive black hole mergers in the final phase of evolution are thought to be a source of low-frequency ($<1$ Hz) gravitational waves in the Universe \citep{smbhwaves}. This low frequency gravitational radiation can be observed via pulsar timing arrays (PTAs; \citealt{pta}) for $M_{\mathrm{SMBH}}\ge10^{8}M_{\odot}$ at $z\le2$, or future missions such as the \textit{Laser Interferometer Space Antenna} (\textit{LISA}; \citealt{lisa}) for $10^{6}M_{\odot}\le M_{\mathrm{SMBH}}\le10^{7}M_{\odot}$ at $z\le20-30$.Importantly, these missions rely on estimates of black hole merger rates to determine detection rates. Most recently, results from PTAs have found evidence for gravitational waves, with oscillations of years to decades, thought to arise from pairs of orbiting SMBHs \citep{NG15yrGWB, PPTA18yrGWB, EPTA10yrGWB, CPTAGWB}. The PTA gravitational wave signal has been compared to simulations of various SMBH binary populations, and based on current measurements, the amplitude of the signal suggests that SMBHs may be (1) more common and/or (2) more massive than previously thought. An important component in breaking this degeneracy is a strong constraint on the overall SMBH coalescence timescale. In particular, the final signal of binaries detected by PTAs is driven by mergers occurring at $z=0.3-0.8$, which correlate with progenitor dual AGN at $<$30 kpc-scale separations at $z=1-3$ (see fig. 12 in \citealt{NG15yrAstro}). Thus, constraining the frequency of dual AGN detections at $z=1-3$, as a function of separation, is vital for future binary SMBH model inferences. 

There has yet to be an X-ray study that quantifies the frequency of dual AGN at high redshift, or as a function of redshift. But, there exist many optical searches for quasar pairs in the high-redshift Universe, where tens of candidates have been identified ($z>1$; e.g., \citealt{Hennawi2006,Myers2008,Hennawi2010,Kayo2012,Eftekharzadeh2017}). Most recently, two of the highest-z dual AGN candidates ($z>5$) were detected via optical spectroscopy and photometery \citep{Yue2021, Yue2023}. However, surveys with wide-area coverage are necessary to find large samples of dual AGN candidates and determine their number density reliably. For example, \cite{Stemo2021} analyzed a catalog of 2585 AGN host galaxies observed with the \emph{Hubble Space Telescope} and spanning a redshift range of $0.2 < z < 2.5$. By identifying AGN host galaxies with multiple stellar bulges, they find 204 offset and dual AGN candidates. 

\par Recently, new observational techniques that leverage the angular resolution of Gaia have provided effective first steps for detecting the dual AGN population at high-z. Varstrometry techniques (see, e.g., \citealt{Shen2019, Hwang2020,Shen2021}) have been used to identify a $z>2$ dual AGN \citep{TChen2023} and the Gaia Multi-peak (GMP) method \citep{Mannucci2022} has been used to detect dual AGN candidates at $z>1$ \citep{Ciurlo2023}.  A handful of large surveys in the optical regime have yielded constraints on the high-z dual AGN fraction, finding the dual AGN fraction at high-$z$ is $<1\%$ with no evolution across redshift \citep{Silverman2020, Shen2023}. However, optical selection techniques are affected by optical extinction and contamination from star formation, which is especially problematic when observing highly-obscured mergers \citep{Kocevski2015, Koss2016, Ricci2017, Weston2017, Blecha2018, DeRosa2018, Koss2018, Lanzuisi2018, TorresAlba2018, Hickox2018}.

A more robust method for directly identifying dual AGN is to observe two X-ray point sources with luminosities consistent with accretion onto a SMBH (i.e., greater than $\sim10^{41}$ erg s$^{-1}$ in the 2$-$10 keV band; \citealt{agnlum}). In particular, \emph{Chandra}'s superb sub-arcsecond half power diameter (HPD) within 3\arcmin~of the optical axis allows separations on the order of a few kpc to be probed at \textit{essentially any} redshift, given sufficiently deep exposures. However, despite the reliability provided by X-ray detections via \emph{Chandra} observations, distinctly resolving two point sources becomes difficult at separations approaching the resolution limit. Furthermore, systems of dual point sources with a large contrast in flux and a low total number counts ($< 100$) can be difficult to resolve even with large physical separations, leading to false positive and false negative identifications \citep{Koss, BAYMAX1}.

To identify dual and multiple AGN in \textit{Chandra} observations at low separations and counts, we have previously developed and employed \BAYMAX (\textbf{B}ayesian \textbf{A}anal\textbf{Y}sis of \textbf{M}ultiple \textbf{A}GN in \textbf{X}-rays; \citealt{BAYMAX0, BAYMAX1, BAYMAX2}). \BAYMAX is a {\tt python} code that carries out a Bayesian analysis to determine whether a given \textit{Chandra} source detection is composed of one or two point sources. Analyses with \BAYMAX increase sensitivity to detecting dual AGN systems (over standard point source detection algorithms such as \textit{wavdetect} in the CIAO analysis package \footnote{\url{https://cxc.cfa.harvard.edu/ciao/threads/wavdetect/}}) for angular separations around or below 1\arcsec, or when the secondary AGN is dim with respect to the primary AGN.

In this paper, we set out to quantify the dual AGN fraction in the redshift range $2.5 < z < 3.5$, as part of a larger effort to measure the dual AGN frequency from $0 < z <3.5$. We examine 62 X-ray sources from publicly available deep \textit{Chandra} fields. Using \BAYMAX, we quantify how likely each source is to be composed of two point sources.

The remainder of the paper is organized into 5 sections. In Section 2, we outline the surveys used to construct the sample and the thresholds enforced in our analysis. In Section 3, we review how \BAYMAX distinguishes between single and dual point sources in the Bayesian paradigm. We also outline the prior densities used for all of the sources in the sample. In Section 4, we present the results of using \BAYMAX on our sample, and a follow-up false positive analyses. In Section 5, we discuss the interpretation of the results and perform follow up analyses on the presence of dual AGN in our sample. We summarize our findings in Section 6. 

\section{Sample} \label{sec:sample}

Our sample of sources is drawn from X-ray point source catalogs created for publicly available deep \textit{Chandra} fields: X-UDS (\textit{Chandra} imaging of the Subaru-XMM Deep/UKIDSS Ultra Deep Survey field; \citet{XUDS}), AEGIS-XD (\textit{Chandra} imaging of the central region of the Extended Groth Strip; \citet{AEGISXD}), CDF-S (\textit{Chandra} Deep Field-South; \citet{CDFS}), and the COSMOS-Legacy survey (\citet{COSMOS}). X-UDS consists of 25 observations covering a total area of 0.33 deg$^2$ with a nominal depth of $\sim$600 ksec in the central 100 arcmin$^2$ and $\sim$200 ksec in the remainder of the field; AEGIS-XD covers a region of approximately 0.29 deg$^2$ with a nominal depth of 800 ksec; CDF-S covers a total area of 484.2 arcmin$^2$ with an effective exposure of 7 Msec; the COSMOS-Legacy survey consists of 56 observations covering an area of 2.2 deg$^2$ with an effective exposure of $\sim$ 160 ksec over the central 1.5 deg$^2$ and $\sim$ 80 ksec of the remaining area. Combined, there are 4574 X-ray point sources across all catalogs.

The X-ray point source catalogs we used supply photometric or spectroscopic redshift information for each source, determined via counterpart-matching from various ground and space-based surveys or computed using SED fitting. Because the C-COSMOS X-ray point source catalogs include spectroscopic classifications, we also filter out any sources from C-COSMOS that have been spectroscopically classified as a star. To create our sample, we enforce the following cuts on each tabulated X-ray point source: $2.5<z<3.5$, $\ge$50 counts between 0.5-8 keV (as determined by the catalog), and the off-axis angle (OAA; the angular distance between the source position and the aim point of the pointing) of every observation must be $< 5$\arcmin. For observations with an OAA above 5\arcmin, modeling \textit{Chandra}'s PSF increases in size and becomes more asymmetric. Likewise, below 50 counts we lose sensitivity to detecting dual point sources at separations below 1\arcsec. For these reasons, we have enforced OAA and count thresholds to remain sensitive to dual AGN at small separations, while lowering the likelihood of false negatives within our analysis. With 50 or more counts, we expect to be sensitive to AGN with luminosities on the order of $10^{43}$ erg s${^{-1}}$ at $z=2.0$.

\begin{figure*}
\begin{minipage}[t]{0.33\textwidth}
  \includegraphics[width=\linewidth]{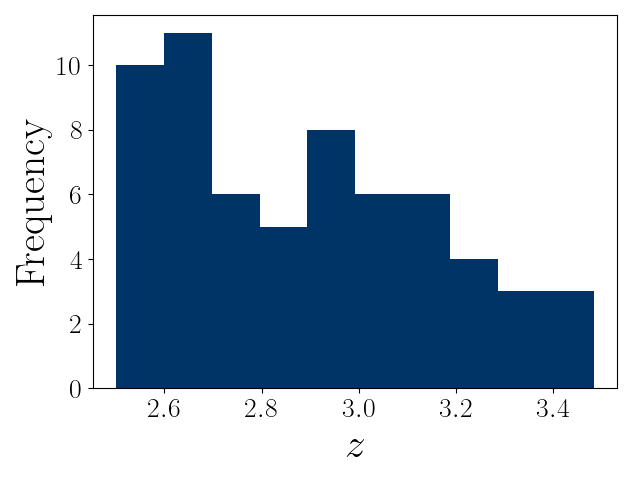}
\end{minipage}%
\begin{minipage}[t]{0.33\textwidth}
  \includegraphics[width=\linewidth]{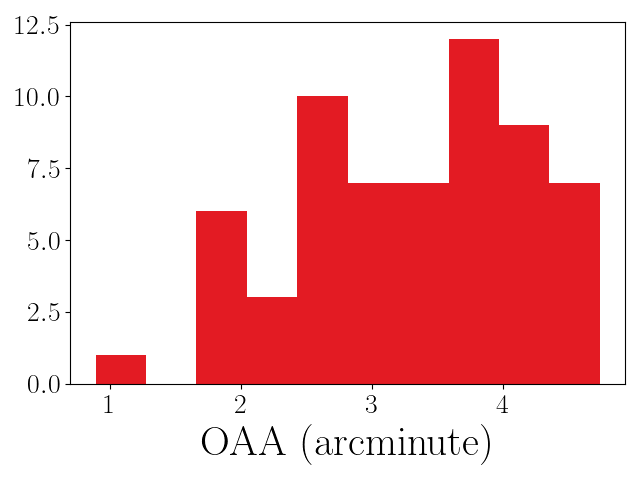}
\end{minipage}%
\begin{minipage}[t]{0.33\textwidth}
  \includegraphics[width=\linewidth]{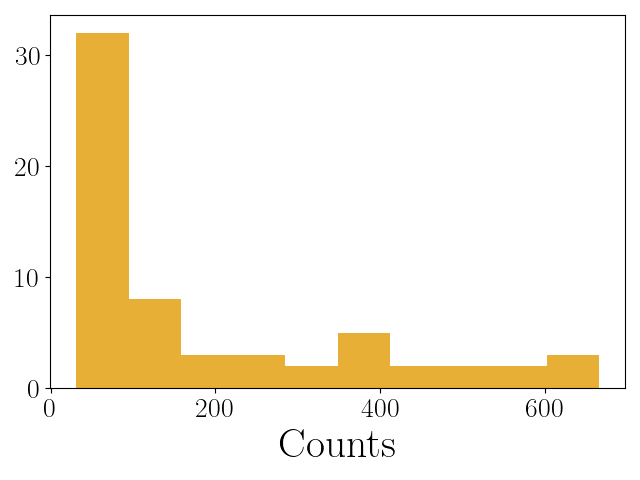}
\end{minipage}%

\textbf{Figure 1:} Distributions of redshift ($\textit{left}$), OAA ($\textit{center}$), and total counts ($\textit{right}$) for the 62 X-ray sources composing the sample. Each X-ray source has photometric or spectroscopic redshift data from its respective catalog. For sources with more than one observation, we calculate a weighted OAA average (via the exposure time). The count number is determined by the number of 0.5 keV$-$8 keV photons in a 20\arcsec $\times$ 20\arcsec ~box centered on the nominal coordinate listed in the source catalog.
\label{fig:first}

\end{figure*}

Due to computational time constraints (see Section 3), we limit our analysis to a maximum of 10 observations per source. Sources that cannot meet our 50 count criterion within their 10 longest observations are cut from the sample. With these considerations, our final sample is comprised of 62 X-ray sources. In Figure \ref{fig:first}, we show the distribution of redshifts, observation-averaged OAA values, and counts in our sample. We note that the majority of sources in our sample have less than 100 counts (34 sources) and an averaged OAA greater than 3\arcmin~(41 sources).

\section{Methodology} \label{sec:methodology}

\BAYMAX is a Bayesian statistical package that estimates the likelihood of multiple point sources within \textit{Chandra} observations. Past studies have shown that for on-axis \textit{Chandra} observations with $\ge$700 counts between 0.5-8 keV, \BAYMAX is sensitive to dual point sources with separations as low as 0.3\arcsec~\citep{BAYMAX0}.

\BAYMAX identifies dual-AGN in an observation by calculating the Bayes factor, hereafter denoted by $\BF$. The $\BF$ is defined as the ratio of the marginal likelihoods corresponding to two hypotheses, namely dual vs. single/point X-ray sources. In this case, the $\BF$ can be written as follows:
\[
\BF = \frac{P(D | M_2)}{P(D | M_1)} = \frac{\int P(D | M_2, \theta_2) P(\theta_2 | M_2) d\theta_2}{\int P(D | M_1, \theta_1) P(\theta_1 | M_1) d\theta_1}
\]
where $D$ is our data, $M_2$ is the dual point source model, $M_1$ is the single point source model, $\theta_1$ is the parameter vector for the single point source model, and $\theta_2$ is the parameter vector for the dual point source model. We see that in calculating the $\BF$, \BAYMAX estimates both the likelihood $P(D | M, \theta)$ and the prior densities $P(\theta | M)$. Generally, a $\BF$ greater or less than 1 indicates which model is favored (see \citealt{BFThresh}); however, in Section \ref{sec:results} we further discuss how to assess the strength of a particular Bayes factor.

\BAYMAX calculates the $\BF$ using nested sampling \citep{nested} via the {\tt{python}} package {\tt{nestle}} \footnote{https://github.com/kbarbary/nestle}. For a thorough description of the statistical techniques used to estimate likelihoods and posterior densities, see \citet{BAYMAX0, BAYMAX1}.

\subsection{Prior Densities}
Here we briefly review the prior densities used for both the single and dual source models. In the single point source model, $\theta_{1}$ is composed of the source's sky coordinate, $\mu$, and the logarithm of the background fraction, $\log f_{\text{bkg}}$. The background fraction is defined as ratio of the number of counts associated with the background versus the number of counts associated with all point sources. For the dual source model, $\theta_{2}$ is composed of the sky coordinates for each source, $\mu_1$ and $\mu_2$, the logarithm of the background fraction, $\log f_{\text{bkg}}$, and the logarithm of the count ratio, $\log f$. The count ratio is defined as the ratio of the number of counts between the secondary and primary X-ray point source. 

For sources with multiple observations, \BAYMAX models the PSF of each observation and calculates the likelihoods for each observation individually. In these cases, $\theta_1$ and $\theta_2$ include the $x,\ y$ components of the astrometric shifts between each additional observation. The shifts are defined with respect to the observation with the longest exposure time.

For all source positions $\mu$, we are using a non-informative prior defined by a continuous uniform distribution. The $x$ and $y$ location priors are defined by a $20\arcsec\times20 \arcsec$ box centered on the nominal X-ray coordinate of the source. The prior distribution for $\log{f_{\text{bkg}}}$ is defined as a truncated Gaussian distribution, with mean $\mu_{\text{bkg}}$ and $\sigma_{\text{bkg}}$. The value for $\mu_{\text{bkg}}$ is estimated using source free-regions within a $50\arcsec\times50 \arcsec$ box centered on the source. The value for $\sigma_{\text{bkg}}$ is set to 0.5, and the Gaussian is truncated at -3 and 0. For the dual point source model, the count ratio $\log f$ prior is defined with a uniform distribution between -4 and 4.

\section{Results} \label{sec:results}
For every source, we have a lower photon energy cut at 0.5 keV and upper energy cut at 8 keV. We analyze the photons in a region equal to the prior distribution for $\mu$. In general, this is a 20\arcsec$\times$ 20\arcsec box on the nominal X-ray coordinate listed in the point source catalogs.

For each source, \BAYMAX outputs a \logBF value and statistical error bars returned from the nested sampling procedure via {\tt nestle}. In the past, we have found that the statistical error bars returned from {\tt nestle} are consistent with the 1$\sigma$ spread in the \logBF values when running \BAYMAX 100 times on a single source \citep{BAYMAX1, BAYMAX2}. In Table \ref{tab:BFs}, we list the \logBF values and their respective error bars.

\begin{center}
\begin{longtable*}{ccc}
\caption{Bayes Factor Results}  \label{tab:BFs} \\
\hline \hline 
\multicolumn{1}{c}{AGN ID} & \multicolumn{1}{c}{Name} & \multicolumn{1}{c}{\logBF} \\ 
\multicolumn{1}{c}{(1)} & \multicolumn{1}{c}{(2)} & \multicolumn{1}{c}{(3)} \\ \hline
\endfirsthead

\multicolumn{3}{c}%
{{\bfseries \tablename\ \thetable{} -- continued from previous page}} \\
\hline \hline \multicolumn{1}{c}{AGN ID} & \multicolumn{1}{c}{Name} & \multicolumn{1}{c}{\logBF}  \\ 

\multicolumn{1}{c}{(1)} & \multicolumn{1}{c}{(2)} & \multicolumn{1}{c}{(3)} \\ \hline 
\endhead

\hline \multicolumn{3}{c}{{Continued on next page}} \\ 
\endfoot

\hline \hline
\multicolumn{3}{@{} p{\textwidth} @{}}{\small {Note. -- Columns: (1) AGN ID; (2) AGN Name; (3) \logBF value in favor of the dual point source model. $\dagger$ denotes a \logBF that is greater than 0 at the 3$\sigma$ level. False positive testing (see Section \ref{sec:falsepositive}) shows that these values are consistent with \logBF values expected from a single point source. This is likely a result of the large OAA of the observations, where the PSF is more difficult to accurately model.}}

\endlastfoot

1 & AEGISXD214.447221+52.586265 & $8.13\pm2.48^{\dagger}$\\ 
2 & AEGISXD214.480367+52.592432 & $0.14\pm2.31$\\ 
3 & AEGISXD214.501556+52.603002 & $0.27\pm2.79$\\ 
4 & AEGISXD214.628453+52.673396 & $0.68\pm2.68$\\ 
5 & AEGISXD214.751858+52.761911 & $-0.28\pm1.85$\\ 
6 & AEGISXD214.809491+52.769021 & $7.70\pm2.72^{\dagger}$\\ 
7 & AEGISXD214.755207+52.836803 & $0.91\pm2.76$\\ 
8 & AEGISXD215.056118+52.939551 & $0.32\pm2.07$\\ 
9 & AEGISXD215.134433+53.078521 & $-0.13\pm1.75$\\ 
10 & CDFS53.033361-27.782539 & $14.34\pm2.40^{\dagger}$\\ 
11 & CDFS53.039401-27.801862 & $4.13\pm2.41^{\dagger}$\\ 
12 & CDFS53.075954-27.878104 & $0.65\pm1.76$\\ 
13 & CDFS53.082561-27.755268 & $-0.13\pm1.95$\\ 
14 & CDFS53.107543-27.855644 & $-0.36\pm2.15$\\ 
15 & CDFS53.108124-27.753992 & $0.77\pm2.64$\\ 
16 & CDFS53.111559-27.767777 & $1.25\pm2.03$\\ 
17 & CDFS53.137971-27.868187 & $0.44\pm1.85$\\ 
18 & CDFS53.161473-27.855948 & $-0.47\pm1.63$\\ 
19 & CDFS53.165266-27.814067 & $0.03\pm2.15$\\ 
20 & CDFS53.178452-27.78402 & $0.94\pm2.10$\\ 
21 & CDFS53.183426-27.776567 & $0.61\pm2.51$\\ 
22 & CDFS53.185805-27.809946 & $0.51\pm2.55$\\ 
23 & COSMOS150.27207+2.230126 & $-0.21\pm1.64$\\ 
24 & COSMOS150.19426+2.106866 & $1.13\pm1.50$\\ 
25 & COSMOS150.38282+2.104631 & $0.11\pm1.45$\\ 
26 & COSMOS150.36472+2.143831 & $-0.55\pm1.76$\\ 
27 & COSMOS150.2477+2.442225 & $0.03\pm1.36$\\ 
28 & COSMOS150.20884+2.48201 & $0.18\pm1.73$\\ 
29 & COSMOS150.29244+2.545221 & $0.24\pm1.46$\\ 
30 & COSMOS150.10389+2.665734 & $0.13\pm1.82$\\ 
31 & COSMOS149.71561+2.016628 & $0.26\pm1.21$\\ 
32 & COSMOS149.63929+2.003248 & $1.18\pm1.51$\\ 
33 & COSMOS149.92258+1.979333 & $0.21\pm1.33$\\ 
34 & COSMOS149.80063+1.870479 & $0.11\pm1.31$\\ 
35 & COSMOS149.97286+1.941684 & $0.70\pm1.27$\\ 
36 & COSMOS149.75631+2.117313 & $-0.32\pm1.37$\\ 
37 & COSMOS150.28557+2.014617 & $-0.19\pm1.44$\\ 
38 & COSMOS149.80849+2.313858 & $0.08\pm1.29$\\ 
39 & COSMOS149.86968+2.294064 & $0.48\pm1.34$\\ 
40 & COSMOS149.89193+2.285167 & $0.10\pm1.50$\\
41 & COSMOS150.0459+2.201258 & $0.59\pm1.61$\\ 
42 & COSMOS150.06453+2.191 & $0.10\pm1.52$\\ 
43 & COSMOS149.9692+2.304833 & $0.00\pm1.33$\\ 
44 & COSMOS149.84815+2.374316 & $-0.06\pm1.36$\\ 
45 & COSMOS149.98156+2.315056 & $3.94\pm1.82^{\dagger}$\\ 
46 & COSMOS149.88247+2.505174 & $0.81\pm1.84$\\ 
47 & COSMOS150.05228+2.369345 & $0.32\pm1.48$\\ 
48 & COSMOS150.41176+2.317611 & $0.41\pm1.20$\\ 
49 & COSMOS149.92304+2.026981 & $0.92\pm1.37$\\ 
50 & COSMOS150.2107+2.391473 & $0.38\pm1.28$\\ 
51 & COSMOS150.28487+2.309435 & $0.43\pm1.53$\\ 
52 & COSMOS150.23548+2.3618 & $-0.58\pm1.45$\\ 
53 & COSMOS149.96966+1.891586 & $0.09\pm1.49$\\ 
54 & COSMOS149.79436+2.073134 & $0.40\pm1.33$\\ 
55 & COSMOS150.31649+1.887004 & $-0.39\pm1.50$\\ 
56 & COSMOS150.01056+2.269484 & $-0.32\pm1.58$\\ 
57 & COSMOS150.18087+2.075997 & $0.59\pm1.51$\\ 
58 & COSMOS149.79304+2.111527 & $0.13\pm1.28$\\ 
59 & XUDS34.526818-5.233219 & $0.60\pm2.07$\\ 
60 & XUDS34.27415-5.227366 & $0.04\pm2.05$\\ 
61 & XUDS34.150063-5.099644 & $0.37\pm1.50$\\ 
62 & XUDS34.528469-5.069177 & $1.26\pm2.03$\\ 
        
\end{longtable*}
\end{center}

We note, because sources AEGISXD214.44+52.58, AEGISXD214.75+52.76, AEGISXD214.80+52.76, AEGISXD215.05+52.93, CDFS53.03-27.78, CDFS53.03-27.80, CDFS53.07-27.87, CDFS53.08-27.75, CDFS53.13-27.86, CDFS53.16-27.85, CDFS53.16-27.81, and CDFS53.17-27.78 required an excessive amount of computing time to analyze their set of 10 observations, only their 5 longest exposure observations were analyzed (where the range of total exposure times is still sufficiently deep, between 183$-$671 ks).
Additionally, sources  XUDS34.52-5.06 and AEGISXD214.93+52.77 contained a second, bright source within the 20\arcsec$\times$ 20\arcsec ~FOV analyzed by \BAYMAX. Both of the additional bright sources were identified in their respective X-ray point source catalogs, and had redshift measurements inconsistent ($> 300$ km s$^{-1}$) with the X-ray point source being analyzed. Therefore, we interpretted these sources as merely projected pairs. For these two sources we ran \BAYMAX with a smaller FOV that avoids the additional bright and nearby source (see Appendix B). We note that all sources still meet our 50 count threshold within this subset.

\subsection{False Positive Analysis}\label{sec:falsepositive}
A source is determined to strongly favor the dual point source model if 1) its \logBF is greater than 0  at the 3$\sigma$ level and 2) the false positive rate of its \logBF is below 10\%. 

The false positive test for a source begins by running \BAYMAX on 100 single point source simulations. The simulations are created via \texttt{MARX} \citep{MARX}, a program designed to simulate the on-orbit performance of the Chandra X-ray Observatory. \texttt{MARX} provides ray-trace simulations of a variety of astrophysical sources and contains detailed models for Chandra's High Resolution Mirror Assembly (HRMA), the HETG and LETG gratings, and all the focal plane detectors (see \citealt{MARX} for more details). 

The input for our \texttt{MARX} simulations are the same detector position, pointing, exposure time, and energy spectrum as the \textit{Chandra} observation(s). This guarantees that the count number and energy spectrum of the simulations closely match the observation. \BAYMAX's analysis on the simulations is carried out with the same prior densities and energy constraints for that particular source. We then compare the measured \logBF to the distribution of \logBF values from our false positive test. We define the false positive rate to be the percentage of \logBF values in the false positive test that are greater than the measured \logBF for that source. The false positive rate represents the probability that \BAYMAX would return a \logBF larger than the measured \logBF if the system were in fact a single point source.

A source is a dual-point source candidate if its false positive rate is 10\% or less. Given this threshold, we find that none of the sources show strong evidence of being a dual point source in the false positive test. Therefore, we cannot conclude that any individual source within our sample has strong evidence of being a dual X-ray point source.

Of the 5 sources with a \logBF greater than zero at the 3$\sigma$ level, all have an average weighted OAA greater than 3\arcmin. Due to the difficulty in modeling the PSF at high OAA, sources with large OAA values are prone to having a \logBF in favor of the dual point source model, with \logBF values increasing as a function of OAA.

To quantify this effect in our analysis, we investigate the distribution of \logBF values that would be expected if our sample were entirely composed of simulated single point sources, determining the results as a function of OAA. To account for the statistical variation in the single point source simulations, for each source we create 100 simulations. These are created following the same procedure for our false positive analysis. We then utilize \BAYMAX to calculate a \logBF for each simulation, for each source.

The results of this investigation are shown in Figure 2, which shows the 90$^{th}$ percentiles of the cumulative distribution function (CDF) of the 100 simulated single point source \logBF values calculated for each source, plotted against the source's OAA. The 5 sources that returned a calculated \logBF greater than zero within error are given red datapoints. We note that for sources with more than one observation, we calculate the weighted OAA average (via the exposure times). By examining the 90$^{th}$ percentiles, we avoid outlier \logBF values which may influence parameters such as the mean.

\section{Discussion}
 
\begin{figure}[t]
    \includegraphics[width=\columnwidth]{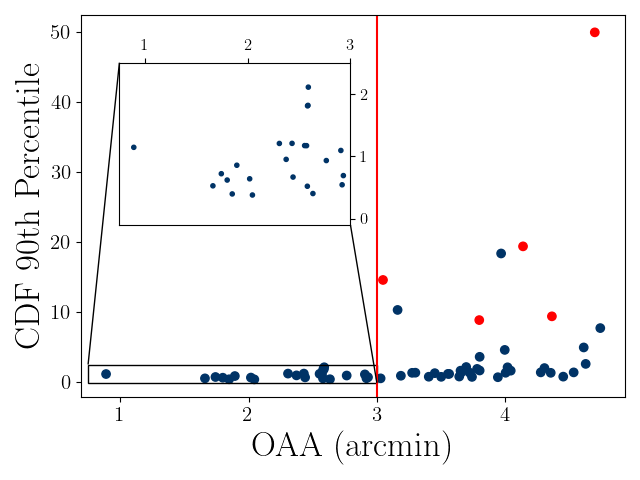}
    \textbf{Figure 2:} We show the 90$^{th}$ percentiles of the CDF of the 100 simulated single point source \logBF values calculated for each source. In red, we show sources with \logBF values greater than 0 at the 3$\sigma$ level. A line is drawn at 3\arcmin, above which we tend to get consistently larger \logBF values. The inset shows the spread of values when restricting our sample to sources with OAA $<$ 3\arcmin. For sources with more than one observation, we calculate a weighted OAA average (via the exposure times).
\label{fig:second}
\end{figure}

\subsection{Aggregate Sample Analysis}
Above OAA values of 3\arcmin, we observe that single point sources are more likely to return larger \logBF values in favor of the dual point source model. This is a result of \emph{Chandra}'s PSF degradation as a function of OAA, and hence becoming more difficult to model. Having no strong evidence for individual dual AGN in our sample, and in light of the spurious \logBF values above an OAA of 3\arcmin, we also tested for the presence of dual AGN by analyzing the distribution of \logBF's after constraining our sample to only include sources with an OAA $< 3$\arcmin. This new OAA threshold reduces our sample from 62 to 21 (see inset in Fig.~\ref{fig:second}).

The aggregate sample analysis is done by comparing our distribution of measured \logBF's to those that one would expect from a sample that is composed entirely of single point sources, and quantifying the differences between the two. For each source in our reduced sample, we utilize the existing 100 simulations of a single point source created during the false positive analysis. We then quantify whether there is evidence for differences in our measured distribution of \logBF's and a distribution expected for a sample of single point sources. We randomly sample a Bayes factor value from each source's suite of single point source simulations to form a simulated distribution. This distribution of simulated Bayes factors represents the expected spread of \logBF for our sample, under the hypothesis that all sources are single point sources. We determine whether our measured distribution and a given distribution of simulated single point sources can be sampled from the same parent distribution via the Kolmogorov-Smirnov (KS) test \citep{KSGen} and the Anderson-Darling (AD) test \citep{ADGen}. Both of these statistical tests compare the equality of two samples under the null hypothesis that both samples are drawn from the same overall distribution. The KS test is more sensitive to differences in the centers of the distributions while the AD test is more sensitive to distribution tails \citep{ADvsKS}.

To account for the statistical variation introduced in the \texttt{MARX} simulation, we repeat this process 10,000 times. Thus, we create a distribution of 10,000 KS test statistics and 10,000 AD test statistics.

\begin{figure}[t] 
        \includegraphics[width=\columnwidth]{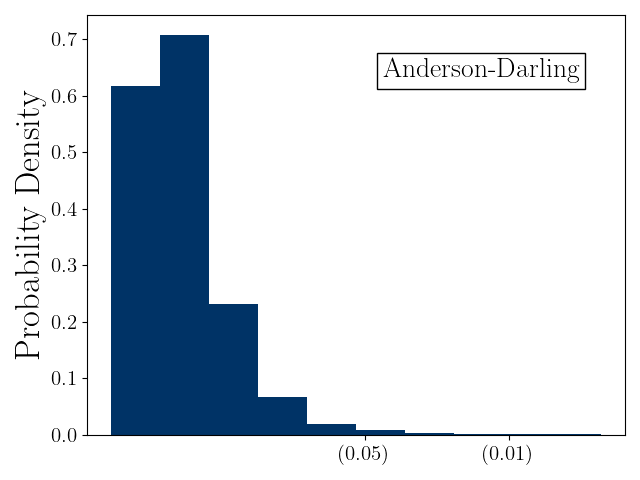}
        \includegraphics[width=\columnwidth]{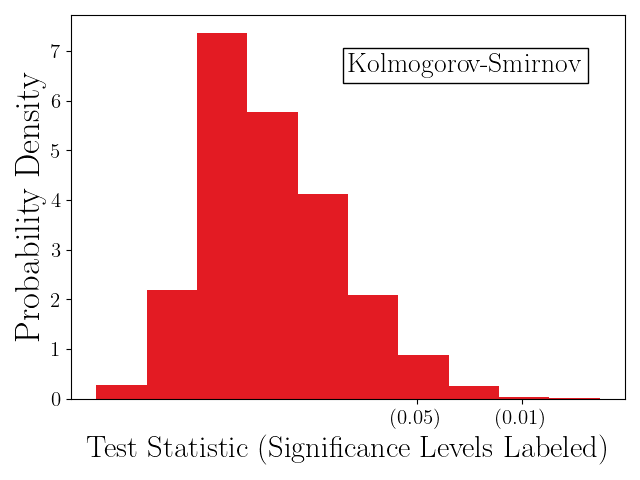}
    \textbf{Figure 3:} We show the test statistic distributions for the AD test (\textit{above}) and the KS test (\textit{below}) when comparing the distribution of measured Bayes factors to the distribution we would expect from a sample composed entirely of single point sources. On the horizontal axis we have labeled the critical values corresponding to the 5\% and 1\% significance levels.
\label{fig:third}
\end{figure}

The distribution of test statistics is shown in Figure \ref{fig:third}, with the corresponding critical values for the 0.05 and 0.01 significance levels marked. For the KS test, 97.48\% of the test statistics lie below the critical value at the 0.05 significance level. For the AD test, 99.44\% of the test statistics lie below the critical value at the 0.05 significance level. Therefore, we conclude that the Bayes factors calculated from our data and the Bayes factors calculated from the single point source simulation are consistent. We find no evidence for the presence of dual AGN sources in the observed sample distribution. We note that the utility of both the AD and the KS tests is largely limited by our sample size of 21 \citep{AD&KSPower}.

\subsection{Comparison to Cosmological Simulations}
Our non-detection of dual AGN activity within the \emph{Chandra} fields at $2.5<z<3.5$ allows us to place a limit on the X-ray dual AGN fraction. Assuming confidence limits for Poisson statistics \citep{Gehrels1986}, we measure a dual AGN fraction upper-limit of 4.8\% at the 95\% confidence level. This upper-limit assumes that there are no dual AGN in our sample.

\par We compare our results to various large-scale cosmological hydrodynamical studies: Magneticum Pathfinder \citep{Steinborn2016}, The Evolution and Assembly of GaLaxies and their Environment (EAGLE) \citep{Rosas-Guevara2019}, Illustris \citep{Kelley2017, PuertoSanchez_inprep}, Horizon-AGN \citep{Volonteri2022}, and ASTRID \citep{Chen2023}. We note that the assumed physics, spatial and mass resolution, as well as selection criteria for dual AGN vary across each simulation. In particular, results from Magneticum (box size = 182 cMpc$^{3}$) resolve SMBH pairs down to 2-5 kpc; EAGLE (box size = 100 cMpc$^{3}$) resolve SMBH pairs down to 5 kpc; Illustris (box size = 106.5 cMpc$^{3}$) resolve SMBH pairs down to 1-2 kpc; Horizon-AGN (box size = 142 cMpc$^{3}$) resolve SMBH pairs down to 4 kpc; and ASTRID (box size = 369 cMpc$^{3}$) resolve SMBH pairs with separations down to 4/(1+z) kpc. Horizon-AGN and ASTRID are the only simulations that include sub-grid dynamical friction modeling. They both model the drag force from gas, while ASTRID additionally includes dynamical friction from stars and dark matter. Therefore, Horizon-AGN and ASTRID allow several BHs to evolve in galaxies, while the other simulations merge BHs immediately after the merger of their respective host galaxies (resulting in galaxies hosting only one BH). All models use a mass cut of $M_{BH} > 10^{7}M_{\odot}$ (corresponding to an Eddington limit of $10^{45}$ erg s$^{-1}$), with the exception of the EAGLE simulations. Furthermore, all models define a dual AGN as SMBHs with bolometric luminosities greater than 10$^{43}$ erg s$^{-1}$ and physical separations below 30 kpc.

\begin{figure}[t] 
    \includegraphics[width=\columnwidth]{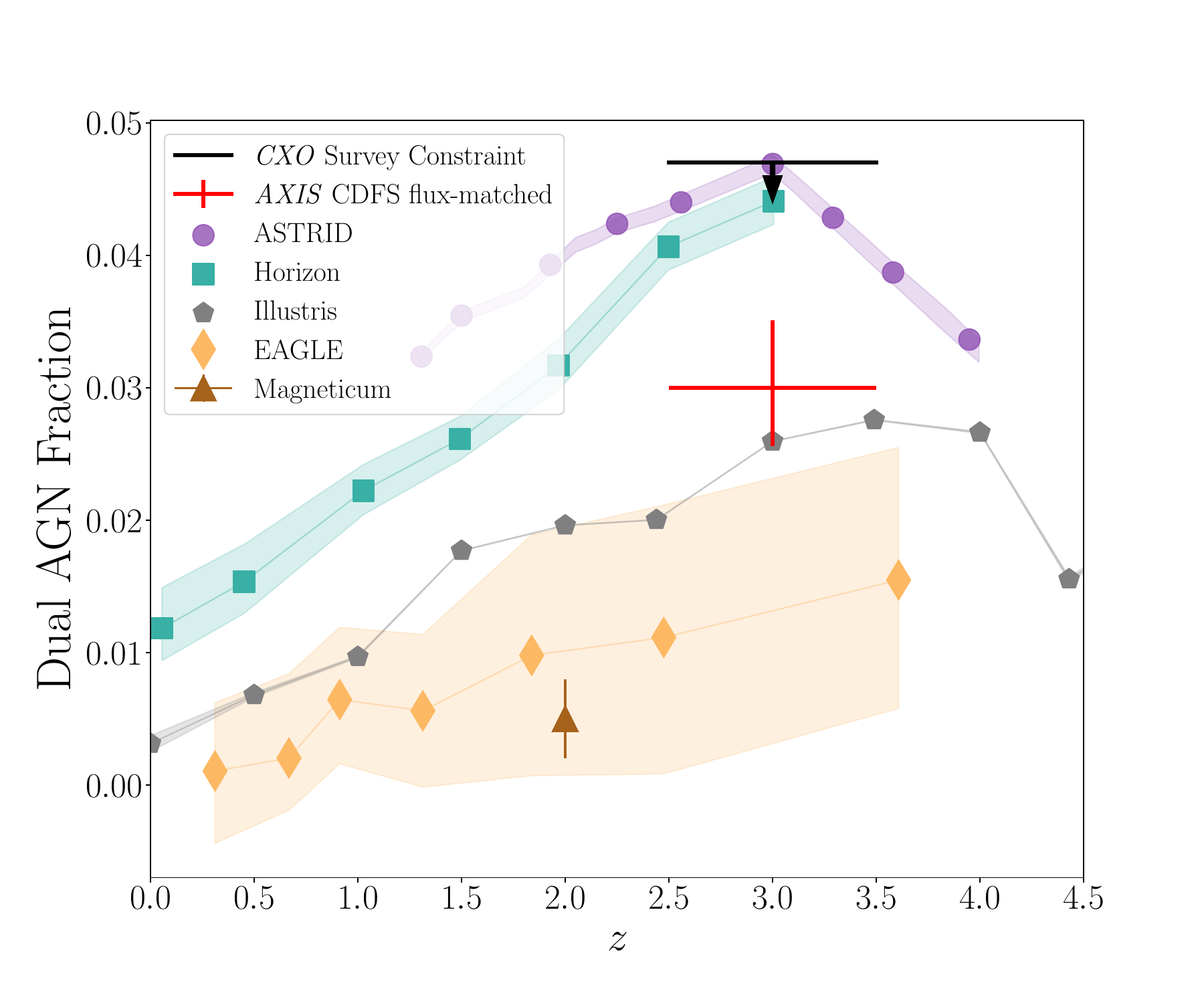}
    \textbf{Figure 4:} Dual AGN fraction measured from our sample (black arrow) at the 95\% confidence interval. We overplot predictions of the dual AGN fraction as a function of redshift from cosmological simulations (Magneticum, EAGLE, Illustris, Horizon-AGN, and ASTRID; \citealt{Steinborn2016, Rosas-Guevara2019, PuertoSanchez_inprep, Volonteri2022, Chen2023}). Albeit an upper-limit, our measurement is consistent with predictions from simulations. We show how well future X-ray mission probe concept \emph{AXIS} could constrain the dual AGN fraction at this same redshift bin (assuming a dual AGN fraction of $\sim3\%$, assuming the flux-limit of CDFS (black solid line). Future X-ray missions with small PSFs, large field-of-views, and high affective areas will find hundreds of new dual AGN, allowing for better statistics and population studies.
\label{fig:cosmosims}
\end{figure}

\par The simulation definitions of dual AGN represent a close comparison to the dual AGN population we are sensitive to. At our lower-count limit (50 counts between $0.5-8$ keV), we expect to be sensitive to AGN with $0.5-8$ keV luminosities down to 10$^{43}$ erg s$^{-1}$ at $z=2$, and we expect to resolve pairs below 20 kpc at all $z$ (folding in \emph{Chandra}'s HPD at our maximum angular diameter distance). Given the range of OAA comprising our archival dataset, we are generally sensitive to larger-separated pairs (the \emph{Chandra} HPD is 1\arcsec~and 2.5\arcsec~at OAA values of 3\arcmin~and 5\arcmin, respectively); note that the Horizon, ASTRID, and Illustris simulations shown in Fig.~\ref{fig:cosmosims} represent the dual AGN fraction for pairs with separations down to $1.5\arcsec$ (via private communication). 

\par Figure~\ref{fig:cosmosims} compares our measured 95\% confidence upper limit on the dual AGN fraction with the predictions from the large-scale cosmological hydrodynamical studies. Our measurement is consistent with predictions from cosmological simulations, where the estimated range of dual AGN activity spans from $\sim0.1-4\%$. In particular, our upper-limit agrees with findings from the Horizon and ASTRID simulations, which have been fine-tuned to better match our sensitivity in separation space. 

\par Overall, we find that the simulation predictions of ASTRID and Horizon are fairly consistent with one-another, across a wide range of redshifts. Both of these simulations include improved prescriptions to measure SMBH dynamics more accurately (and with increased physical realism) via the inclusion of dynamical friction in subgrid models. This approach replaces the `repositioning` schemes used by previous large cosmological volume simulations, which lead to instantaneous mergers of dual AGN at kpc distances (and thus suppresses the measured dual AGN population). 

\subsection{Quantifying Our Sensitivity}
\par We note the caveats associated with our results, mainly that (1) we are not sensitive to the faintest and mostly closely separated dual AGN, and our measurements represent the dual AGN fraction for the most luminous and largely separated systems, and (2) our upper-limit assumes there are no dual AGN in our sample, at the separations and flux-limits we are sensitive to (i.e., dual AGN rates below 4.8\% can statistically result in detecting 0 dual AGN out of our sample of 62). Regarding (2), our results show that the asymmetric \emph{Chandra} PSF at high OAA values impacts our measurements. For example, if we place a similar constraint on the dual AGN fraction using only the subset of observations with average OAA values below 3\arcmin, where no source had a measured \logBF greater than 0 at the 3$\sigma$ level (a sample size of 21), our upper-limit increases to $\sim$14\% at the 95\% confidence level. 

\par Quantifying the incompleteness of our measurement is complicated by the unknown underlying distributions of the flux ratios and separations of X-ray dual AGN in our redshift bin. There has yet to be a large sample of detected dual AGN beyond $z>2.5$ for which population statistics can be measured. A recent analysis using the Near-InfraRed Spectrograph (NIRSpec) on the James Webb Space Telescope (JWST) have claimed to find a dual AGN fraction of $\approx23\%$ between $3.0<z<5.5$, where rest-frame optical AGN diagnostics (namely, ``Baldwin, Phillips \& Terlevich'', or BPT; \citealt{Kewley2006}) were used to flag dual AGN candidates \citep{Perna2023}. Out of a sample of 17 AGN, 4 ``multiple AGN'' candidates were found (3 dual AGN candidates and 1 triple AGN candidate). All 4 candidates have observations in either COSMOS or CDFS, however they did not meet our sample criteria due to the off-axis angle of their observations, their redshift, and/or the low number of X-ray counts (1/4 sources has no X-ray detection in the COSMOS field). 
\par At face-value, the measurements presented in \cite{Perna2023} are inconsistent with ours -- assuming a dual AGN fraction of 23\%, we statistically expect to measure $>$1 dual AGN in our sample. To better understand if there exists a tension between our results and those presented in \cite{Perna2023}, we test whether or not we would be sensitive to detecting these 4 systems if their \emph{Chandra} observations more closely matched the average observation in our sample. We assume that their COSMOS or CDFS observations have OAA values within our selection criteria and that the exposure times match the average exposure time of the AGN in our sample, evaluated individually for COSMOS and CDFS. We create 100 simulations of each multiple AGN via \texttt{MARX}. We assume the measured (or, upper-limit) X-ray flux and hydrogen column density ($N_{H}$) as calculated in \cite{Marchesi2016} or \cite{Liu2017}, the separation and location of each AGN as presented in \cite{Perna2023}, and the average exposure time of 49 ks for COSMOS sources and 1267 ks for CDFS sources. We assign an OAA value to each simulation, between $0-3\arcmin$ via inverse transform sampling. We then run \BAYMAX on each simulation, and analyze the distribution of \logBF values in favor of the dual point source model.  
\par For all sources, we find that we are insensitive to strongly identifying them as multiple AGN. Modeling the \logBF distributions with a Gaussian profile, we find that all distributions are centered at values $<0$ and are consistent with $0$ within their 1$\sigma$ spread. Given the X-ray flux values of the sources, most (3/4) have simulations with $\sim$50 counts or less between $0.5-8$ keV counts. The exception is the triple AGN candidate, where the source has on average $\sim$200 X-ray counts. However, the separation between the primary and secondary AGN is $<1\arcsec$, which is difficult to probe at OAA$>1\arcmin$, and the third AGN is extremely dim in X-rays (where the upper-limit on the X-ray flux predicts less than 3 X-ray counts associated with the source). Thus, dual AGN sources with similar flux and separation values as presented in \cite{Perna2023} would likely be missed in our analysis, further emphasizing that the dual AGN fraction we present in this paper represents the brightest and most largely separated systems. 
\par An important note is that the dual AGN fraction presented in \cite{Perna2023} likely represents a different population of AGN than ours. The majority have redshifts greater than 3.5 (11/17; likely contributing to the low-count and/or non-detections in the X-ray datasets). Of the remaining 6, only 1 is a dual AGN candidate. Additionally, the majority have angular separations $\sim$1\arcsec or less, which we are insensitive to detecting for most of our high OAA observations. Overall, the frequency of dual AGN at high-$z$ and low-separation are likely to be enhanced with respect to their lower-$z$ and larger-separation counterparts.  

\subsection{Future X-ray Observatories}
\par The biggest limitation of our analysis is the difficulty in modeling the $\emph{Chandra}$ PSF at OAA values greater than 3\arcmin, as the loss of sensitivity in detecting duals increases as the asymmetries and size of the PSF increases. Future observations with X-ray missions such as \emph{AXIS} \citep{AXIS} or \emph{Lynx} \citep{Lynx} will revolutionize the study of observational dual AGN studies via improved PSFs and increased sensitivity. 
\par In particular, \emph{AXIS} is expected to greatly increase the sample size of known dual AGN at redshifts beyond $z=3$. Although the proposed on-axis angular resolution of \emph{AXIS} (PSF Half Energy Width$=$1.5\arcsec) is marginally larger than \emph{Chandra}'s on-axis angular resolution (PSF Half Energy Width$=$0.8\arcsec), the field-of-view average PSF is stable as a function of increasing off-axis angle (1.6\arcsec~up to OAA$=$7.5\arcmin) and is significantly smaller than \emph{Chandra}'s field-of-average ($\sim$5\arcsec~up to OAA$=$7.5\arcmin ~on ACIS-I). The \emph{AXIS} PSF, coupled with the effective area ($A_{eff}$) at 1 and 6 keV ($A_{eff\mathrm{,~1~keV}}=4200$ cm$^2$ and $A_{eff\mathrm{,~6~keV}}=830$ cm$^2$, as compared to ACIS at launch with $A_{eff\mathrm{,~1~keV}}=500$ cm$^2$ and $A_{eff\mathrm{,~6~keV}}=200$ cm$^2$), and 24\arcmin~diameter field of view (compared to ACIS-I with 16\arcmin~square field of view) will significantly increase the number of confirmed dual X-ray AGN. 
\par For example, a single 300 ks exposure with AXIS yields a sample size of 1000 AGN for which blind dual AGN searches down to 1.5\arcsec~can be carried out. In comparison, with a 300 ks ACIS-I observation, less than 20 AGN are expected to be detected within the field that has a PSF $<$1.5\arcsec. Assuming the dual fraction at $2.5<z<3.5$ is 3\%, we calculate how well \emph{AXIS} can constrain the dual AGN fraction assuming a single deep stare, matching the flux-limit of the deepest survey in our dataset (CDFS; $F_{2-7~\mathrm{keV}} = 8.7\times10^{-17}$ at 20\% completeness between 2$-$7 keV; \citealt{CDFS}). At this flux-limit, AXIS detects $>$1000 AGN with $L_{X}>10^{42}$ erg s$^{-1}$ within $2.5<z<3.5$, corresponding to approximately 40 dual AGN detections and tightly constraining the dual AGN fraction (within $<1\%$, see Fig.~\ref{fig:cosmosims}). The proposed mission's deep 5 Ms survey will detect sources that are an order of magnitude fainter (4.3$\times10^{-18}$ erg s$^{-1}$), and combined with additional wide-surveys, the probe is expected to detect hundreds to thousands of new dual AGN across the redshift range $0<z<4$. 

\section{Conclusions} \label{sec:conclusions}
We have presented a statistically rigorous search for dual AGN in deep \textit{Chandra} observations. This is done using \BAYMAX, a Bayesian statistical package that calculates a Bayes factor to determine whether a given \textit{Chandra} observation is better described by a single or dual point source. The results and observations of our study are summarized as follows:

\begin{enumerate}
\item We analyze 62 X-ray point sources identified in the fields X-UDS, AEGIS-XD, CDF-S, and COSMOS-Legacy. We enforce the following criteria when creating our sample from archival data: sources with $2.5 < z < 3.5$,  observations with OAA $< 5$\arcmin, and $\geq 50$ counts with energies between $0.5 - 8$ keV (across all observations). For C-COSMOS, we also exclude sources that have been spectroscopically identified as a star.
\item After carrying out false-positive tests, which quantifies the strength of the Bayes factor in favor of the dual point source model for a specific source, we find no strong evidence that any source in our sample is a dual point source candidate.
\item We test for the presence of dual X-ray point sources across our sample as a whole. We do this for sources where the PSF is most accurately modeled (sources with average OAA across all observations $<$ 3\arcmin). We compare the results from our sample to a distribution of measured Bayes factors for a similar sample composed of only single point sources. We find that the two distributions are consistent with one-another via both the Kolmogorov-Smirnov and Anderson-Darling tests.
\item Our non-detection of dual AGN activity allows us to place an upper-limit on the dual AGN fraction between $2.5<z<3.5$. Assuming confidence limites for Poisson statistics \citep{Gehrels1986}, we measure an upper-limit of 4.8\% at the 95\% confidence level. This upper-limit is in agreement with recent cosmological simulations, although we note our sensitivity in separation space is not uniform across our entire sample due to our sample's large range of OAA values. 
\end{enumerate}

Our ability to find and detect high-redshift dual AGN is largely limited by the difficulty in accurately modeling the \emph{Chandra} PSF at high OAA, the low number of counts associated with most high-redshift sources observed in these fields, and the loss of sensitivity to closely-separated dual X-ray point sources with increasing OAA values. We emphasize, however, that we have utilized one of the largest catalogs of high-redshift X-ray point sources available; moving forward, this type of analysis will continue to improve as future X-ray missions with highersensitivity and a more stable PSF than \emph{Chandra}, such as \emph{AXIS} or \emph{Lynx}. In a series of upcoming papers, we are extending this study to lower redshifts as a part of a larger effort to measure the dual AGN frequency across a wide range of redshift.

\section*{Data Availability}
All \textit{Chandra} X-ray data used in this work are available from the Chandra Data Archive at \url{https://cxc.harvard.edu/cda/}.

\section*{Acknowledgements}

The scientific results reported in this article are based on observations
made by the Chandra X-ray Observatory and data obtained from the Chandra Data Archive. Support for this work was provided by the National Aeronautics and Space Administration through Chandra Awards Number AR1-22007X, issued by the Chandra X-ray Center, which is operated by the Smithsonian Astrophysical Observatory for and on behalf of the National Aeronautics Space Administration under contract NAS8-03060. The work of A.S. was supported by the National Science Foundation MPS-Ascend Postdoctoral Research Fellowship under Grant No. 2213288. This research has made use of software provided by the Chandra X-ray Center (CXC) in the application packages CIAO.

\section*{Software}
 \noindent {\tt CIAO} (v4.15; \citealt{Fruscione2006}), \newline XSPEC (v12.9.0; \citealt{Arnaud1996}), \newline {\tt nestle}~(https://github.com/kbarbary/nestle), \newline {\tt PyMC3} \citep{Salvatier2016}, \newline {\tt MARX} (v5.3.3; \citealt{Davis2012})

\appendix

\section{Sample Information}
\begin{deluxetable*}{ccccccc}[ht!]
\tablenum{2}
\tablecaption{Sample Properties\label{tab:sample}}
\tablehead{
\colhead{Name} & \colhead{RA} & \colhead{Dec} & \colhead{Redshift} & \colhead{Observation ID} & \colhead{Exposure} & \colhead{Off-Axis Angle}\\
& \colhead{(deg)} & \colhead{(deg)} & & &\colhead{(ks)} & \colhead{(arcmin)}
}
\startdata
AEGISXD214.447221+52.586265 & 214.447221 & +52.586265 & 2.745 & 9459 & 69.55&3.797\\
\dots & \dots & \dots & \dots & 9738 & 61.39&3.797\\
 \dots & \dots & \dots & \dots & 9736 & 49.48&3.797\\
 \dots & \dots & \dots & \dots & 9737 & 49.48&3.797\\
 \dots & \dots & \dots & \dots & 9734 & 49.47&3.797\\
 \dots & \dots & \dots & \dots & 9735 & 49.47&3.797\\
\dots & \dots & \dots & \dots & 9739 & 42.59&3.797\\
 \dots & \dots & \dots & \dots & 10769 & 26.68&3.797\\
 \dots & \dots & \dots & \dots & 9461 & 23.73&3.797\\
 \dots & \dots & \dots & \dots & 10896 & 23.29&3.797\\
 \enddata

\tablecomments{Columns: (1) Galaxy Name; (2) the central R.A. of the X-ray AGN; (3) the central declination of the X-ray AGN; (4) the redshift of the X-ray AGN; (5) the \emph{Chandra X-ray Observatory} observation ID used in analysis; (6) the exposure time of the observation ID in kiloseconds; (7) the off-axis angle between the AGN coordinate and the nominal pointing for each observation ID. Table 2 is published in its entirety in the machine-readable format. A portion is shown here for guidance regarding its form and content.}
\end{deluxetable*}

\begin{figure*}[h!]
     \centering
    \begin{subfigure}
         \centering
         \includegraphics[width=0.33\textwidth]{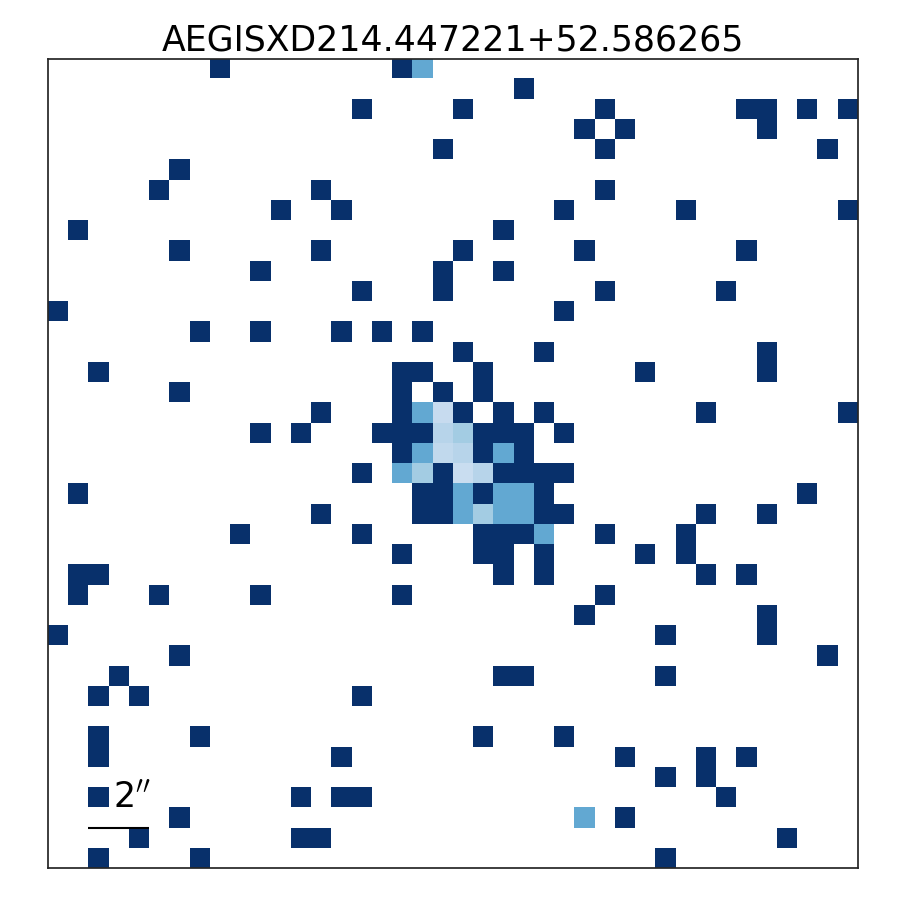}
    \end{subfigure}
    \hspace{-1cm}
    \hfill
    \begin{subfigure}
         \centering
         \includegraphics[width=0.33\textwidth]{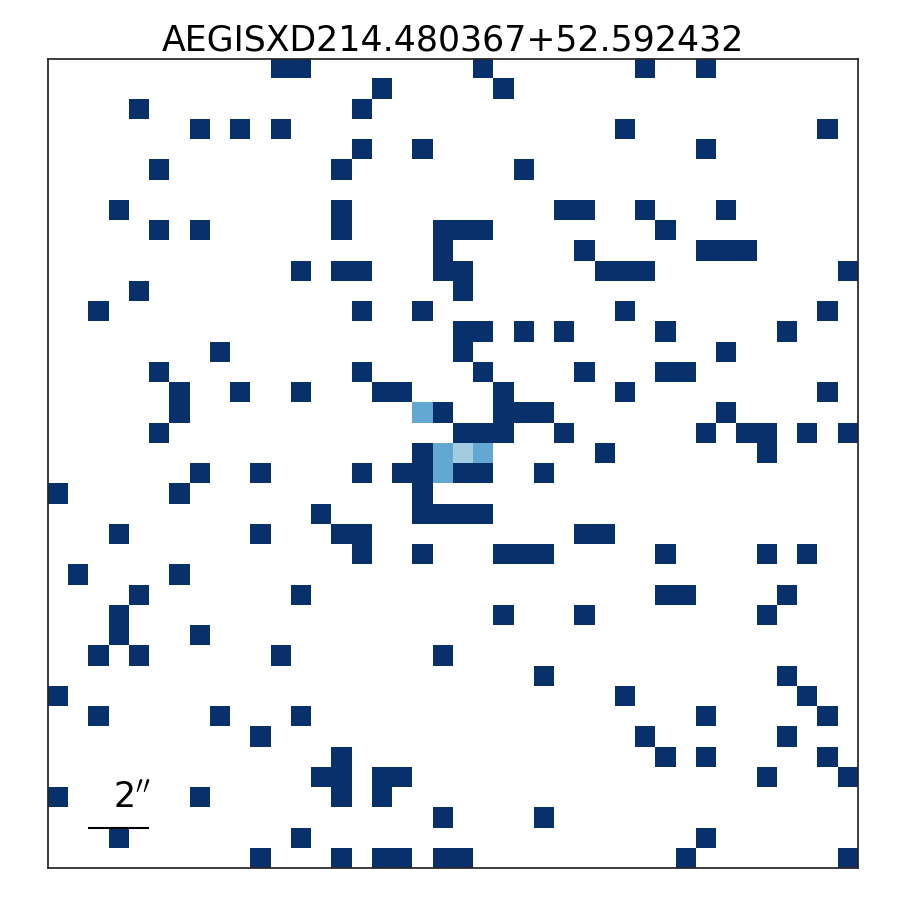}
    \end{subfigure}
    \hspace{-1cm}
    \hfill
    \begin{subfigure}
         \centering
         \includegraphics[width=0.33\textwidth]{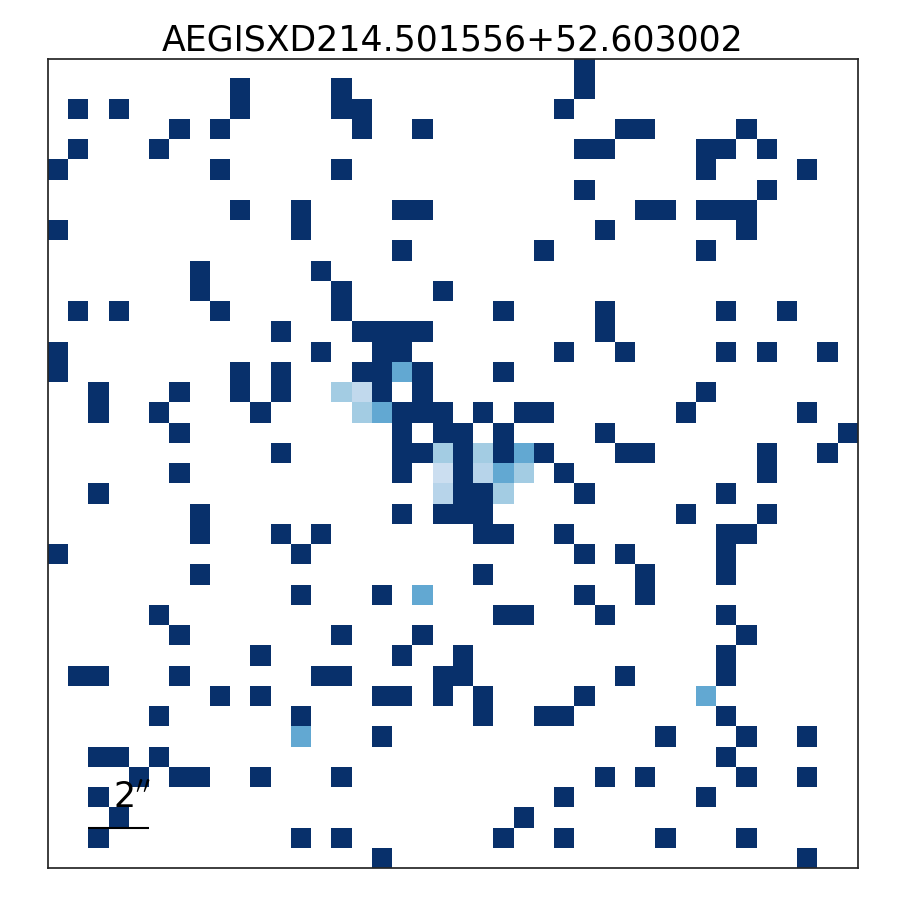}
     \end{subfigure}
        \hfill
    
    \vspace{-0.3cm}
         (1)$\dagger$ \hspace{5.3cm} (2)  \hspace{5.3cm} (3)  

    \begin{subfigure}
         \centering
         \includegraphics[width=0.33\textwidth]{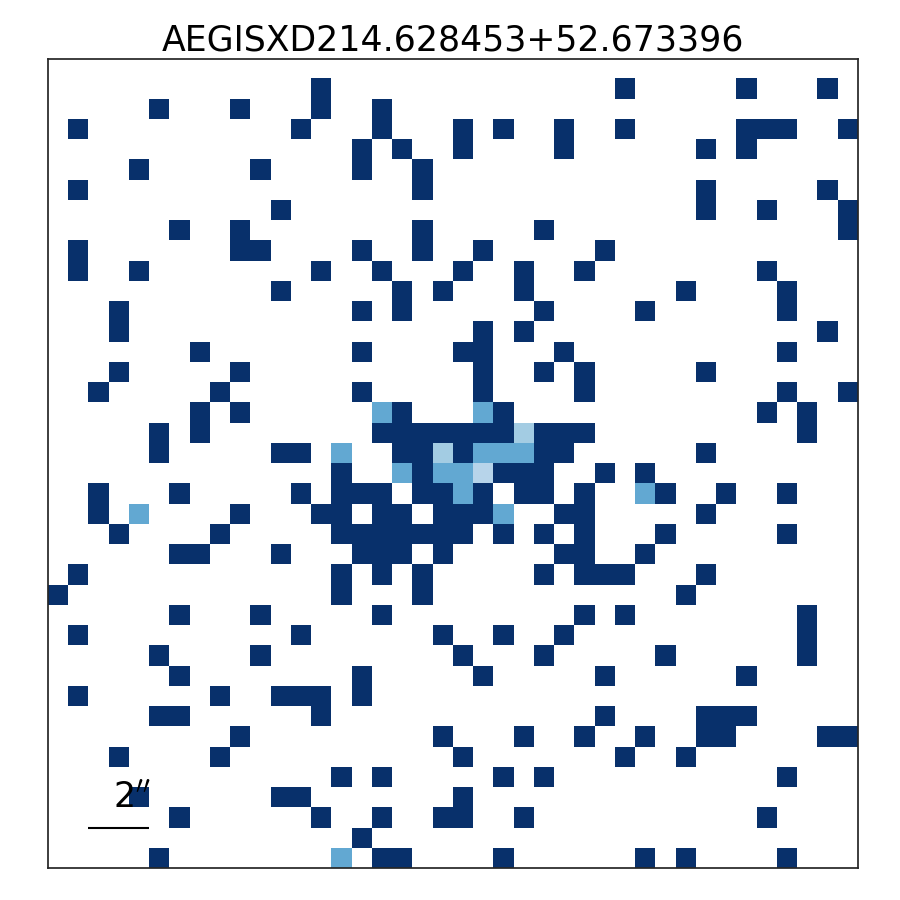}
    \end{subfigure}
    \hspace{-1cm}
    \hfill
    \begin{subfigure}
         \centering
         \includegraphics[width=0.33\textwidth]{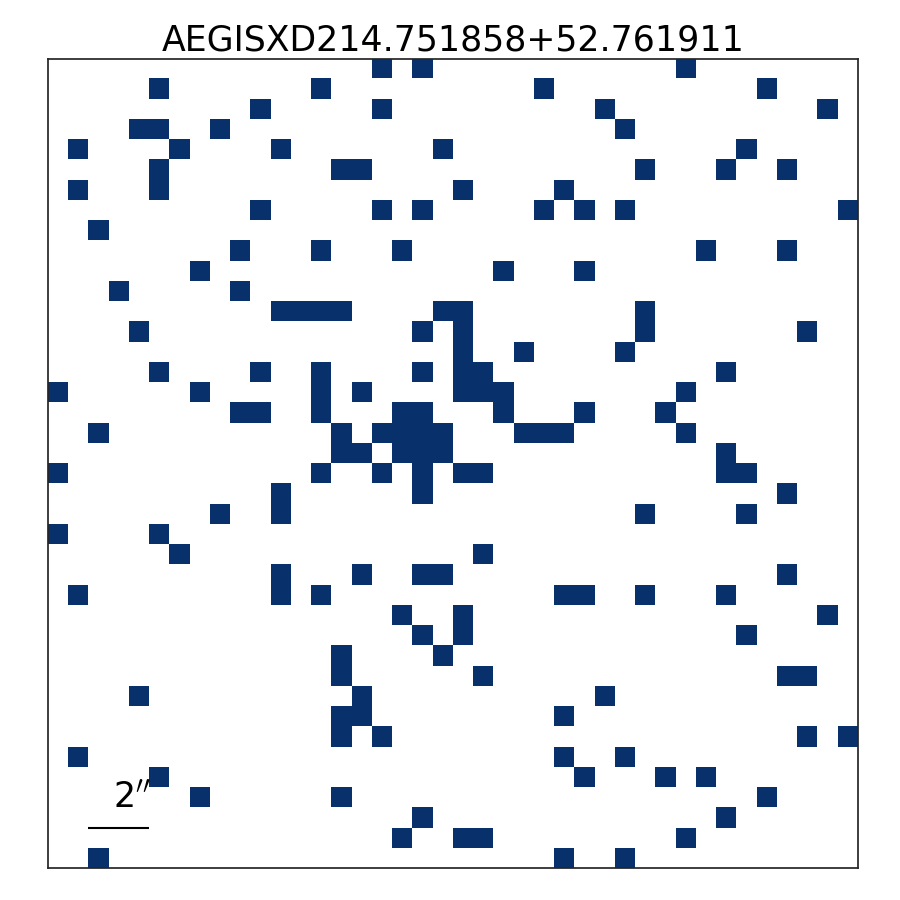}
    \end{subfigure}
    \hspace{-1cm}
    \hfill
    \begin{subfigure}
         \centering
         \includegraphics[width=0.33\textwidth]{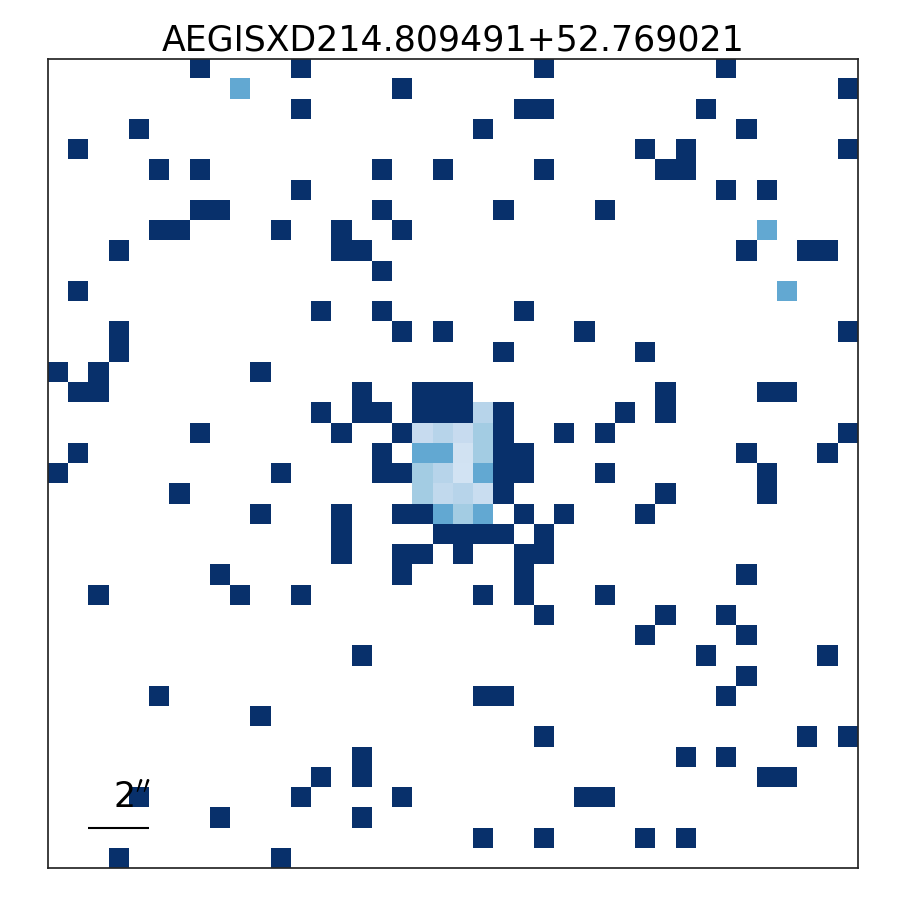}
     \end{subfigure}
        \hfill

    \vspace{-0.3cm}
    (4) \hspace{5.3cm} (5)  \hspace{5.3cm} (6)$\dagger$  
    
    \begin{subfigure}
         \centering
         \includegraphics[width=0.33\textwidth]{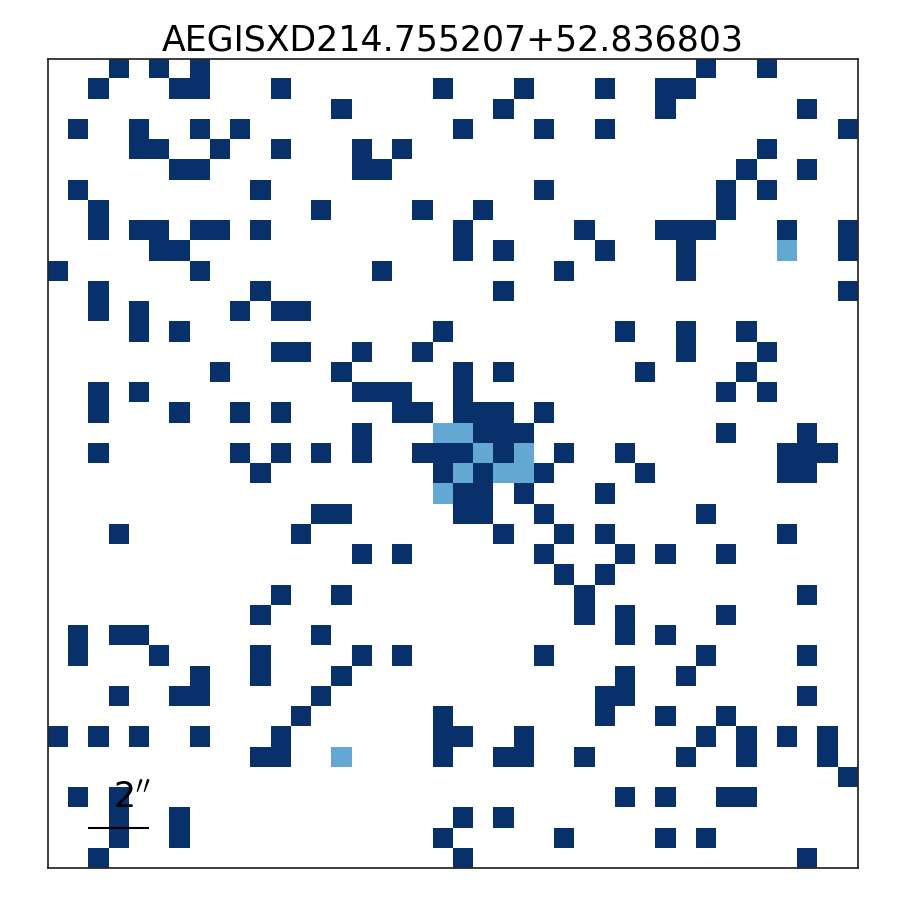}
    \end{subfigure}
    \hspace{-1cm}
    \hfill
    \begin{subfigure}
         \centering
         \includegraphics[width=0.33\textwidth]{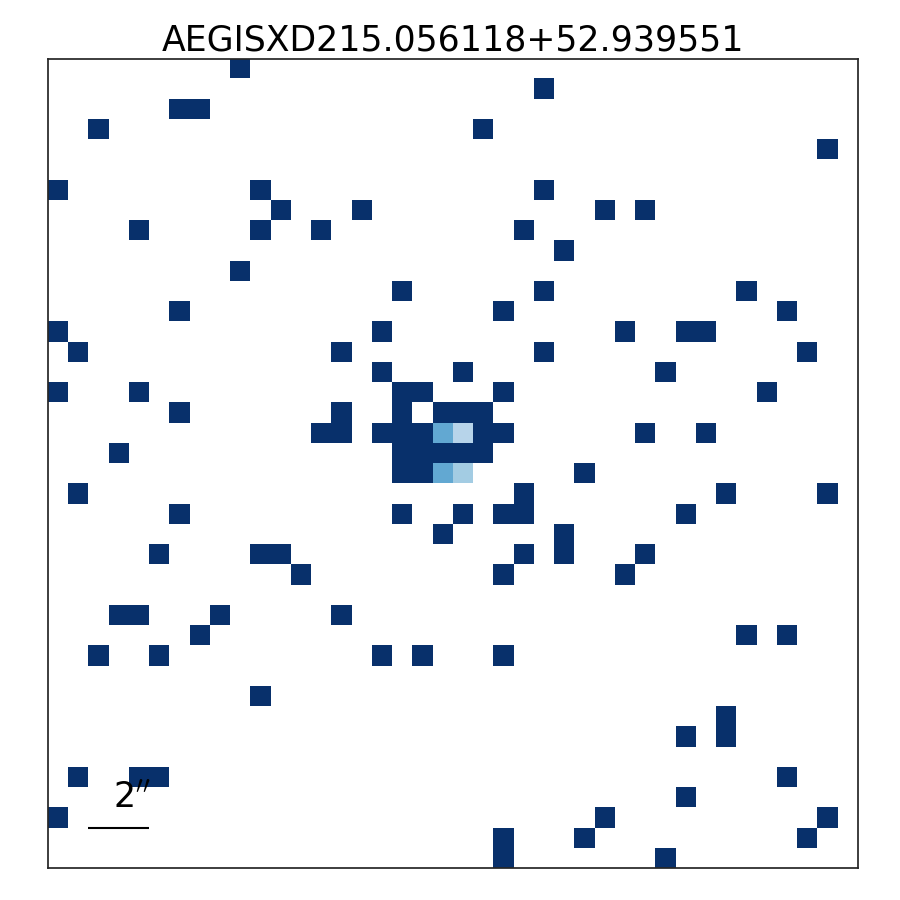}
    \end{subfigure}
    \hspace{-1cm}
    \hfill
    \begin{subfigure}
         \centering
         \includegraphics[width=0.33\textwidth]{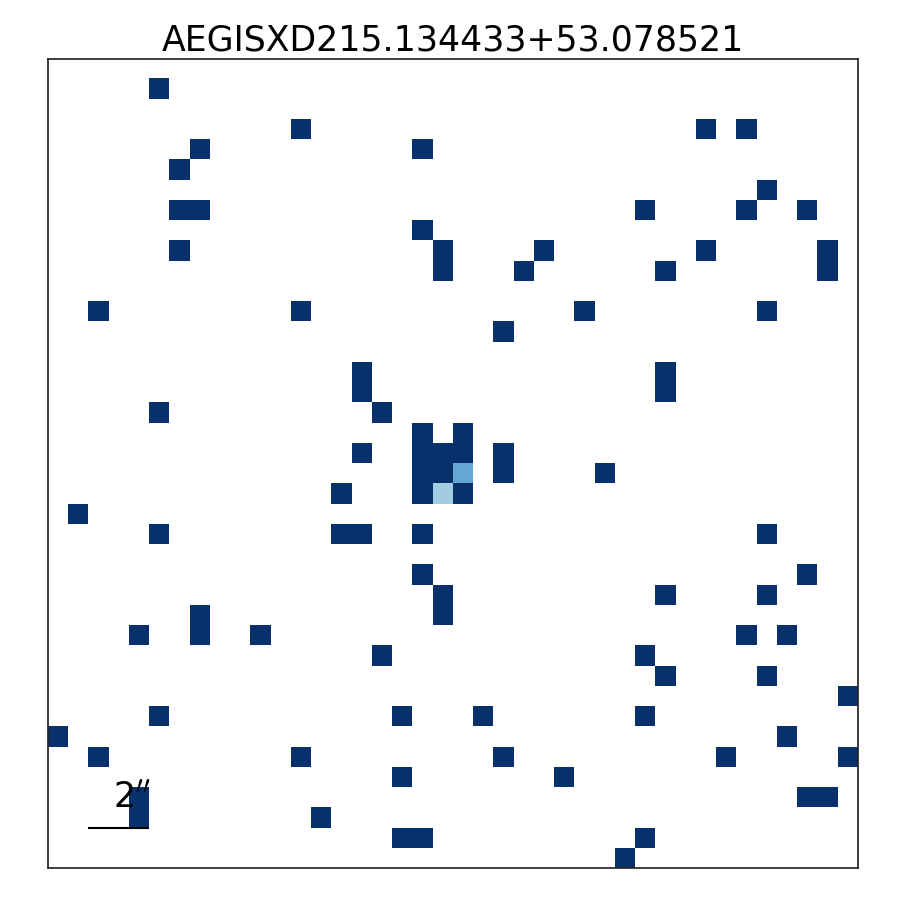}
     \end{subfigure}
        \hfill
        
    \vspace{-0.3cm}
    (7) \hspace{5.3cm} (8)  \hspace{5.3cm} (9) 

\RaggedRight{\textbf{Figure 5:} The stacked 0.5$-$8 keV \emph{Chandra} observations analyzed for each AGN. We analyze the photons in a 20$\arcsec$ $\times$ 20$\arcsec$ box on the nominal X-ray coordinate listed in the point source catalogs, with the exception of AEGISXD214.93+52.77 (7) and XUDS34.52-5.06 (64). These sources contained a second, bright source within the 20$\arcsec$ $\times$ 20$\arcsec$ field of view with a redshift measurement inconsistent ($>$ 300 km$^{-2}$) with the primary AGN being analyzed. For these two sources, we analyzed a smaller FOV that avoids these secondary nearby sources. We denote sources that have \logBF that is greater than 0 at the 3$\sigma$ level with $\dagger$ (see Table 1). False positive testing shows that these values are consistent with \logBF values expected from a single point source (see Section 4.1).} 
\label{fig:binned}
\end{figure*}

\begin{figure*}
     \centering
    \begin{subfigure}
         \centering
         \includegraphics[width=0.33\textwidth]{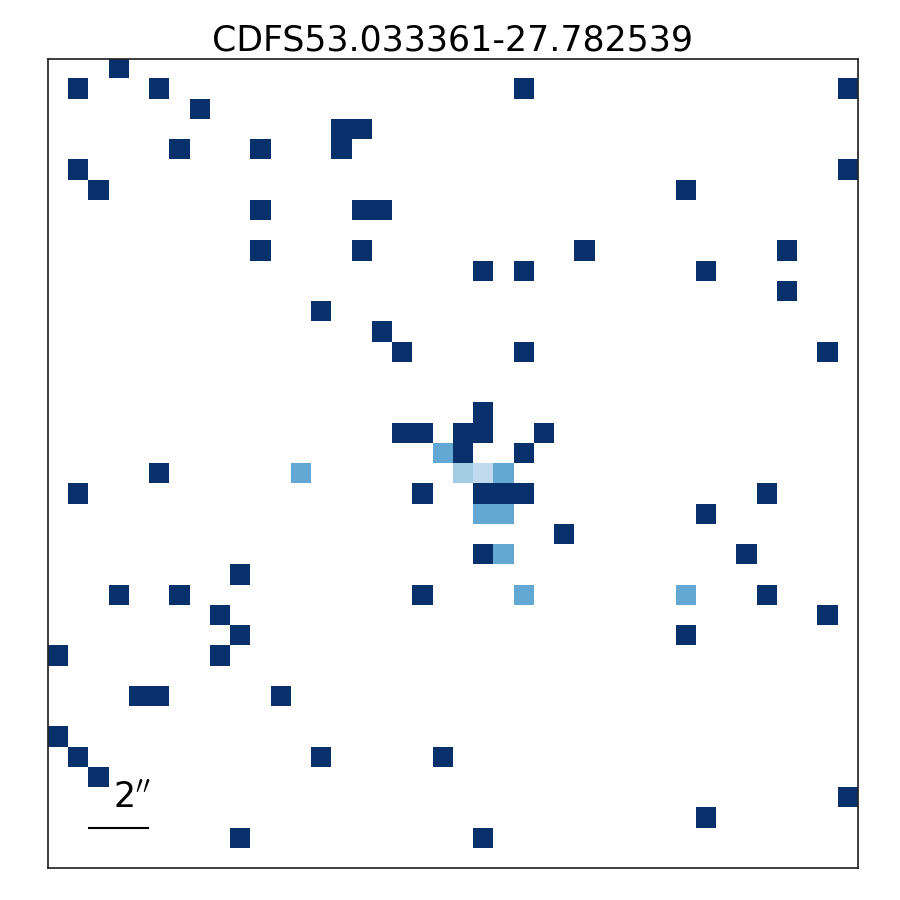}
    \end{subfigure}
    \hspace{-1cm}
    \hfill
    \begin{subfigure}
         \centering
         \includegraphics[width=0.33\textwidth]{CDFS1_binnedCXO.png}
    \end{subfigure}
    \hspace{-1cm}
    \hfill
    \begin{subfigure}
         \centering
         \includegraphics[width=0.33\textwidth]{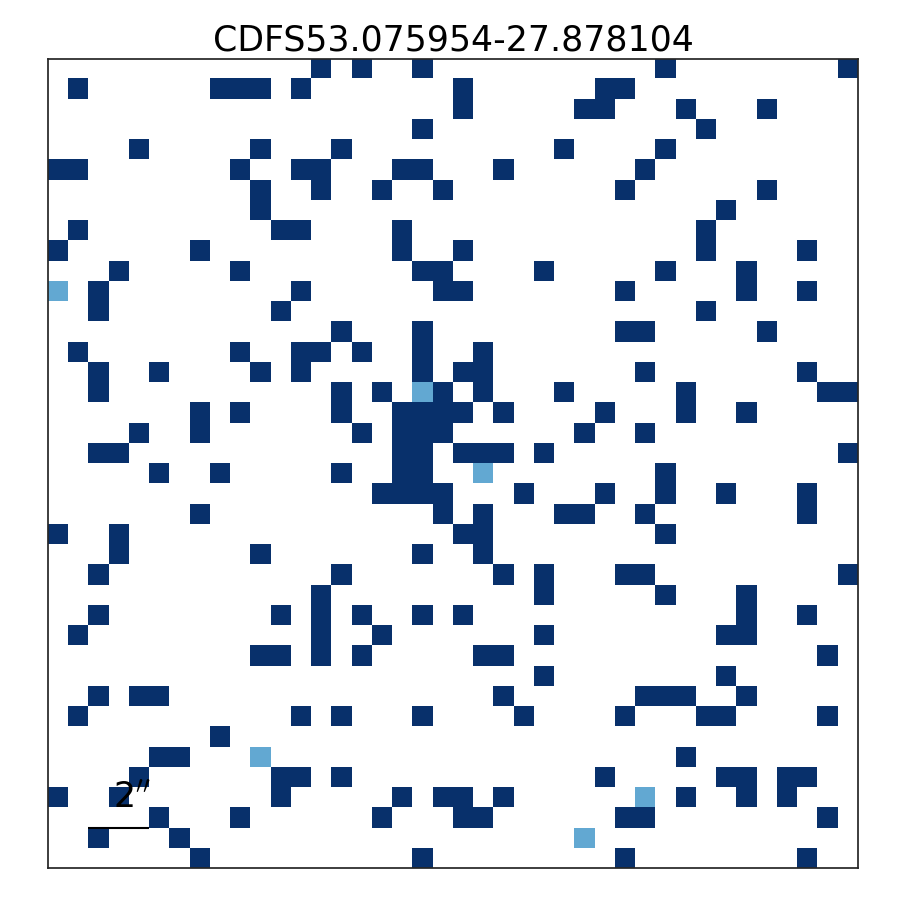}
     \end{subfigure}
        \hfill

    \vspace{-0.3cm}
    (10)$\dagger$  \hspace{5.3cm} (11)$\dagger$  \hspace{5.3cm} (12)

    \begin{subfigure}
         \centering
         \includegraphics[width=0.33\textwidth]{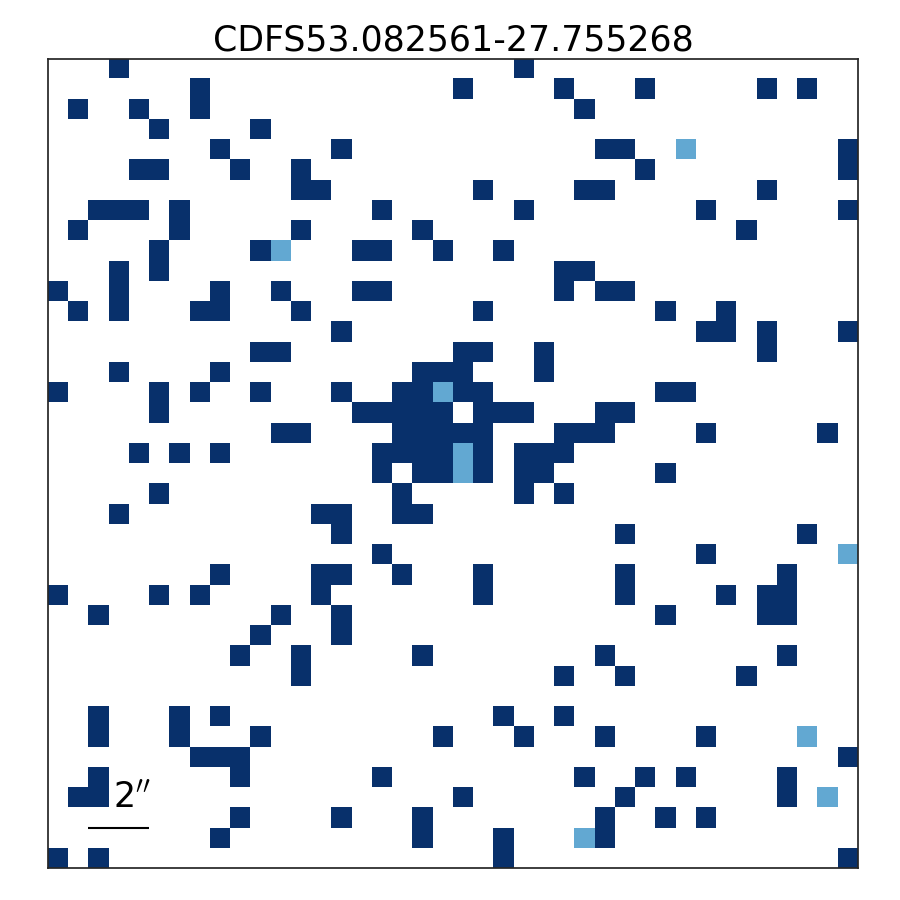}
    \end{subfigure}
    \hspace{-1cm}
    \hfill
    \begin{subfigure}
         \centering
         \includegraphics[width=0.33\textwidth]{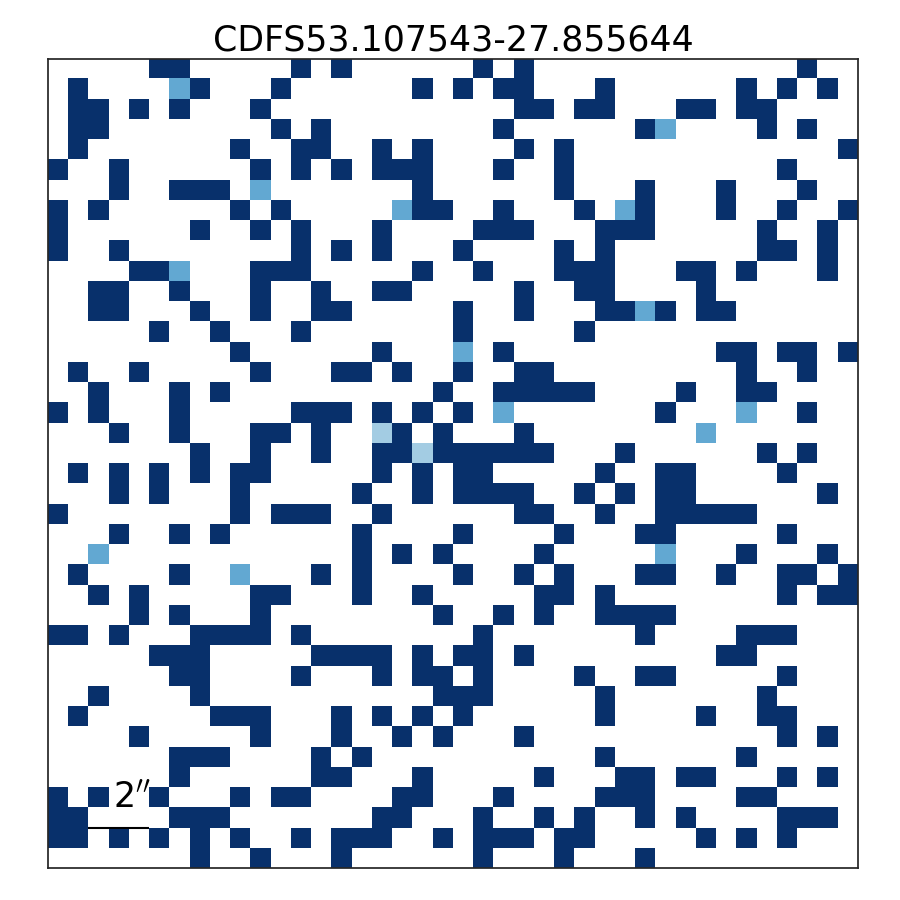}
    \end{subfigure}
    \hspace{-1cm}
    \hfill
    \begin{subfigure}
         \centering
         \includegraphics[width=0.33\textwidth]{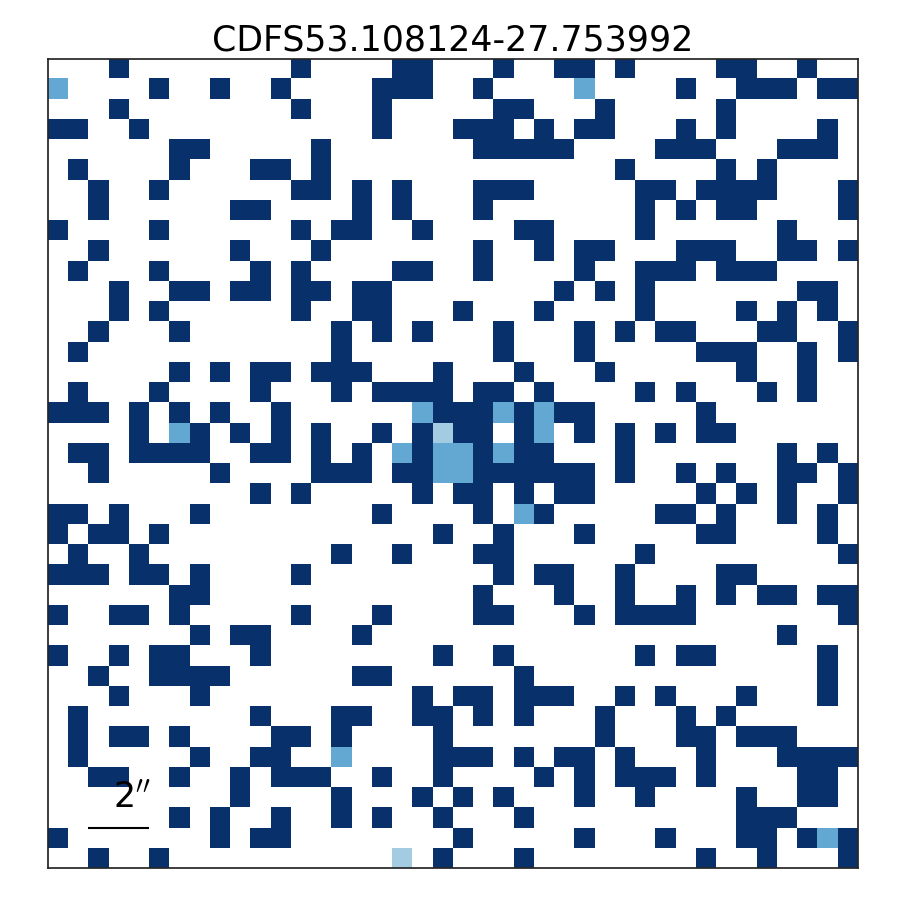}
     \end{subfigure}
        \hfill

    \vspace{-0.3cm}
    (13) \hspace{5.3cm} (14)  \hspace{5.3cm} (15)
    
    \begin{subfigure}
         \centering
         \includegraphics[width=0.33\textwidth]{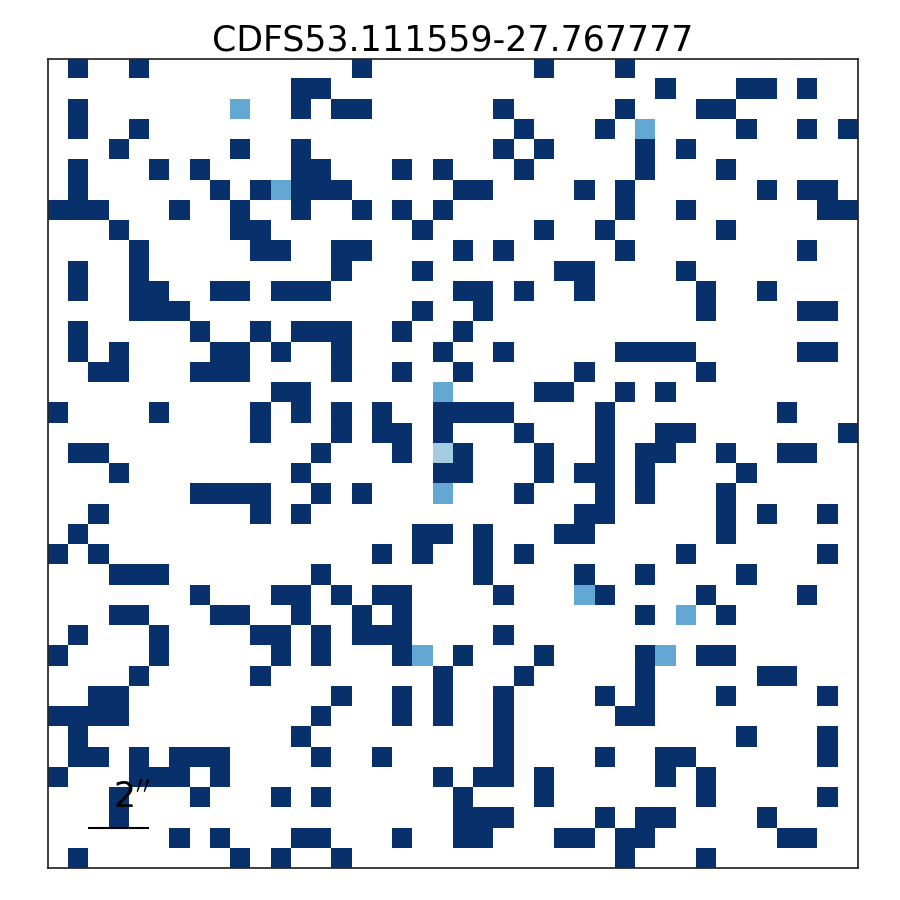}
    \end{subfigure}
    \hspace{-1cm}
    \hfill
    \begin{subfigure}
         \centering
         \includegraphics[width=0.33\textwidth]{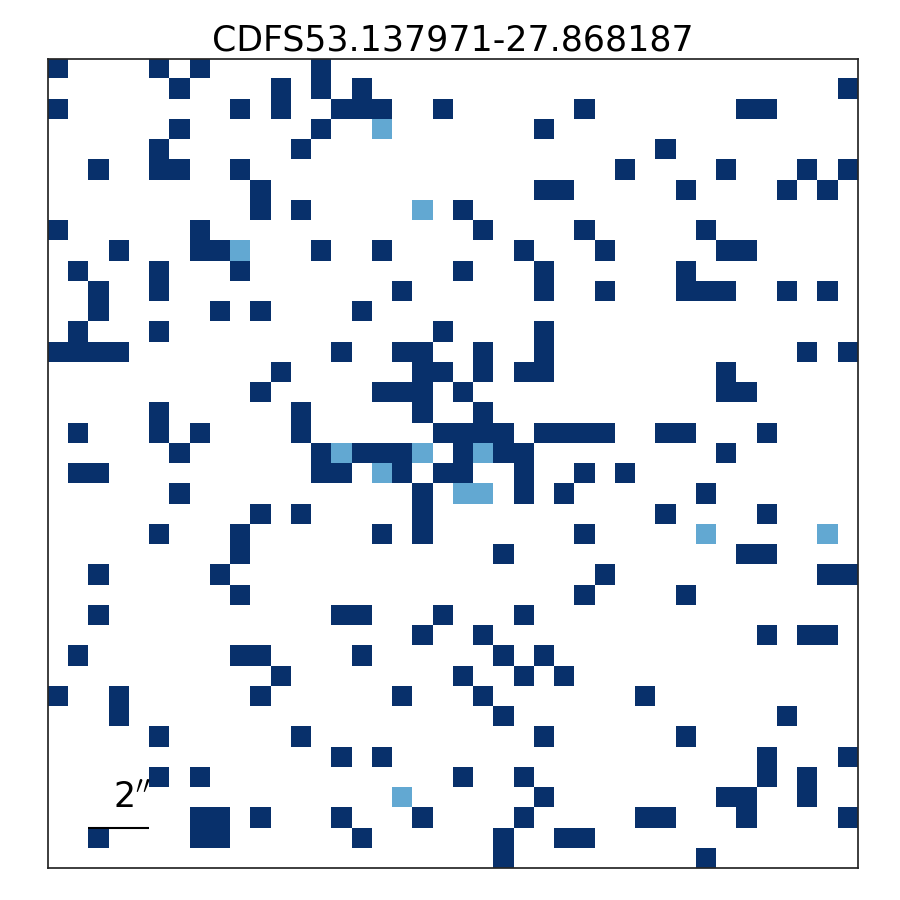}
    \end{subfigure}
    \hspace{-1cm}
    \hfill
    \begin{subfigure}
         \centering
         \includegraphics[width=0.33\textwidth]{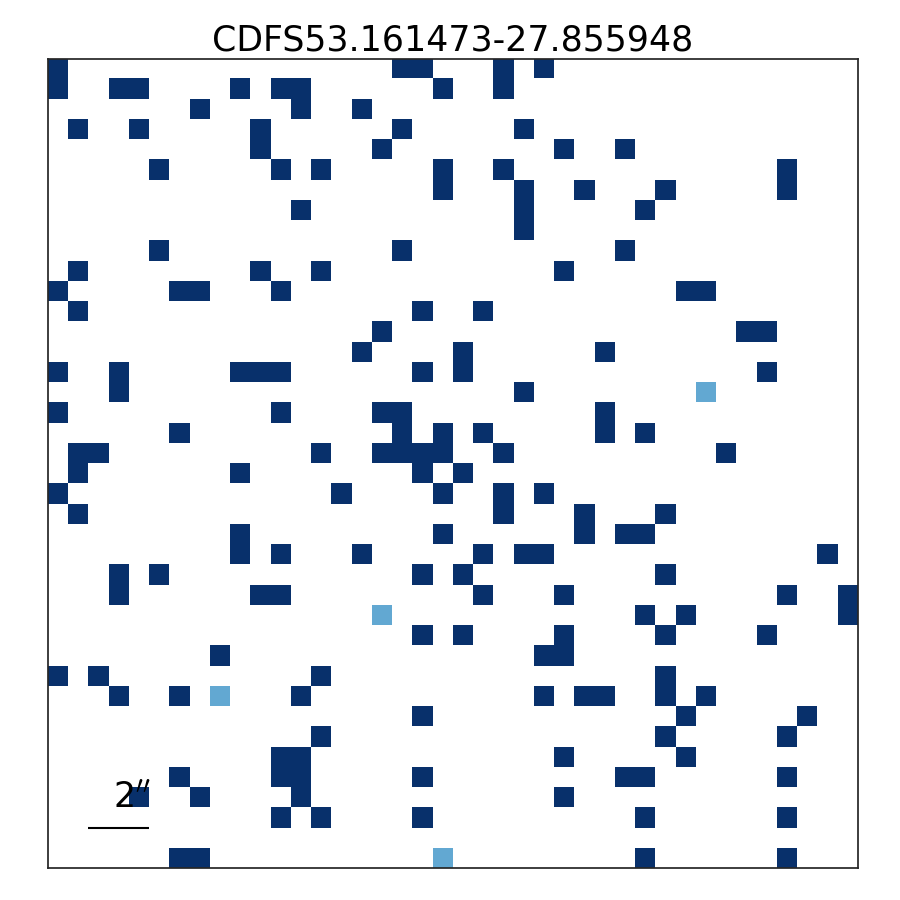}
     \end{subfigure}
        \hfill

    \vspace{-0.3cm}
    (16) \hspace{5.3cm} (17)  \hspace{5.3cm} (18)
    
    \begin{subfigure}
         \centering
         \includegraphics[width=0.33\textwidth]{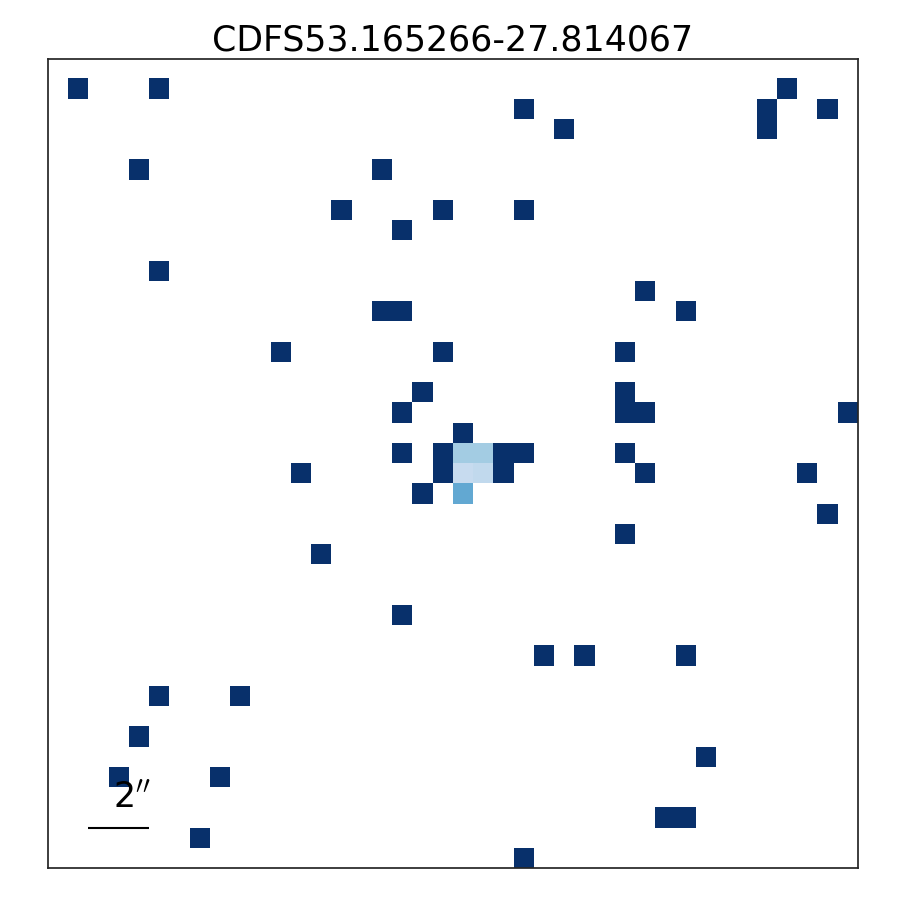}
    \end{subfigure}
    \hspace{-1cm}
    \hfill
    \begin{subfigure}
         \centering
         \includegraphics[width=0.33\textwidth]{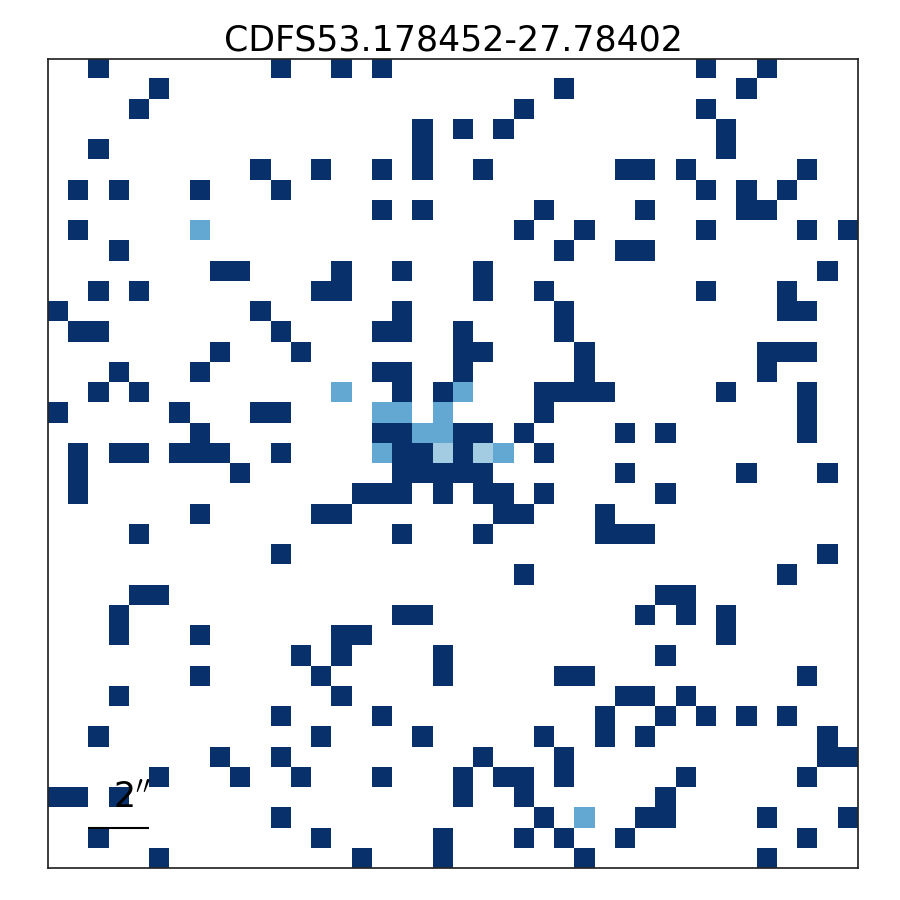}
    \end{subfigure}
    \hspace{-1cm}
    \hfill
    \begin{subfigure}
         \centering
         \includegraphics[width=0.33\textwidth]{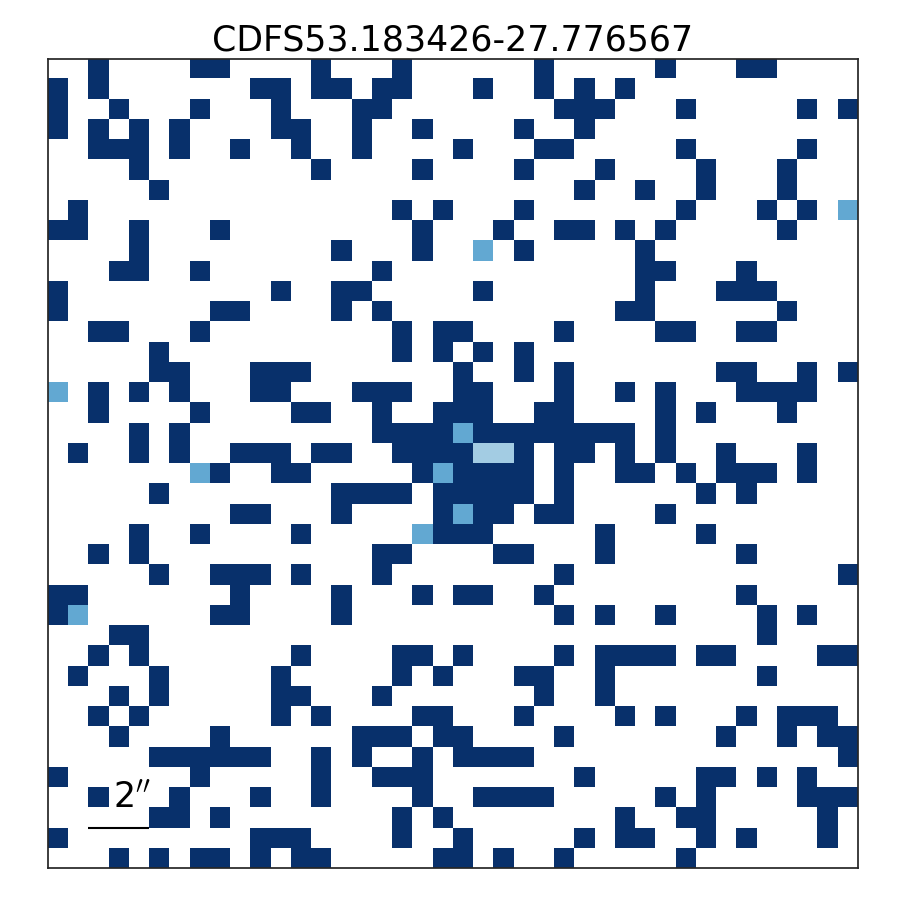}
     \end{subfigure}
        \hfill
        
    \vspace{-0.3cm}
    (19) \hspace{5.3cm} (20)  \hspace{5.3cm} (21) \\
    
\RaggedRight{\textbf{Figure 5.} (continued)} 
\label{}
\end{figure*}

\begin{figure*}
     \centering
     
    \begin{subfigure}
         \centering
         \includegraphics[width=0.33\textwidth]{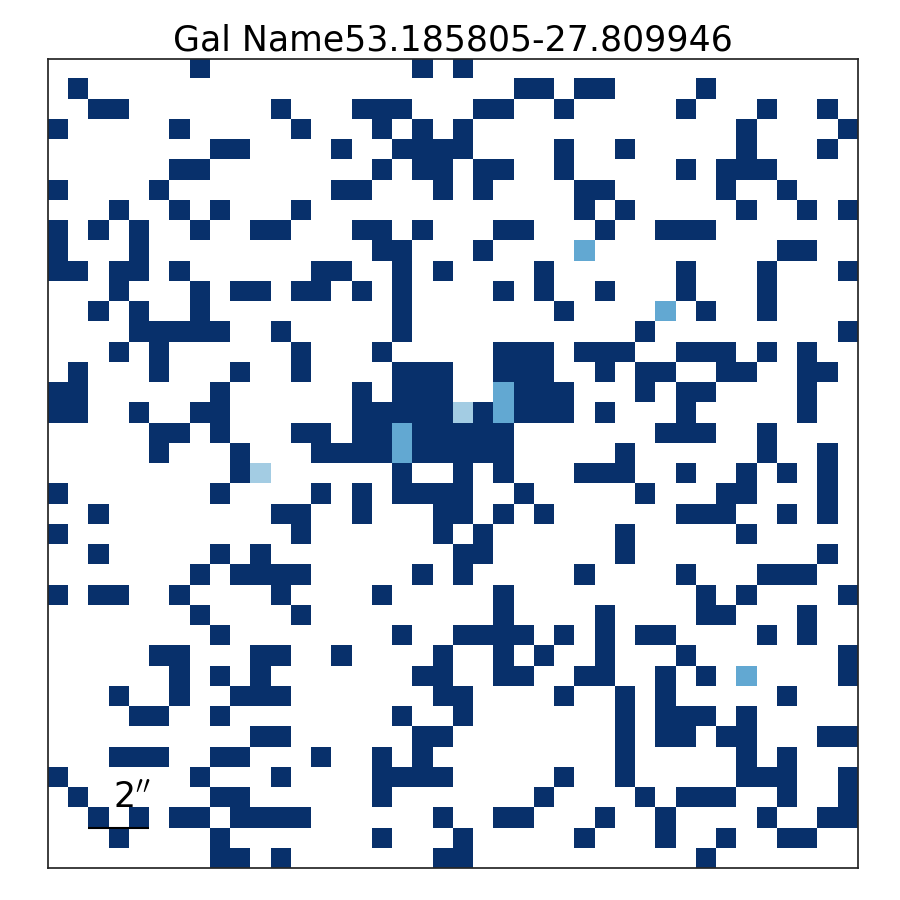}
    \end{subfigure}
    \hspace{-1cm}
    \hfill
    \begin{subfigure}
         \centering
         \includegraphics[width=0.33\textwidth]{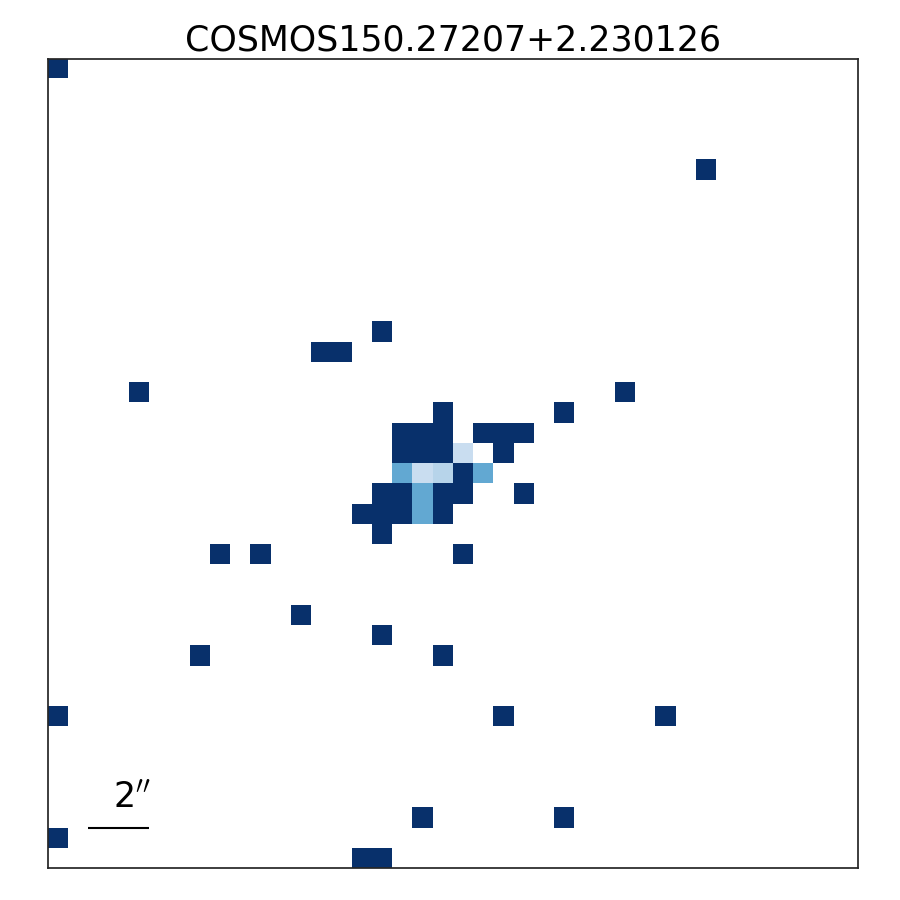}
    \end{subfigure}
    \hspace{-1cm}
    \hfill
    \begin{subfigure}
         \centering
         \includegraphics[width=0.33\textwidth]{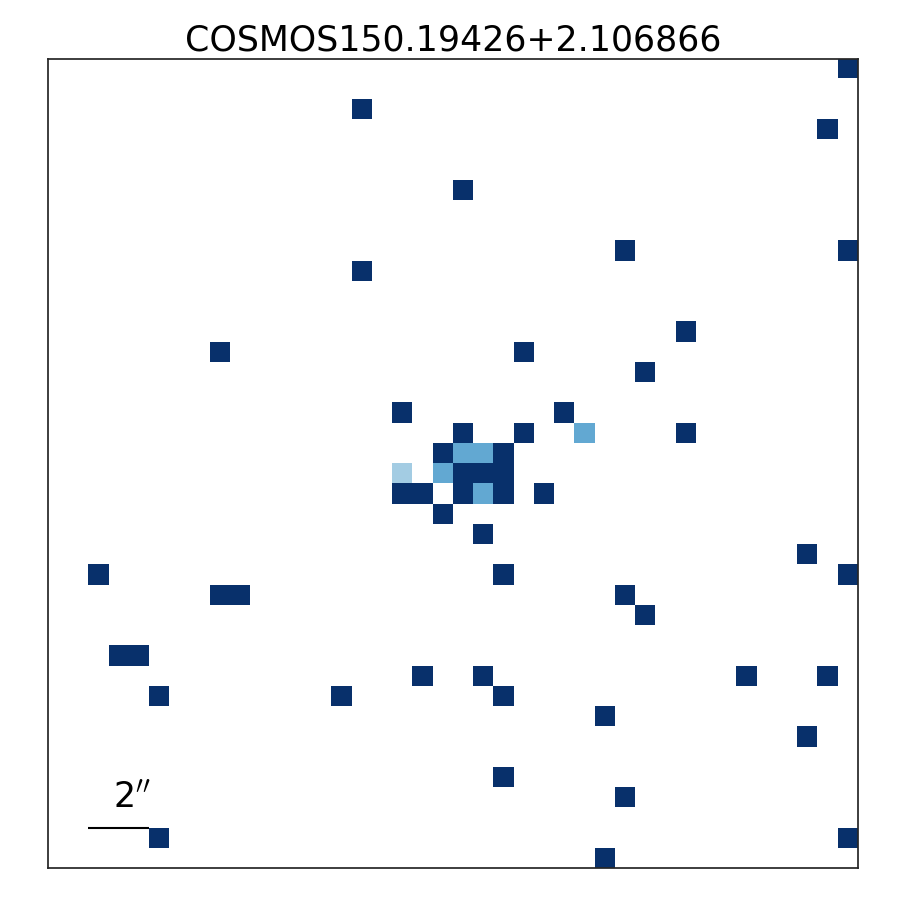}
     \end{subfigure}
        \hfill

    \vspace{-0.3cm}
    (22) \hspace{5.3cm} (23)  \hspace{5.3cm} (24)
    
    \begin{subfigure}
         \centering
         \includegraphics[width=0.33\textwidth]{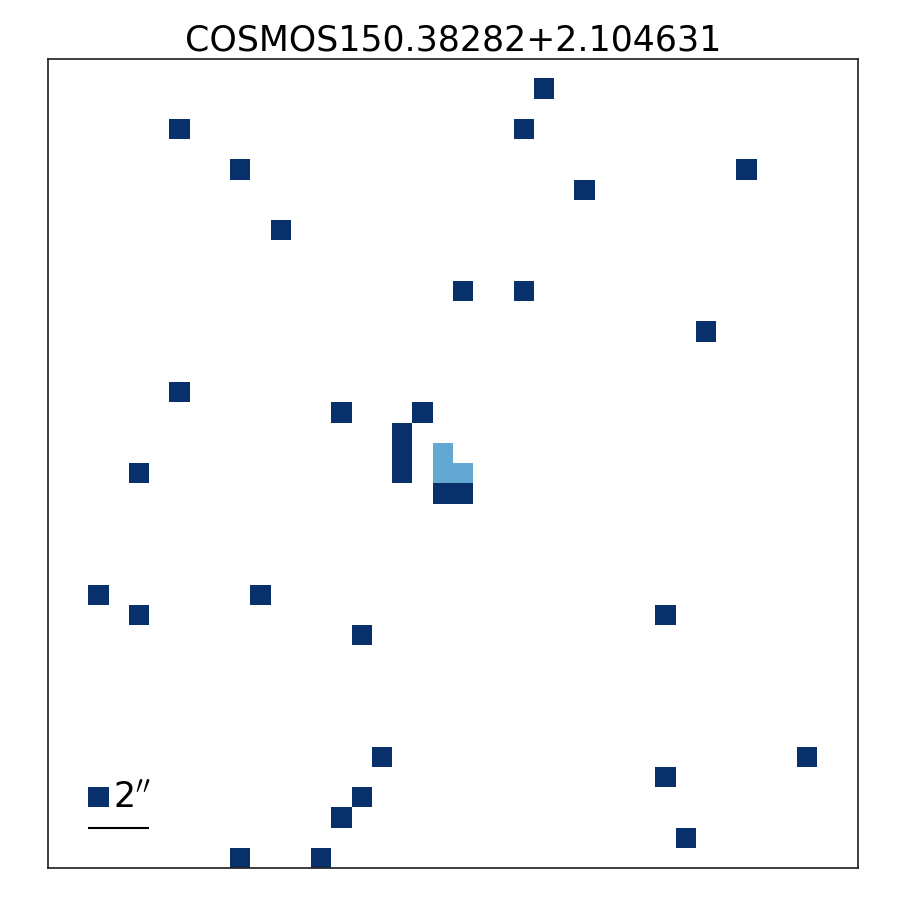}
    \end{subfigure}
    \hspace{-1cm}
    \hfill
    \begin{subfigure}
         \centering
         \includegraphics[width=0.33\textwidth]{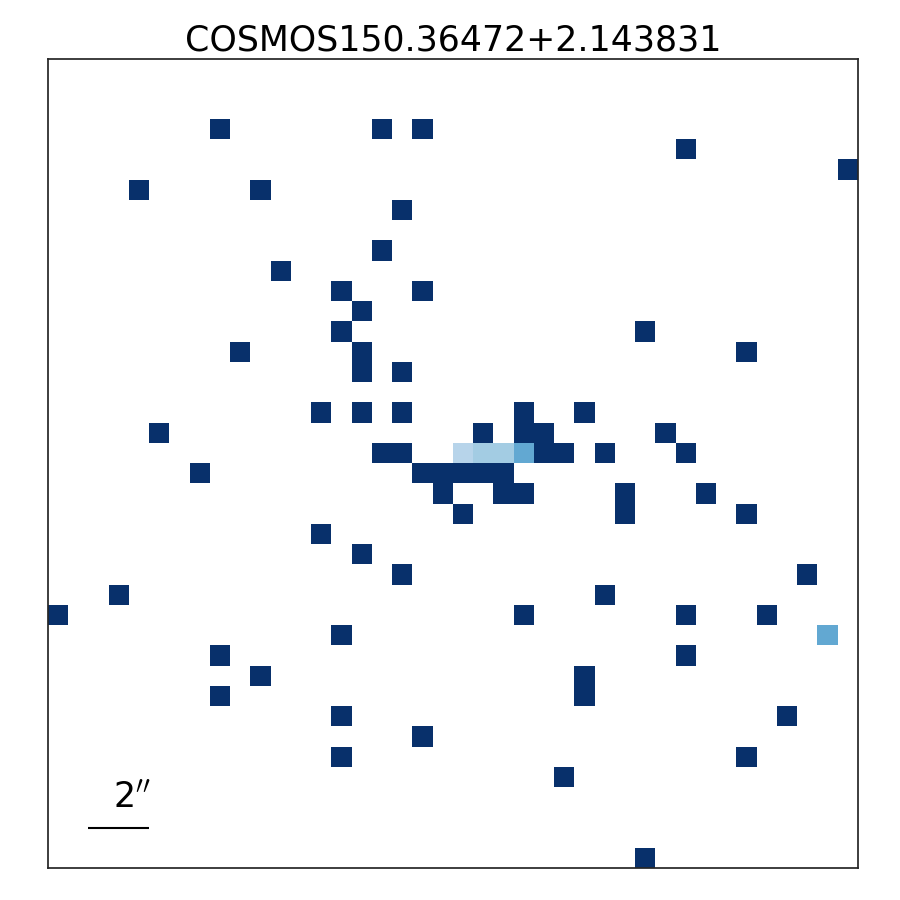}
    \end{subfigure}
    \hspace{-1cm}
    \hfill
    \begin{subfigure}
         \centering
         \includegraphics[width=0.33\textwidth]{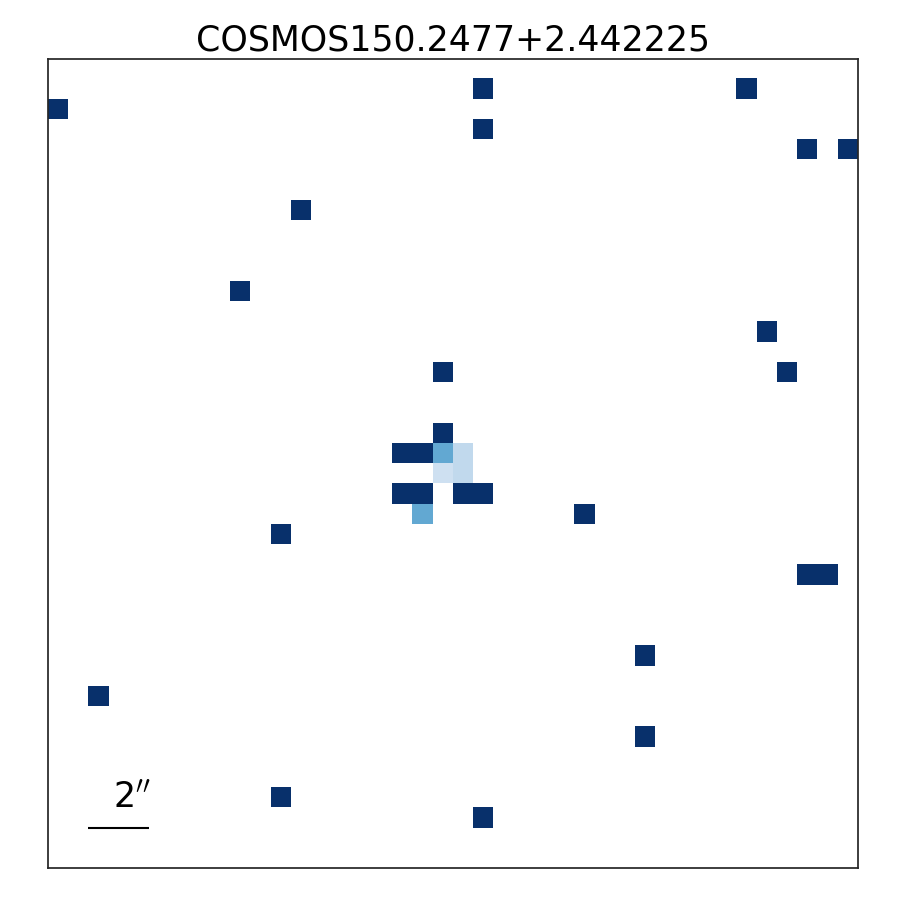}
     \end{subfigure}
        \hfill

    \vspace{-0.3cm}
    (25) \hspace{5.3cm} (26)  \hspace{5.3cm} (27)

       \begin{subfigure}
         \centering
         \includegraphics[width=0.33\textwidth]{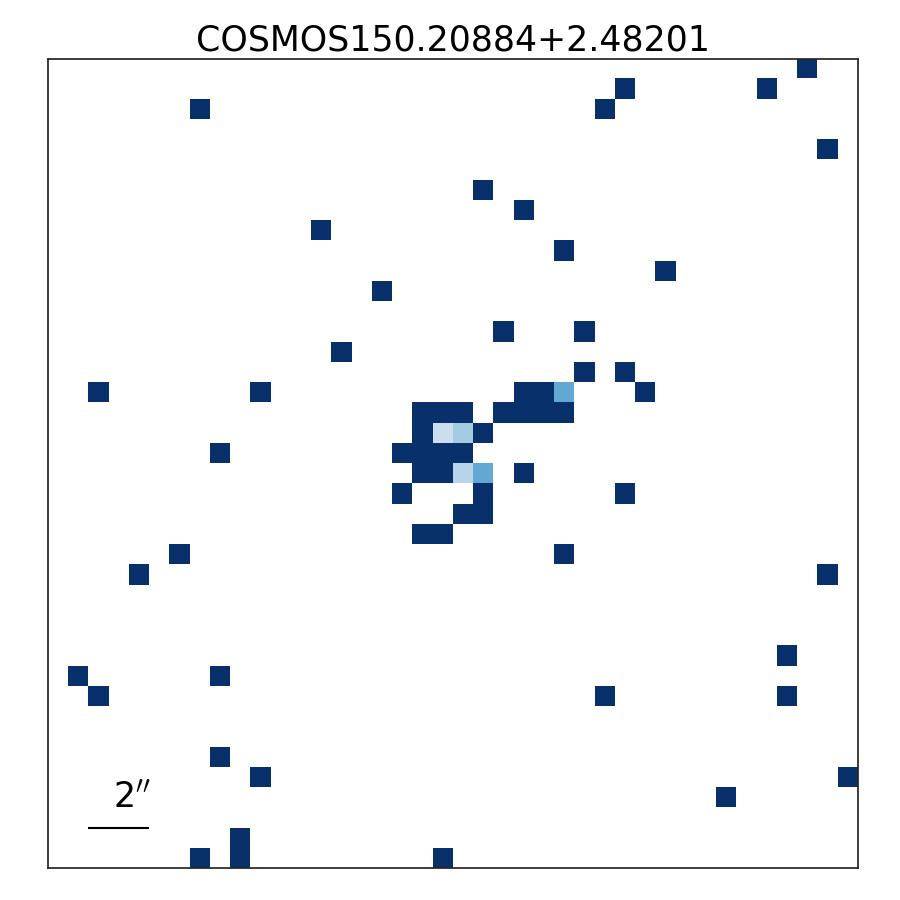}
    \end{subfigure}
    \hspace{-1cm}
    \hfill
    \begin{subfigure}
         \centering
         \includegraphics[width=0.33\textwidth]{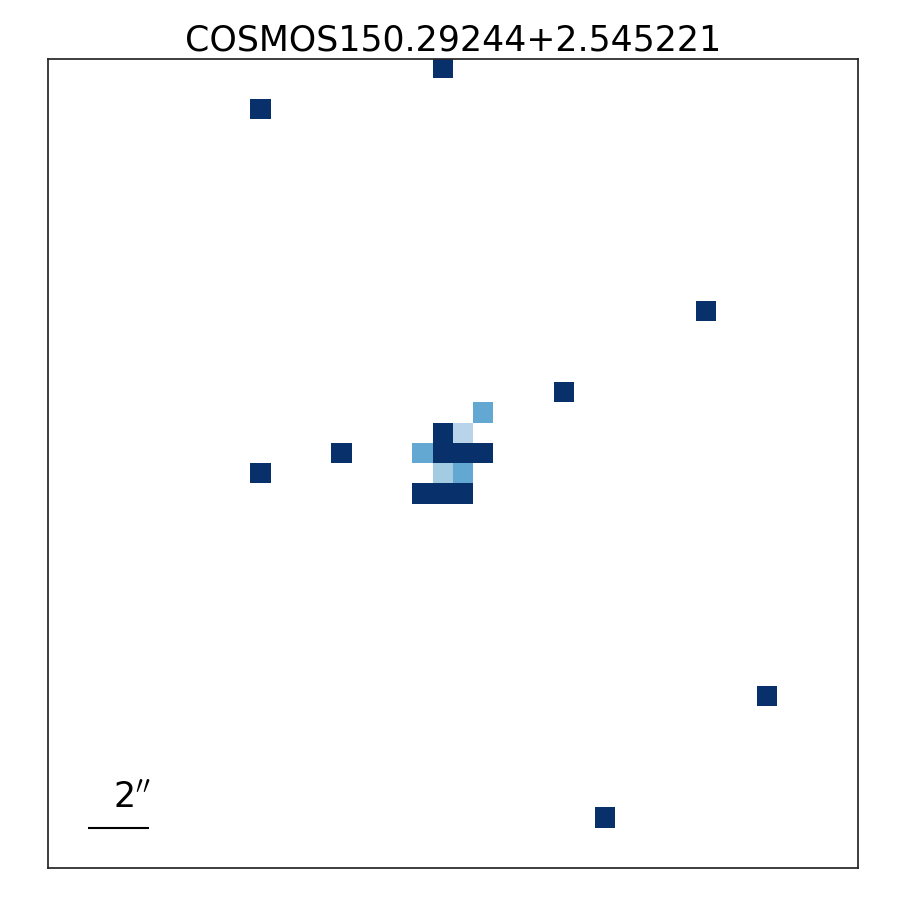}
    \end{subfigure}
    \hspace{-1cm}
    \hfill
    \begin{subfigure}
         \centering
         \includegraphics[width=0.33\textwidth]{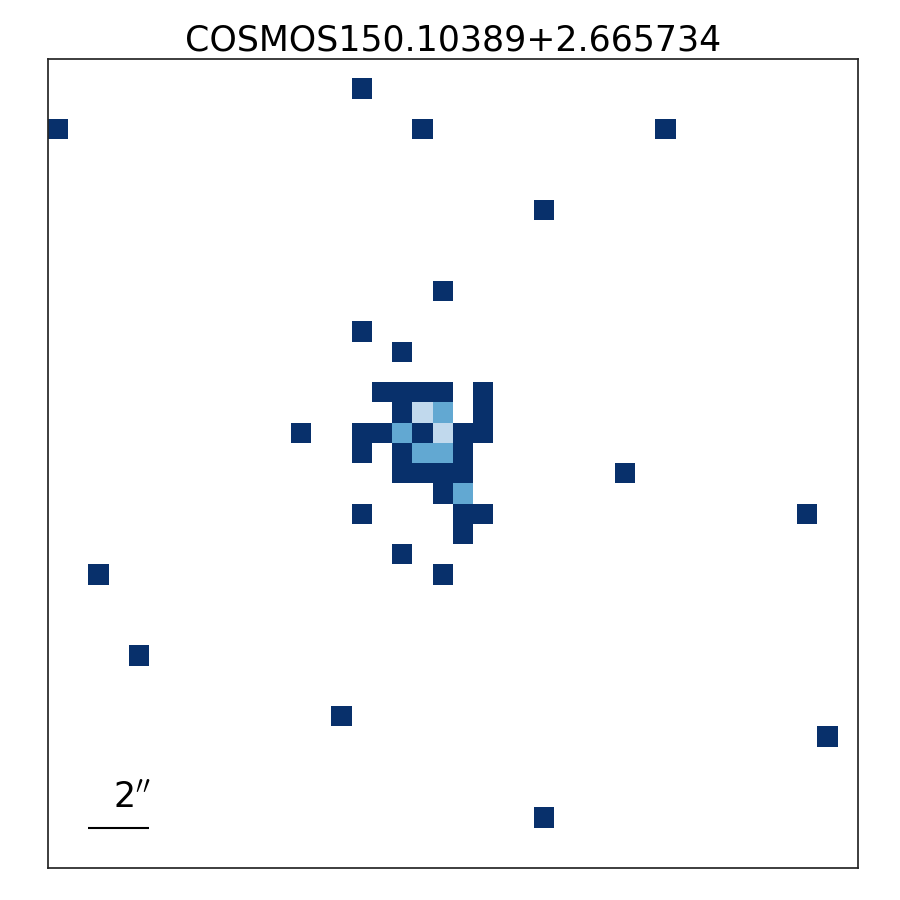}
     \end{subfigure}
        \hfill

    \vspace{-0.3cm}
    (28) \hspace{5.3cm} (29)  \hspace{5.3cm} (30)

    \begin{subfigure}
         \centering
         \includegraphics[width=0.33\textwidth]{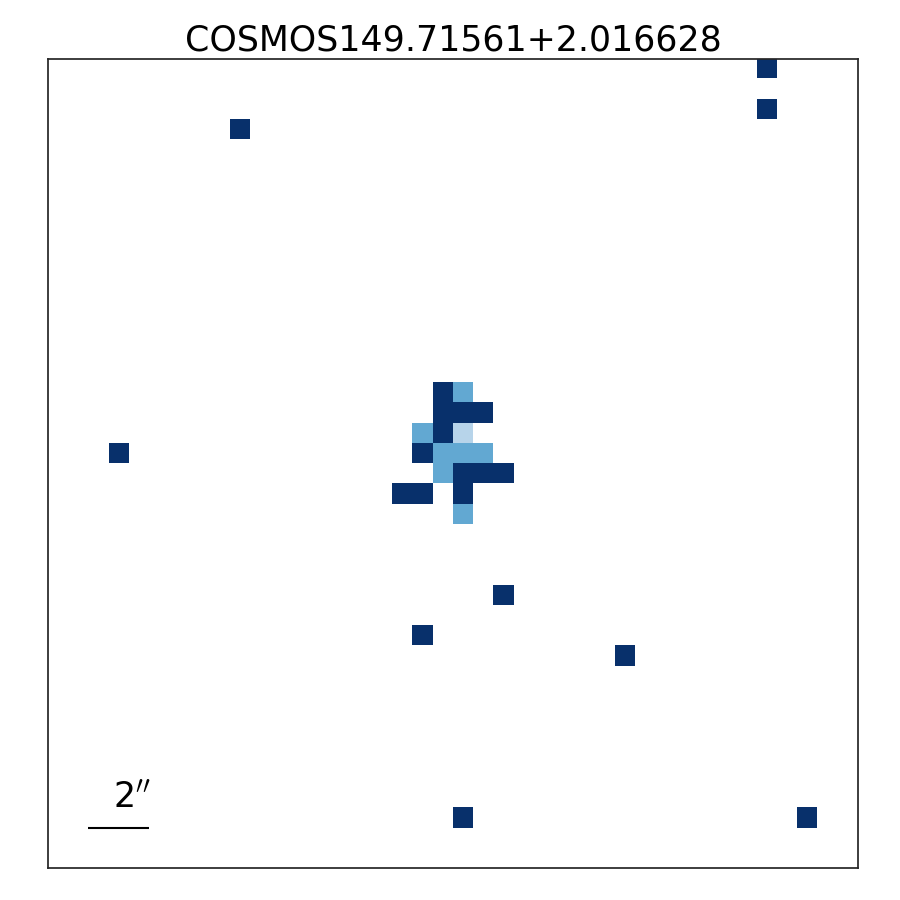}
    \end{subfigure}
    \hspace{-1cm}
    \hfill
    \begin{subfigure}
         \centering
         \includegraphics[width=0.33\textwidth]{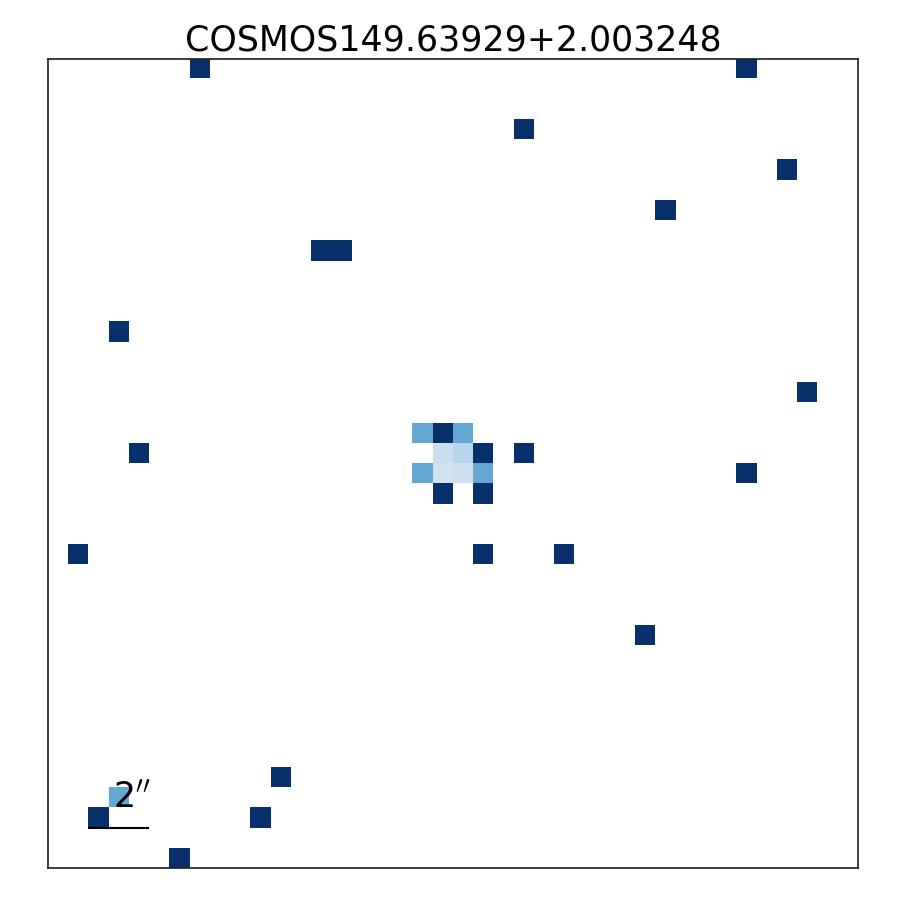}
    \end{subfigure}
    \hspace{-1cm}
    \hfill
    \begin{subfigure}
         \centering
         \includegraphics[width=0.33\textwidth]{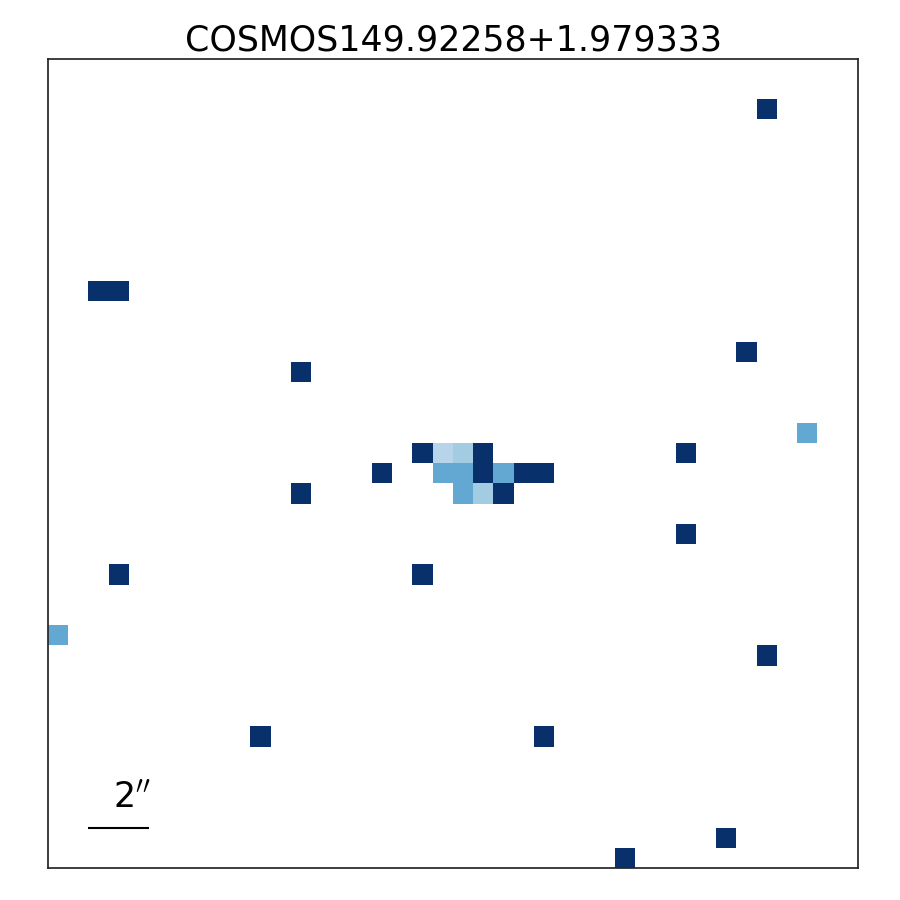}
     \end{subfigure}
        \hfill     

    \vspace{-0.3cm}
    (31) \hspace{5.3cm} (32)  \hspace{5.3cm} (33) \\
     
\RaggedRight{\textbf{Figure 5.} (continued)} 
\label{}
\end{figure*}

\begin{figure*}
     \centering
    
    \begin{subfigure}
         \centering
         \includegraphics[width=0.33\textwidth]{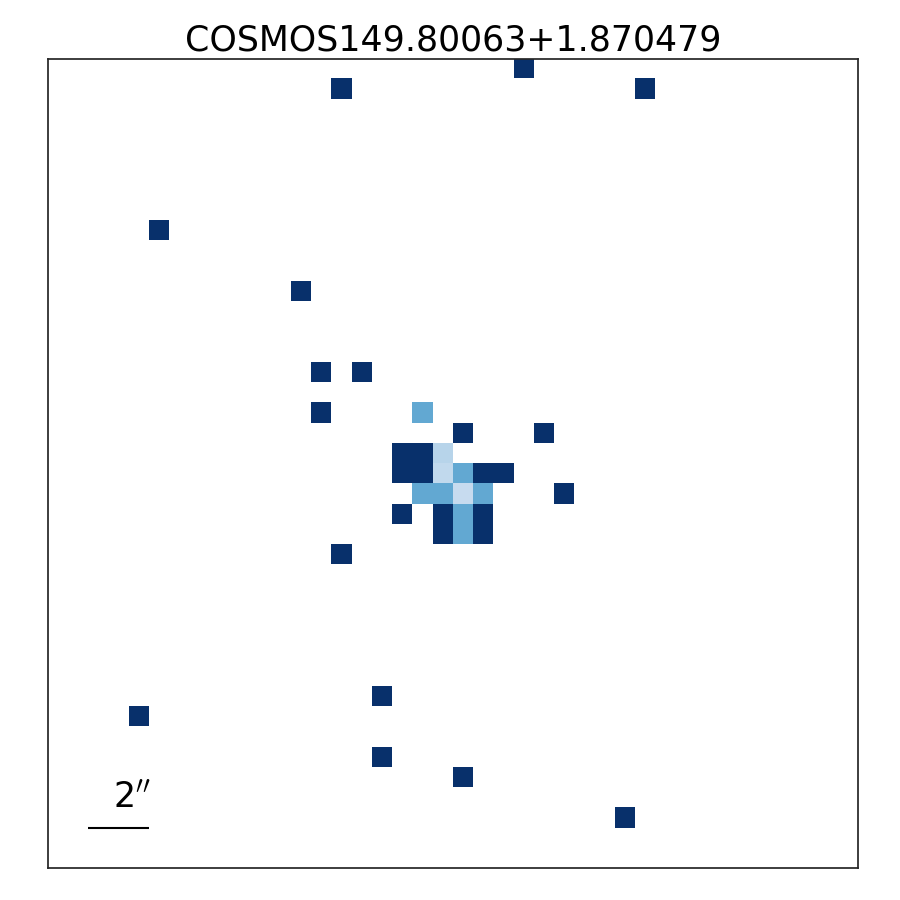}
    \end{subfigure}
    \hspace{-1cm}
    \hfill
    \begin{subfigure}
         \centering
         \includegraphics[width=0.33\textwidth]{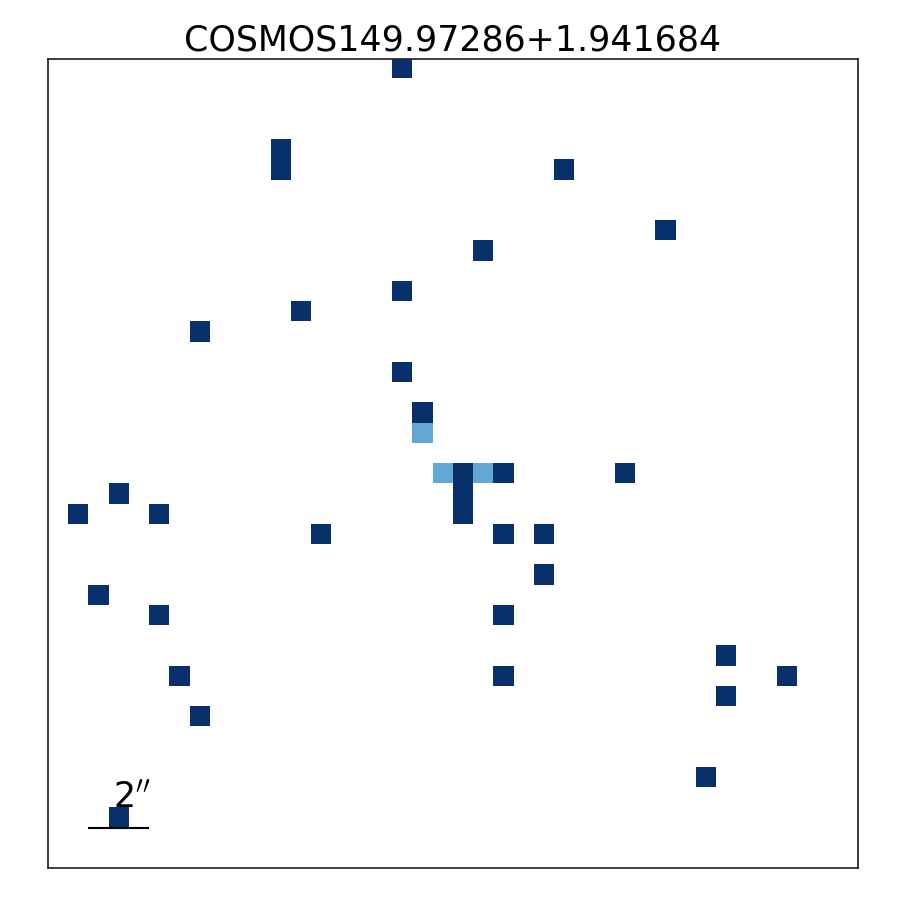}
    \end{subfigure}
    \hspace{-1cm}
    \hfill
    \begin{subfigure}
         \centering
         \includegraphics[width=0.33\textwidth]{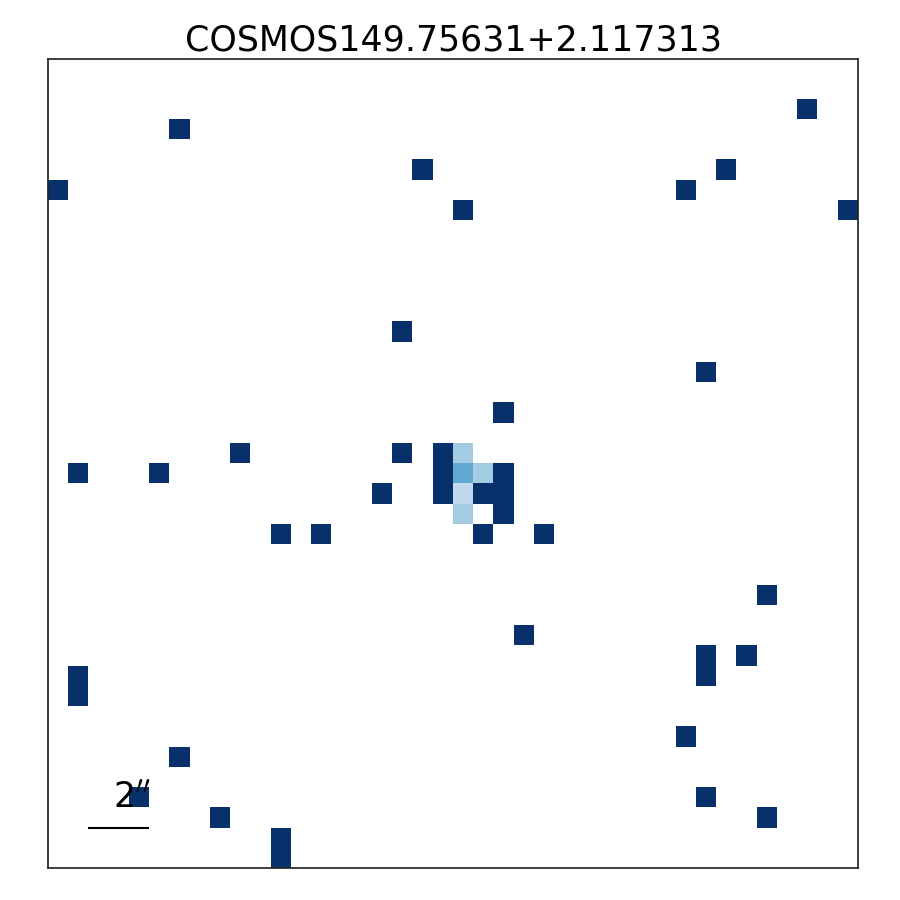}
     \end{subfigure}
        \hfill

    \vspace{-0.3cm}
    (34) \hspace{5.3cm} (35)  \hspace{5.3cm} (36)
        
    \begin{subfigure}
         \centering
         \includegraphics[width=0.33\textwidth]{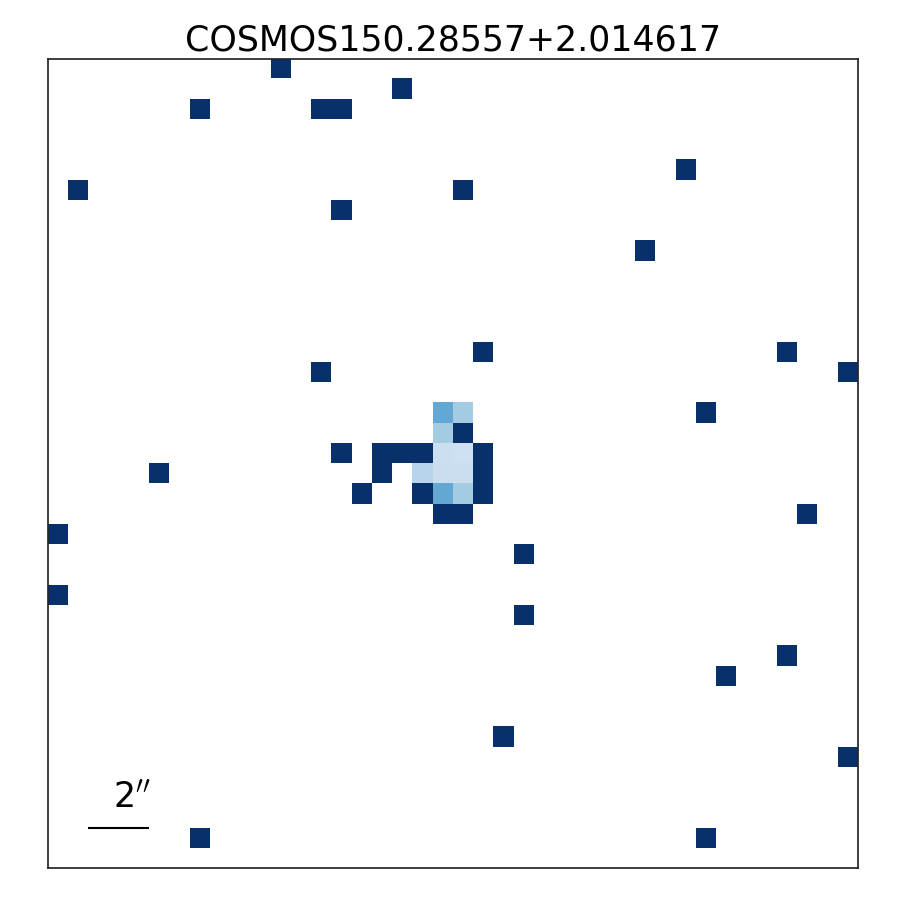}
    \end{subfigure}
    \hspace{-1cm}
    \hfill
    \begin{subfigure}
         \centering
         \includegraphics[width=0.33\textwidth]{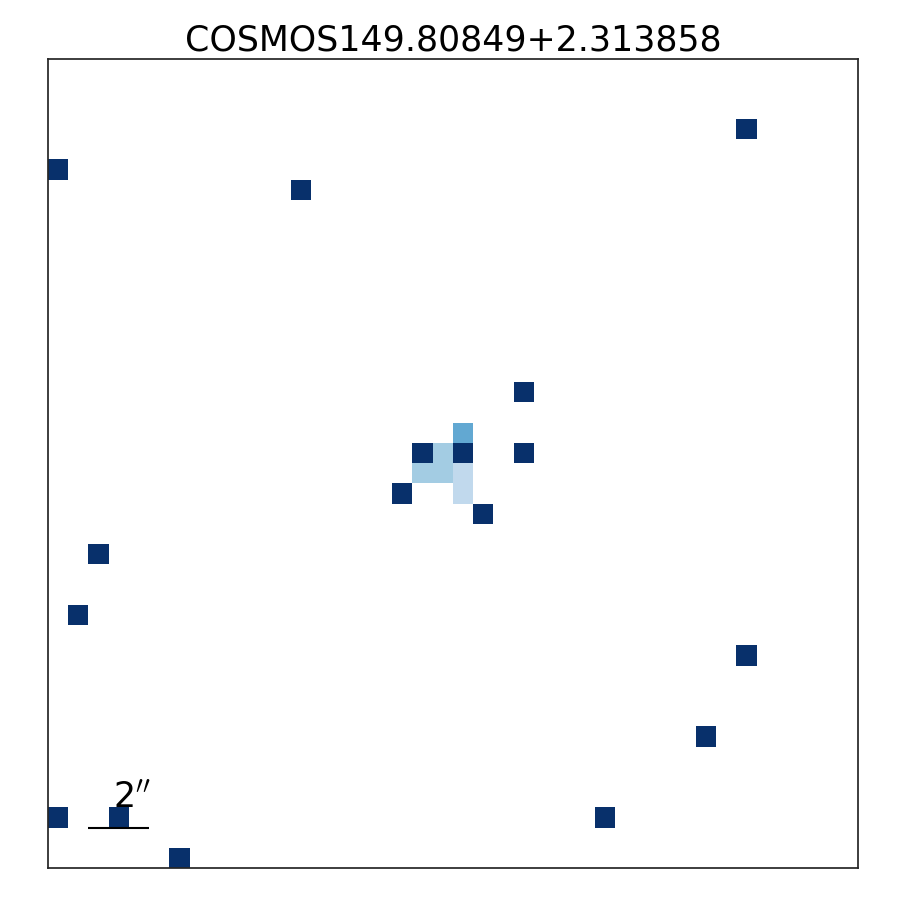}
    \end{subfigure}
    \hspace{-1cm}
    \hfill
    \begin{subfigure}
         \centering
         \includegraphics[width=0.33\textwidth]{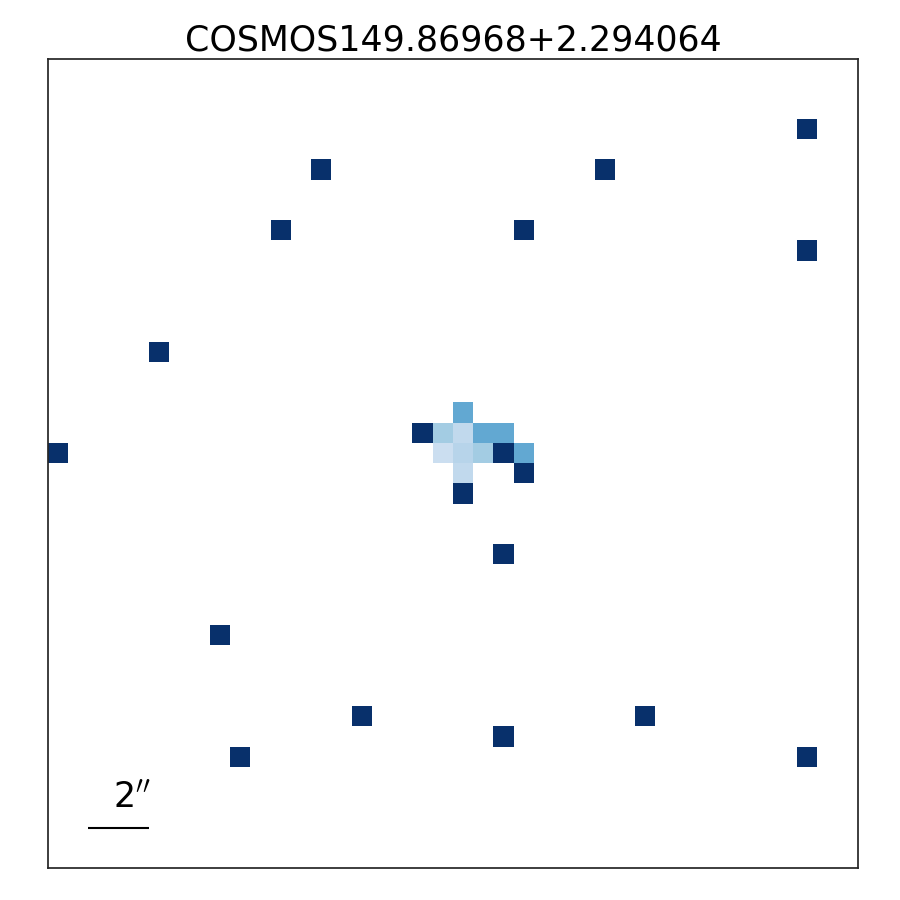}
     \end{subfigure}
        \hfill

        \vspace{-0.3cm}
    (37) \hspace{5.3cm} (38)  \hspace{5.3cm} (39)    

    \begin{subfigure}
         \centering
         \includegraphics[width=0.33\textwidth]{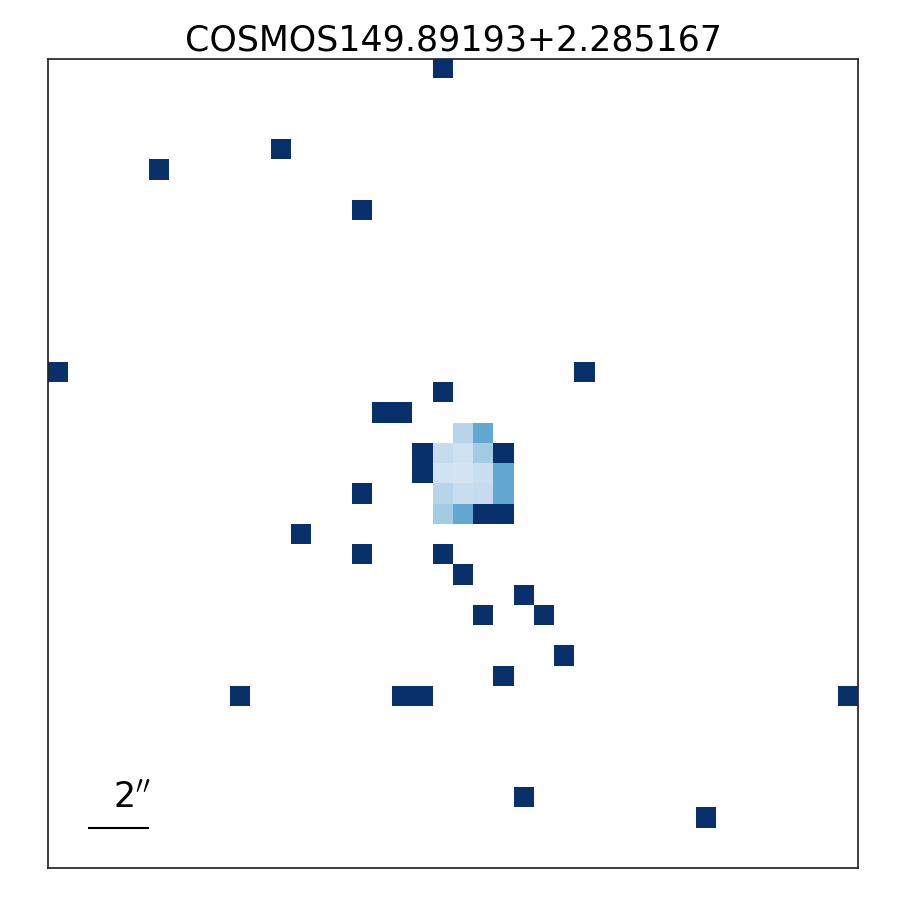}
    \end{subfigure}
    \hspace{-1cm}
    \hfill
    \begin{subfigure}
         \centering
         \includegraphics[width=0.33\textwidth]{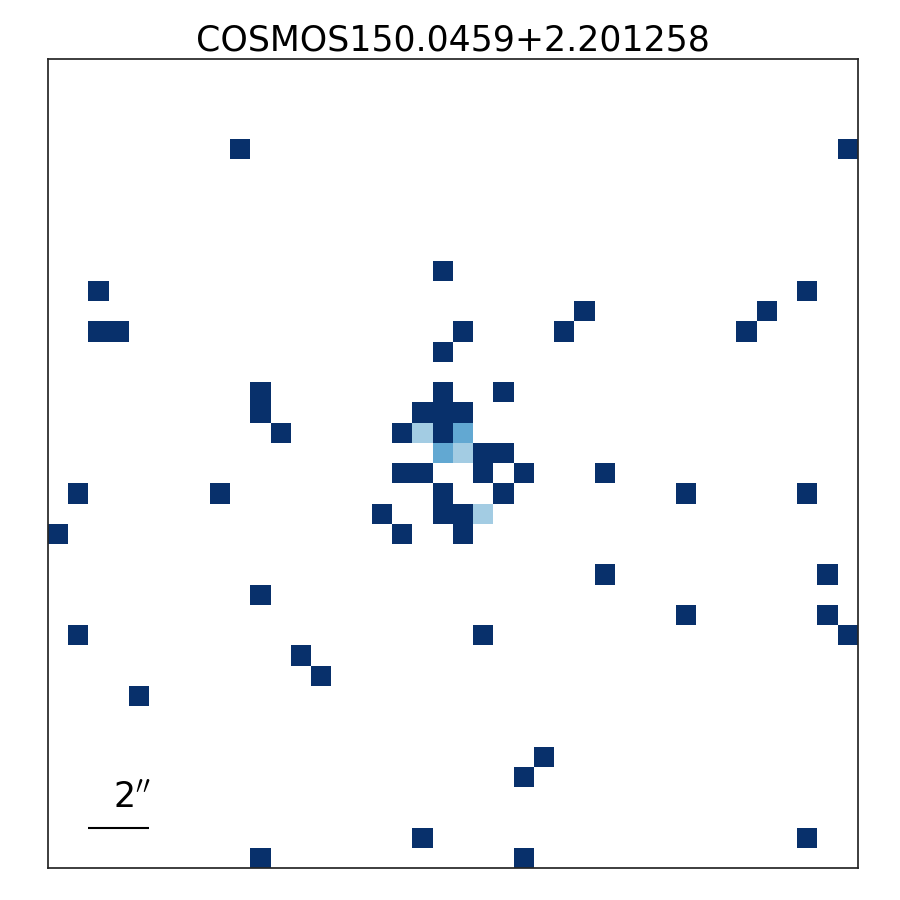}
     \end{subfigure}
        \hfill
    \hspace{-1cm}
    \begin{subfigure}
         \centering
         \includegraphics[width=0.33\textwidth]{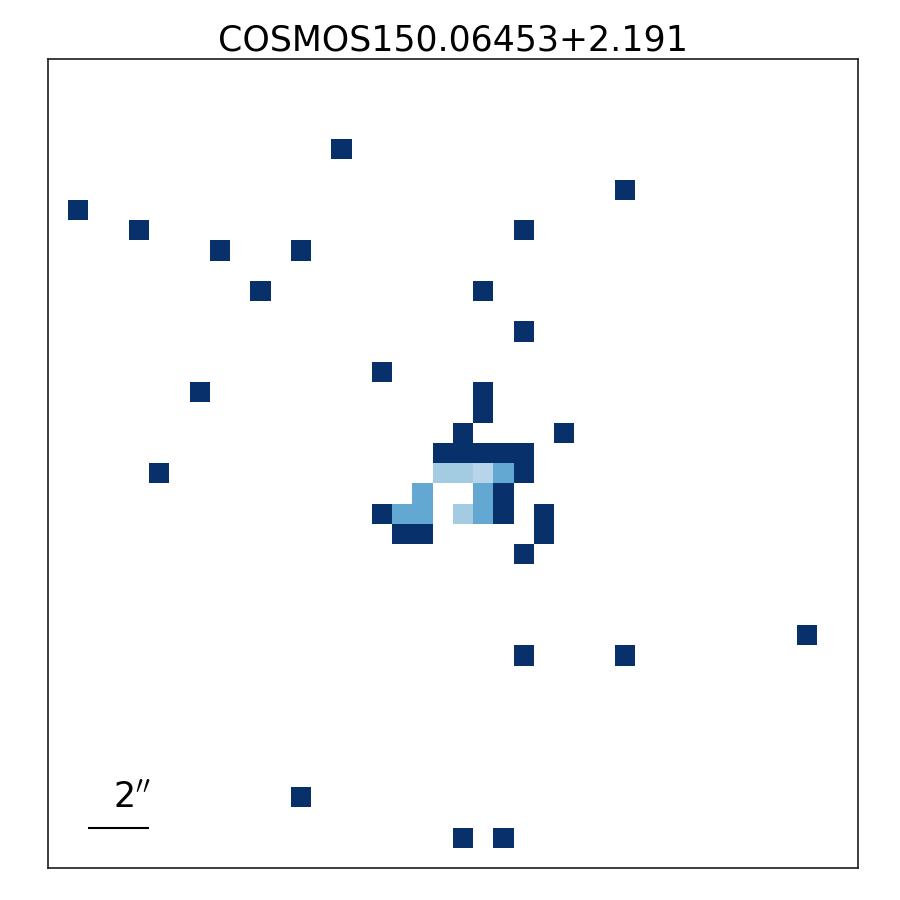}
    \end{subfigure}
    \hfill

        \vspace{-0.3cm}
    (40) \hspace{5.3cm} (41)  \hspace{5.3cm} (42)
    
    \begin{subfigure}
         \centering
         \includegraphics[width=0.33\textwidth]{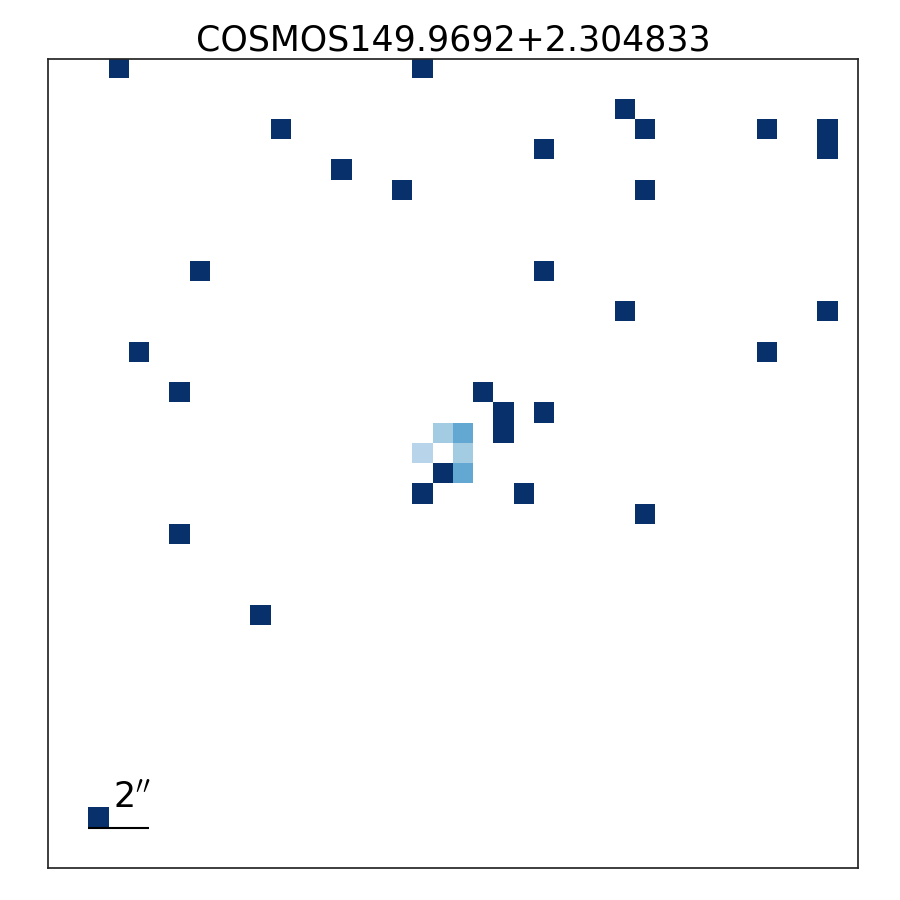}
    \end{subfigure}
    \hspace{-1cm}
    \hfill
    \begin{subfigure}
         \centering
         \includegraphics[width=0.33\textwidth]{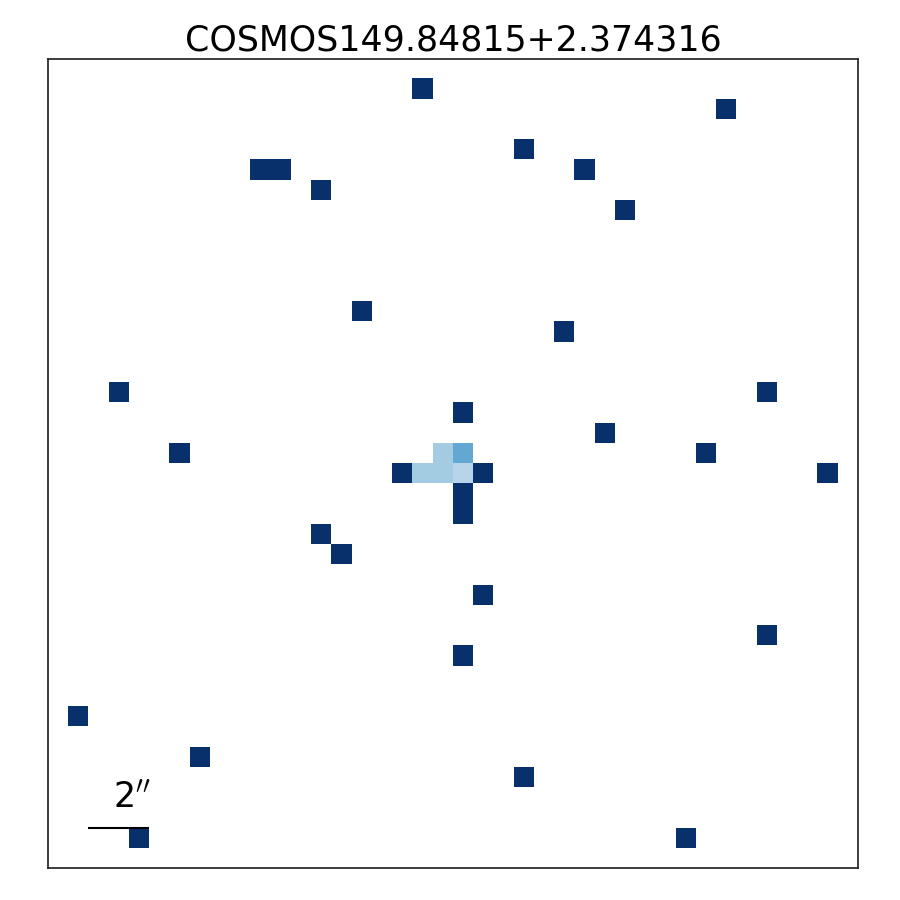}
    \end{subfigure}
    \hspace{-1cm}
    \hfill
    \begin{subfigure}
         \centering
         \includegraphics[width=0.33\textwidth]{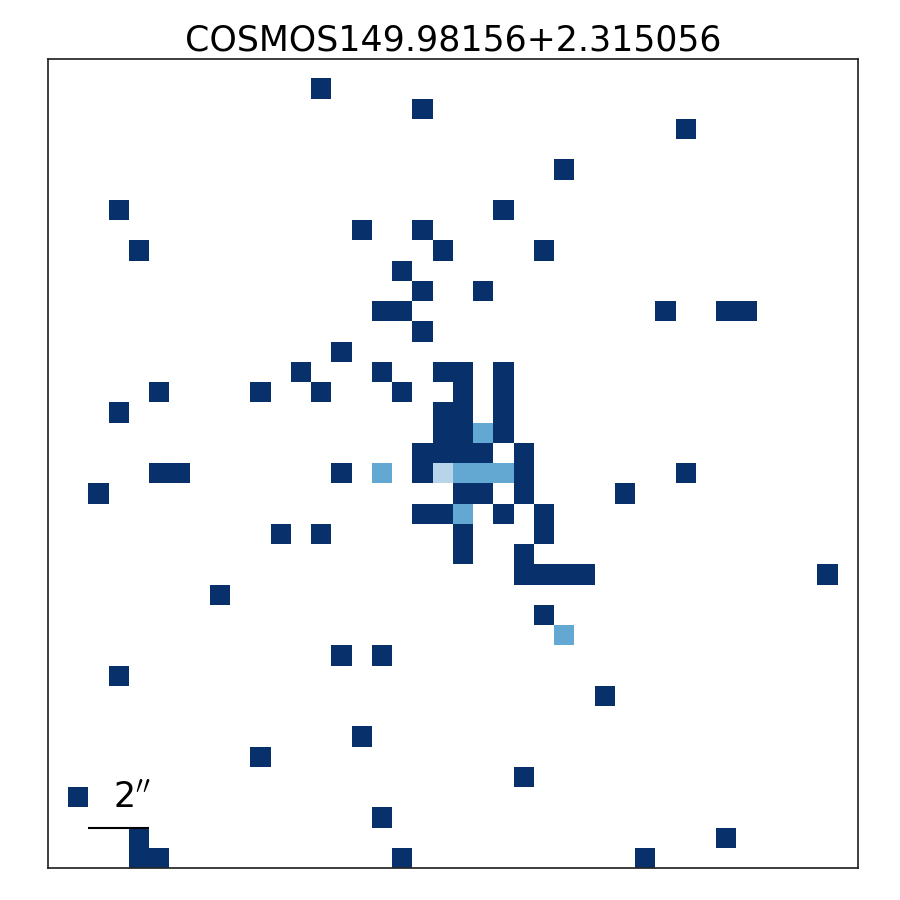}
    \end{subfigure}
    \hfill

      \vspace{-0.3cm}
    (43) \hspace{5.3cm} (44)  \hspace{5.3cm} (45)$\dagger$   \\
     
\RaggedRight{\textbf{Figure 5.} (continued)} 
\label{}
\end{figure*}

\begin{figure*}
     \centering
    \begin{subfigure}
         \centering
         \includegraphics[width=0.33\textwidth]{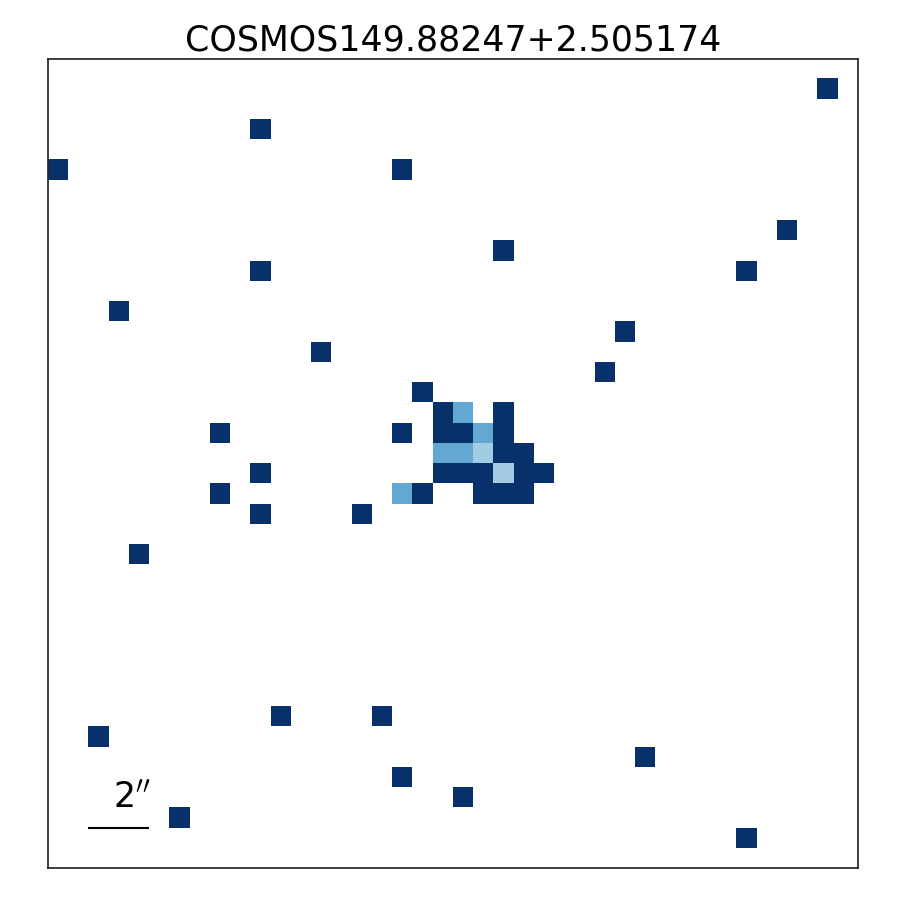}
    \end{subfigure}
    \hspace{-1cm}
    \hfill
    \begin{subfigure}
         \centering
         \includegraphics[width=0.33\textwidth]{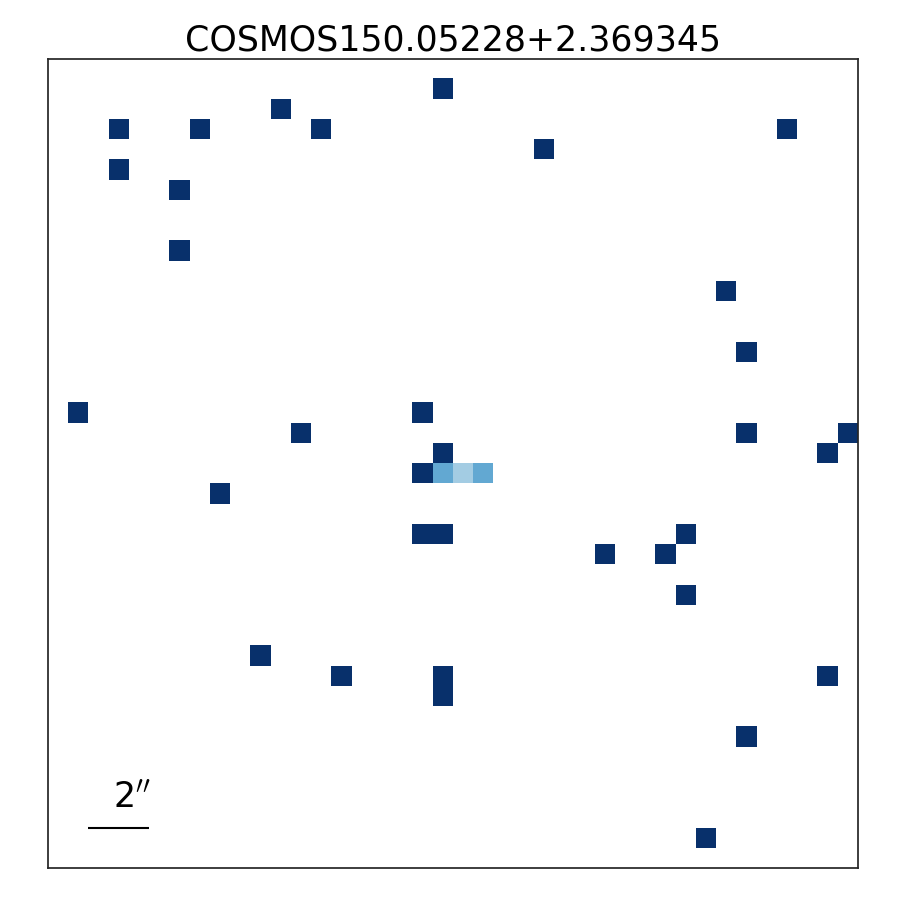}
    \end{subfigure}
    \hspace{-1cm}
    \hfill
    \begin{subfigure}
         \centering
         \includegraphics[width=0.33\textwidth]{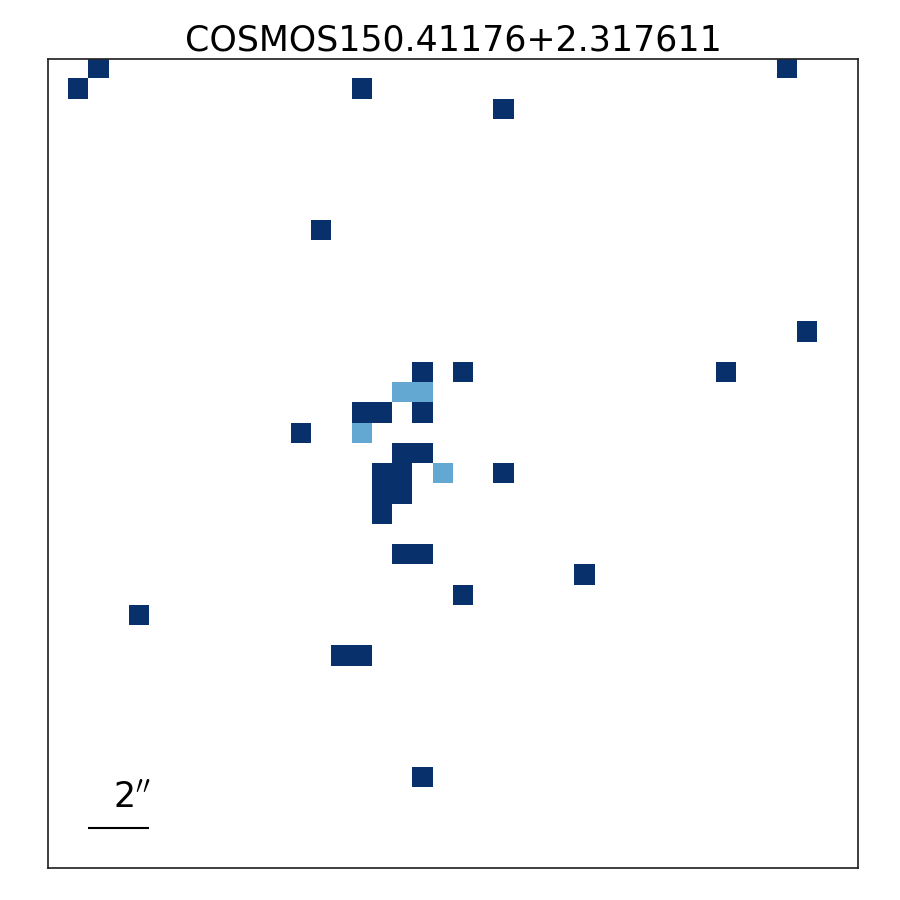}
     \end{subfigure}
        \hfill
    
    \vspace{-0.3cm}
    (46) \hspace{5.3cm} (47)  \hspace{5.3cm} (48)

    \begin{subfigure}
         \centering
         \includegraphics[width=0.33\textwidth]{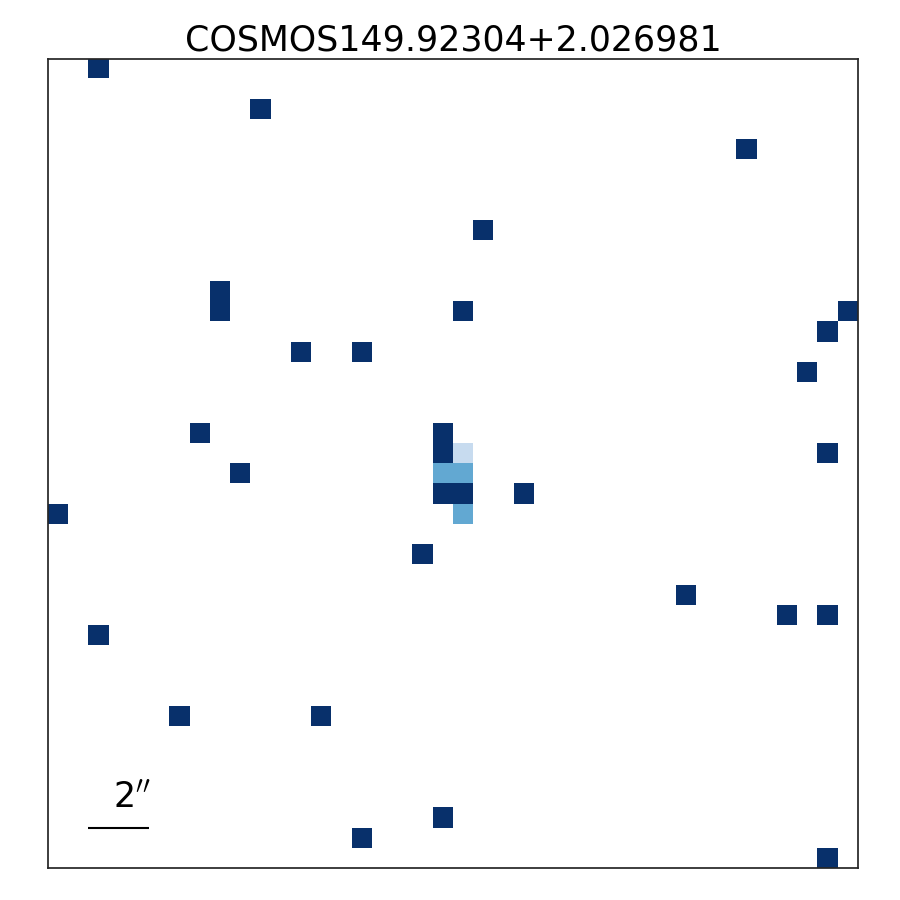}
    \end{subfigure}
    \hspace{-1cm}
    \hfill
    \begin{subfigure}
         \centering
         \includegraphics[width=0.33\textwidth]{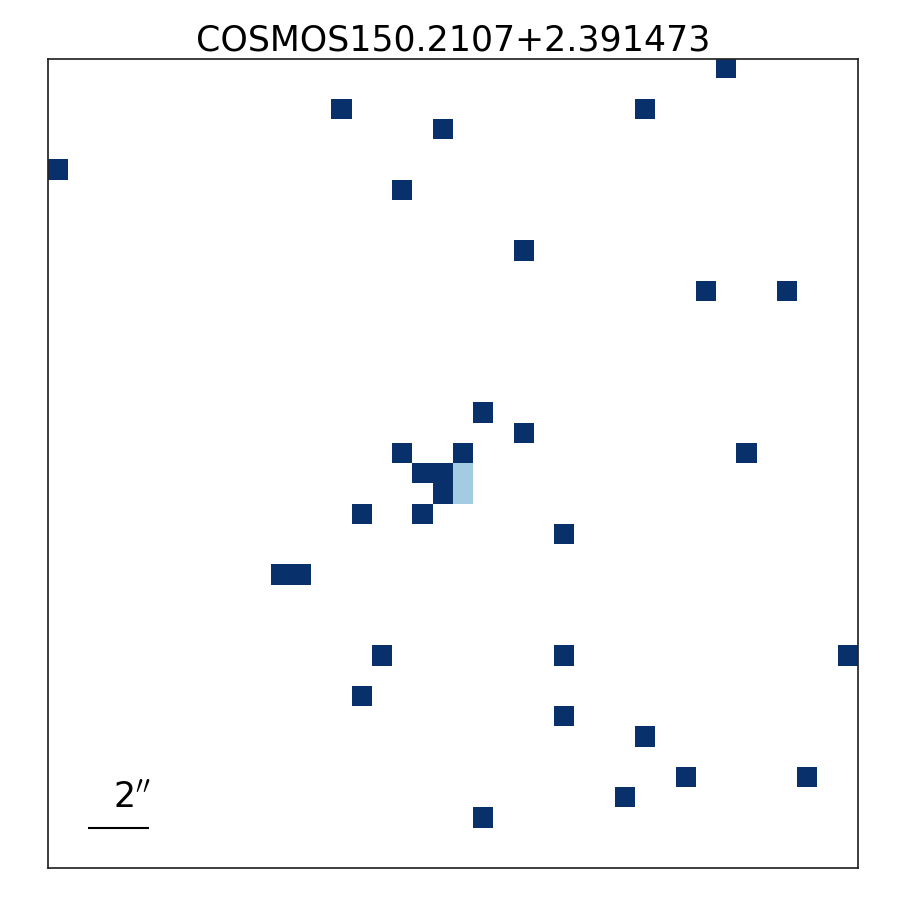}
    \end{subfigure}
    \hspace{-1cm}
    \hfill
    \begin{subfigure}
         \centering
         \includegraphics[width=0.33\textwidth]{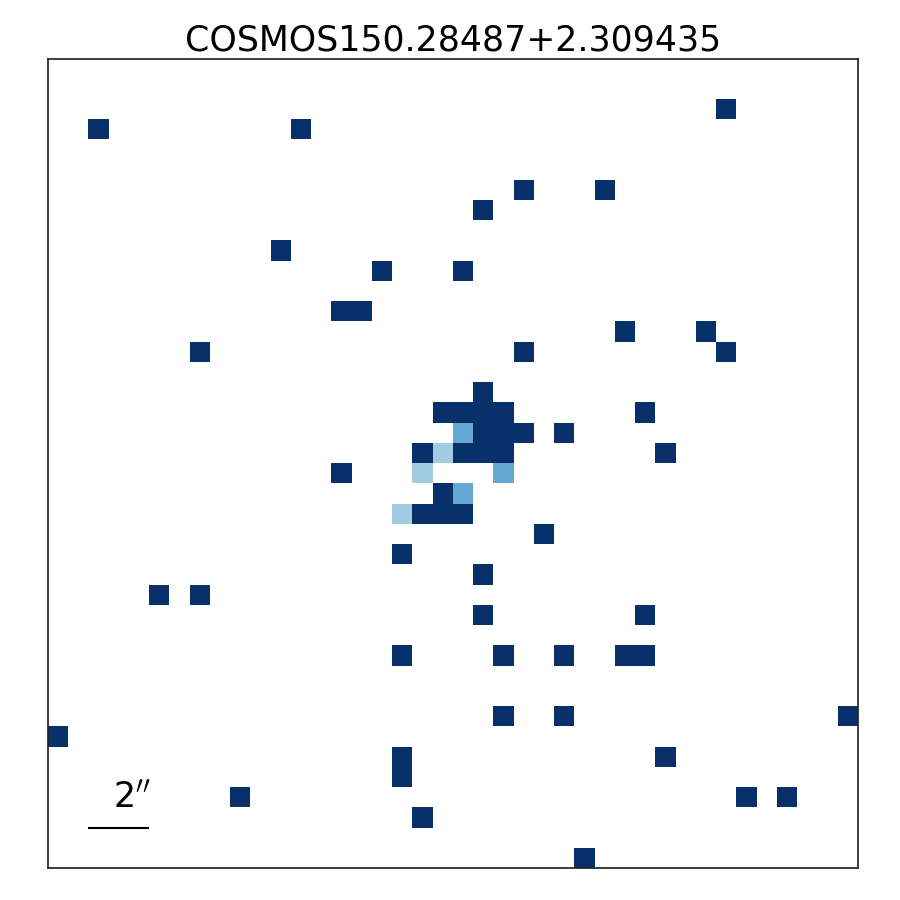}
     \end{subfigure}
        \hfill
    
    \vspace{-0.3cm}
    (49) \hspace{5.3cm} (50)  \hspace{5.3cm} (51)
    
    \begin{subfigure}
         \centering
         \includegraphics[width=0.33\textwidth]{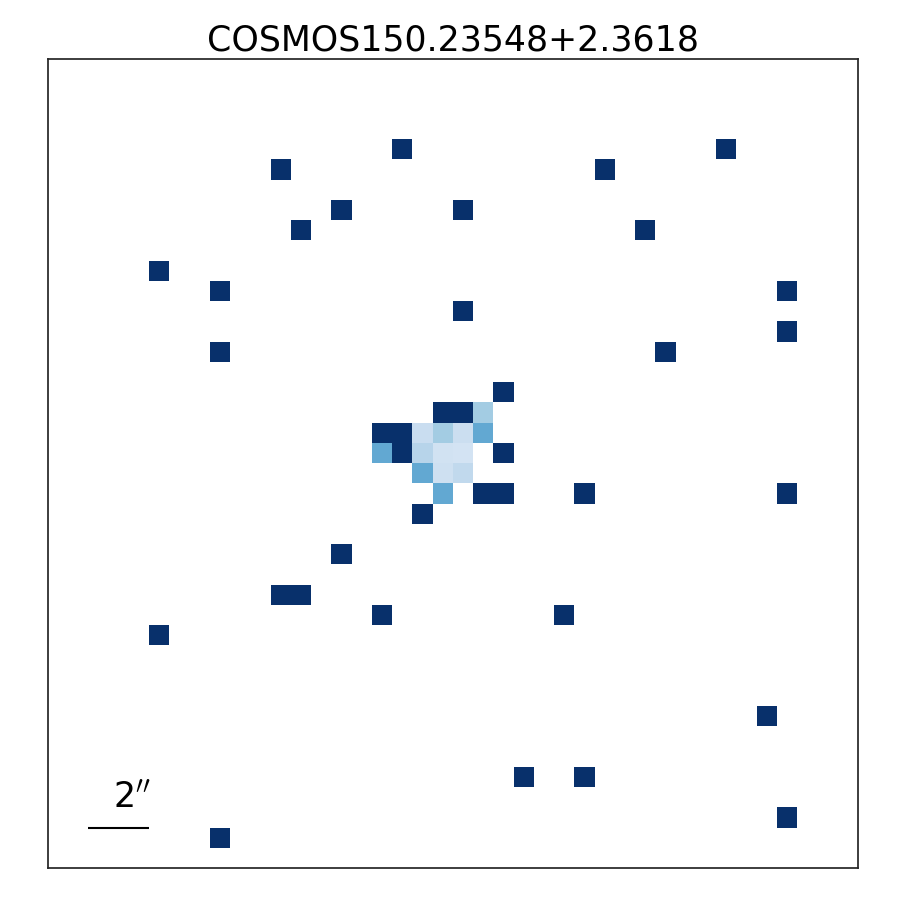}
    \end{subfigure}
    \hspace{-1cm}
    \hfill
    \begin{subfigure}
         \centering
         \includegraphics[width=0.33\textwidth]{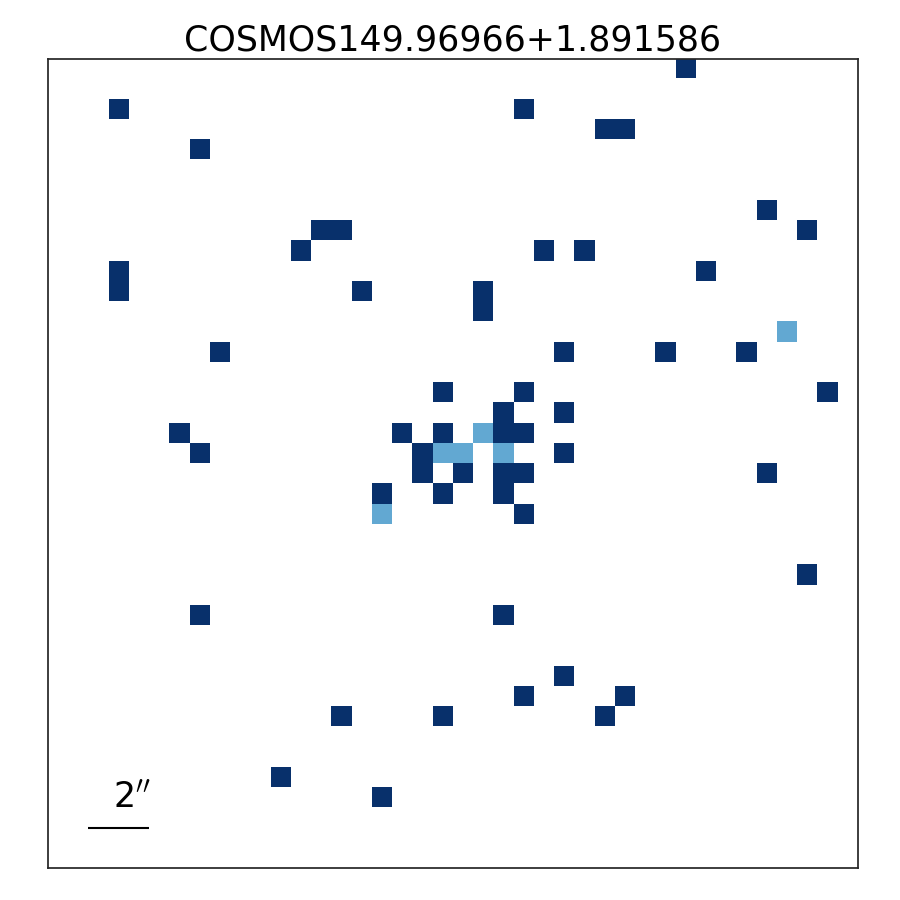}
    \end{subfigure}
    \hspace{-1cm}
    \hfill
    \begin{subfigure}
         \centering
         \includegraphics[width=0.33\textwidth]{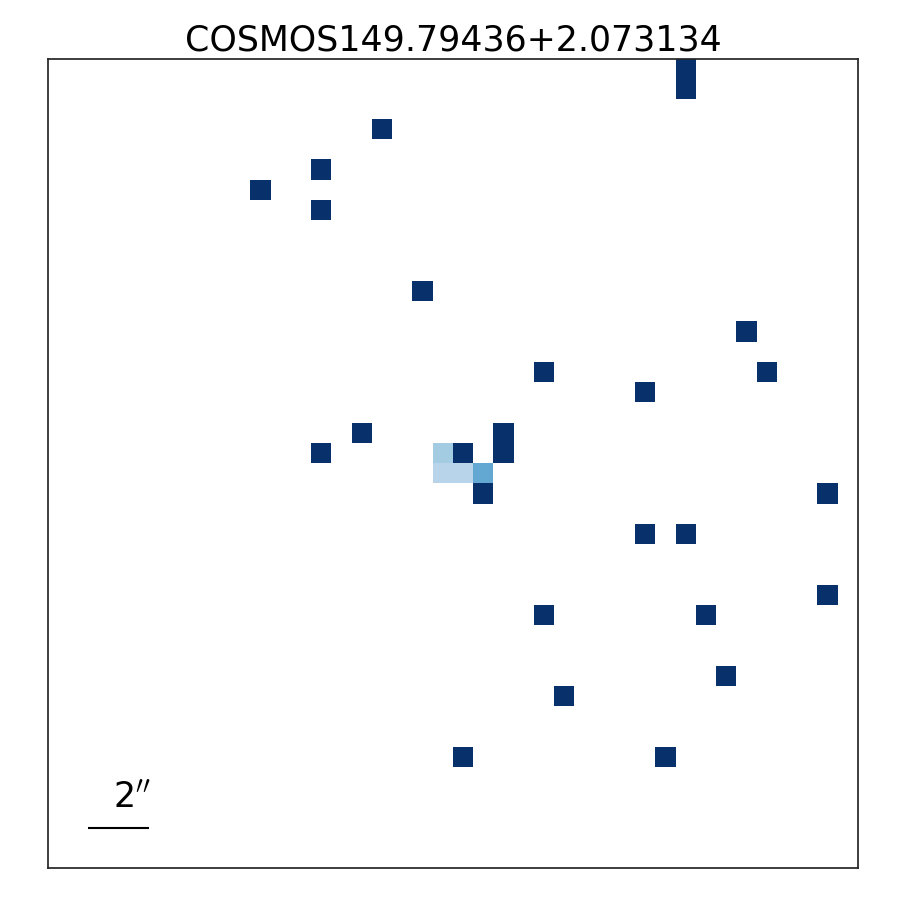}
    \end{subfigure}
    \hfill

    \vspace{-0.3cm}
    (52) \hspace{5.3cm} (53)  \hspace{5.3cm} (54)

    \begin{subfigure}
         \centering
         \includegraphics[width=0.33\textwidth]{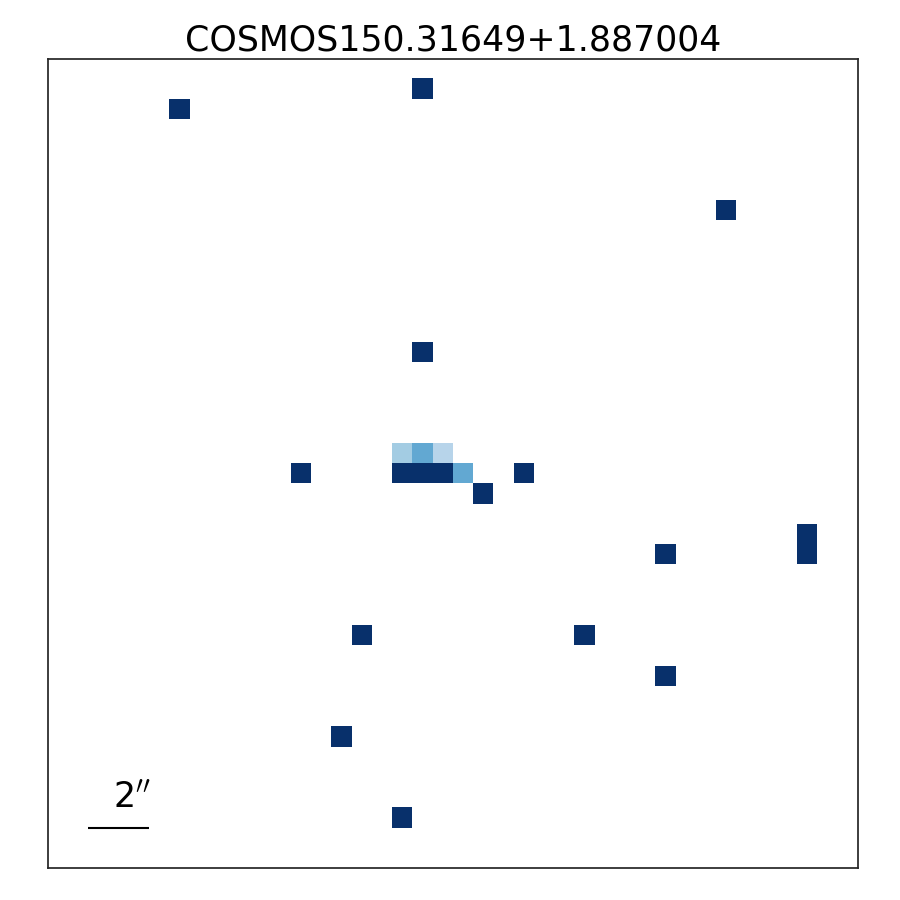}
    \end{subfigure}
    \hspace{-1cm}
    \hfill
    \begin{subfigure}
         \centering
         \includegraphics[width=0.33\textwidth]{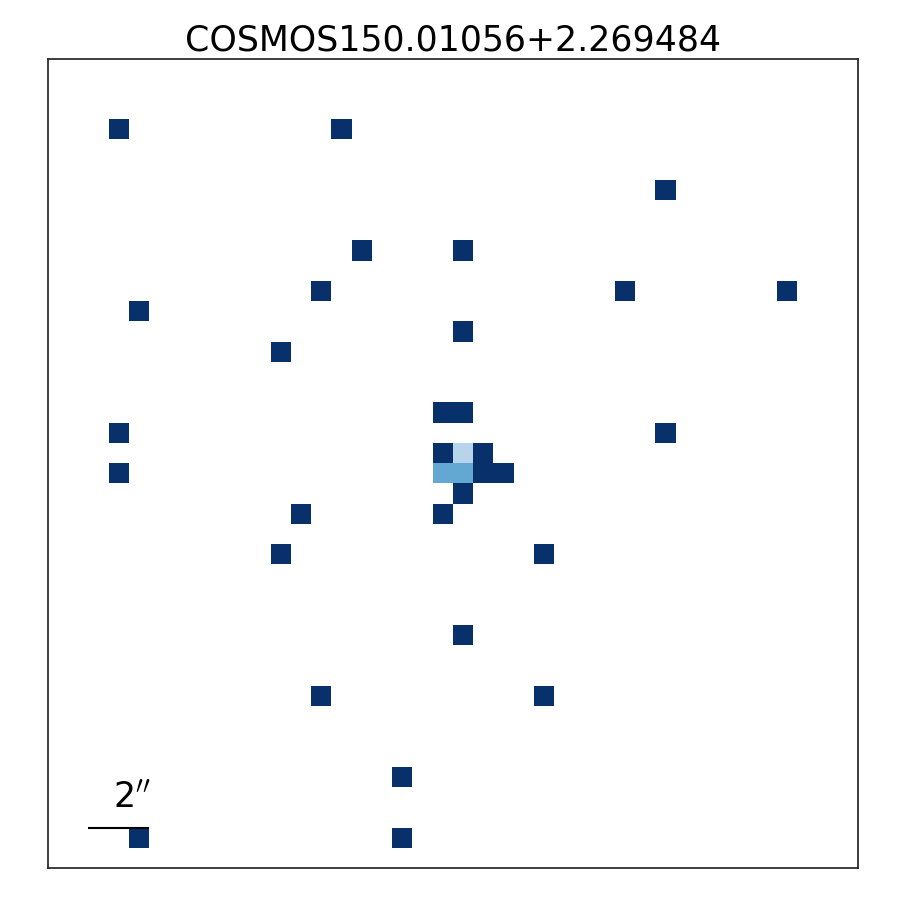}
    \end{subfigure}
    \hspace{-1cm}
    \hfill
    \begin{subfigure}
         \centering
         \includegraphics[width=0.33\textwidth]{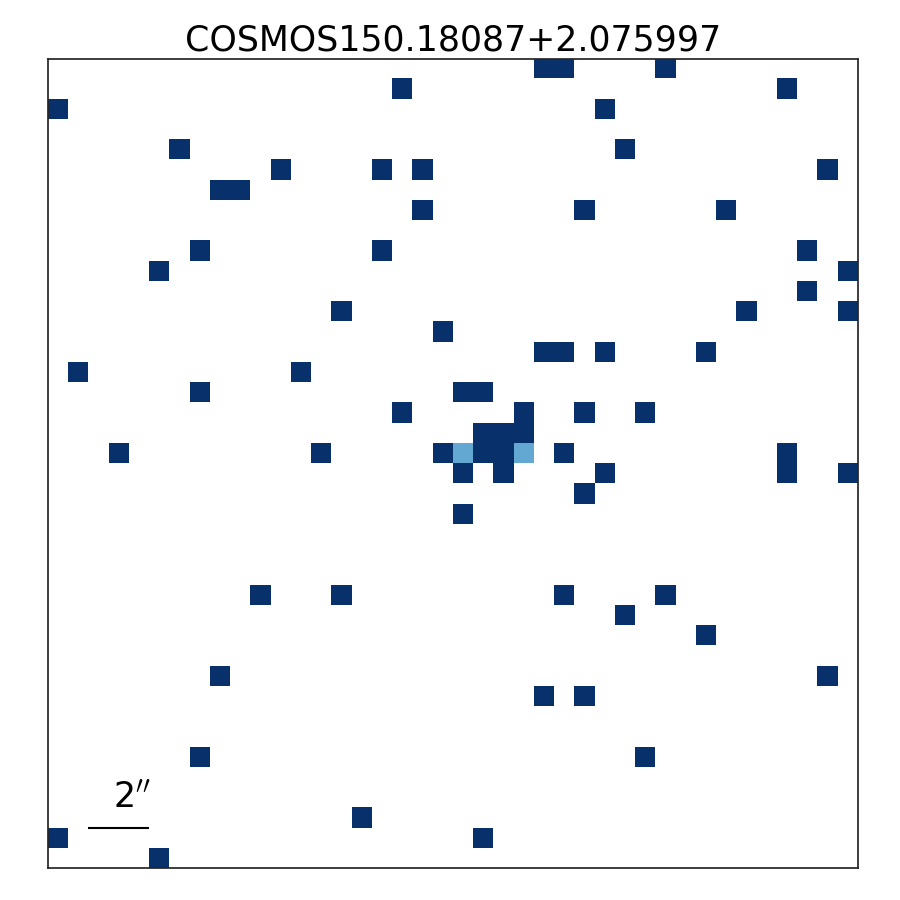}
    \end{subfigure}
    \hfill

    \vspace{-0.3cm}
    (55) \hspace{5.3cm} (56)  \hspace{5.3cm} (57)  \\
    
\RaggedRight{\textbf{Figure 5.} (continued)} 
\label{}
\end{figure*}

\begin{figure*}
     \centering
    \begin{subfigure}
         \centering
         \includegraphics[width=0.33\textwidth]{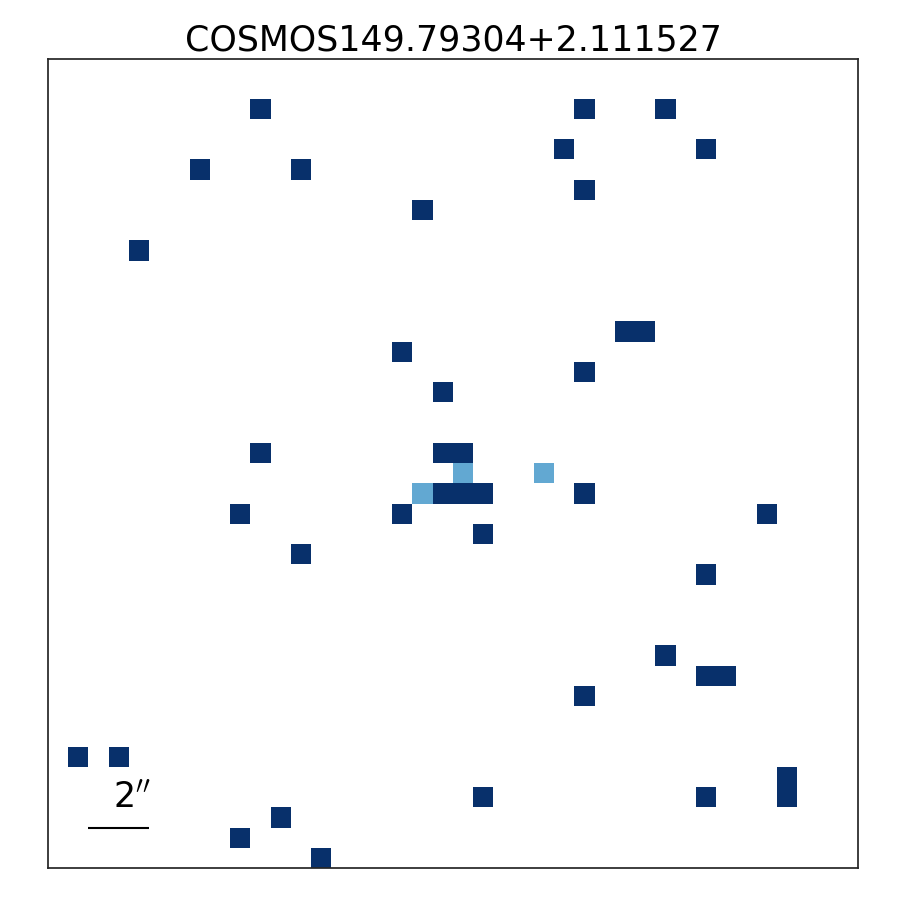}
    \end{subfigure}
    \hspace{-1cm}
    \hfill
    \begin{subfigure}
         \centering
         \includegraphics[width=0.33\textwidth]{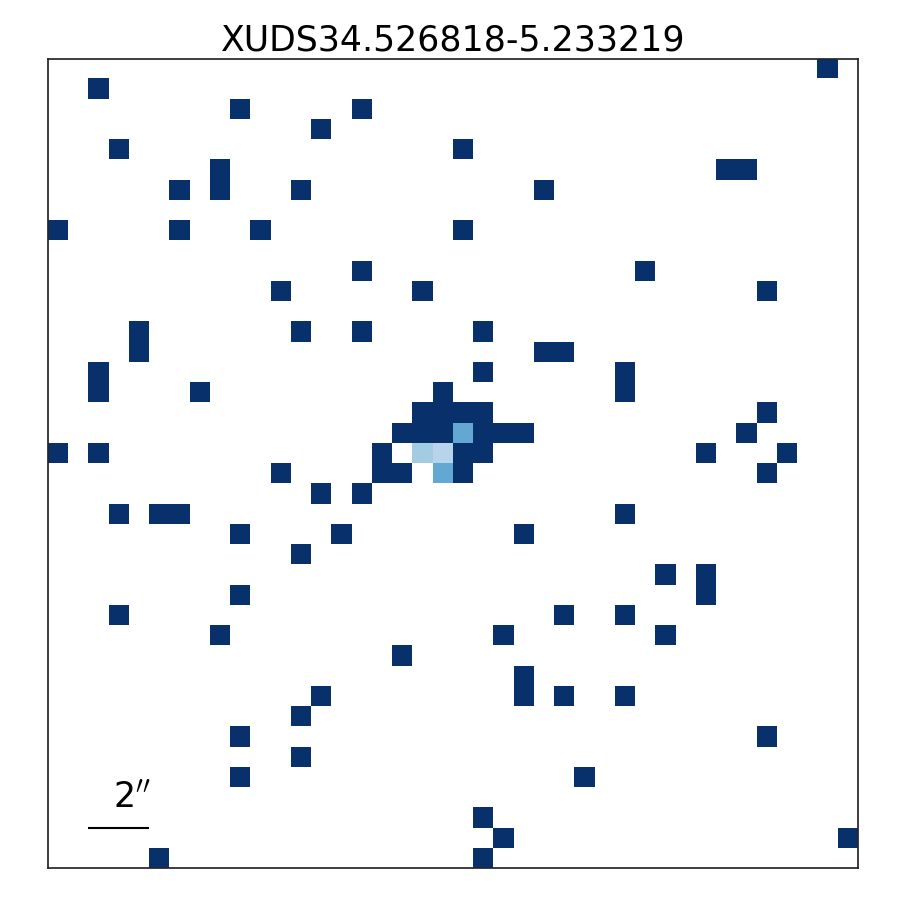}
    \end{subfigure}
    \hspace{-1cm}
    \hfill
    \begin{subfigure}
         \centering
         \includegraphics[width=0.33\textwidth]{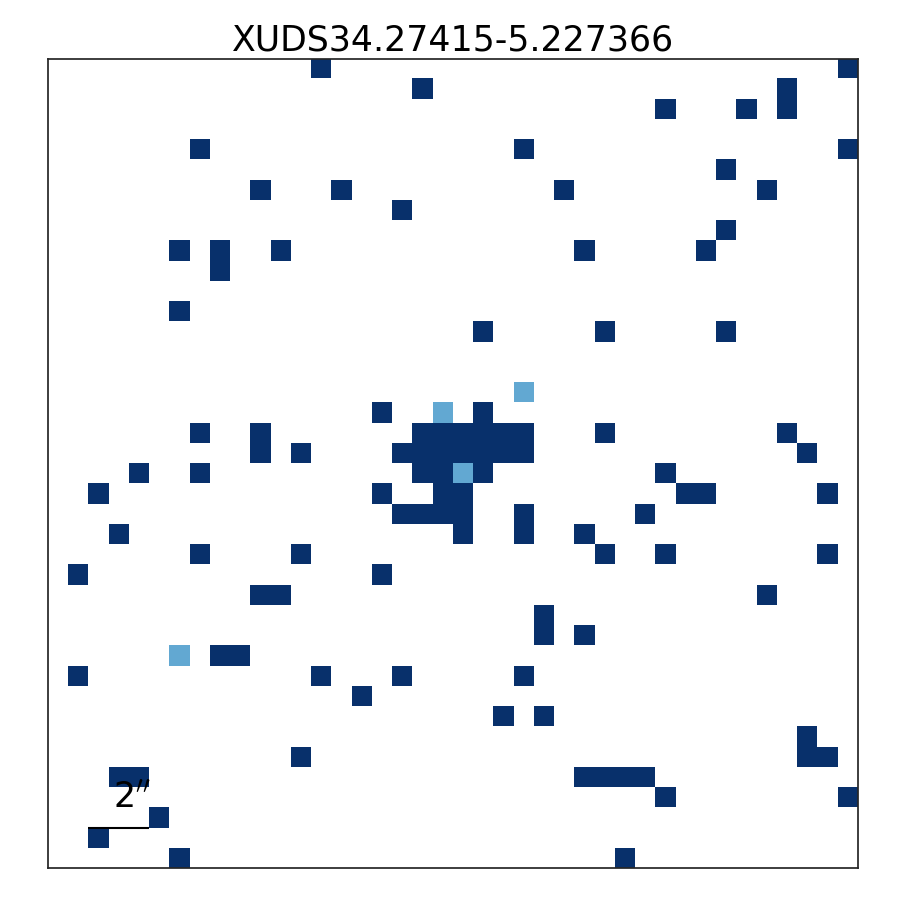}
     \end{subfigure}
        \hfill

    \vspace{-0.3cm}
    (58) \hspace{5.3cm} (59)  \hspace{5.3cm} (60)

    \begin{subfigure}
         \centering
         \includegraphics[width=0.33\textwidth]{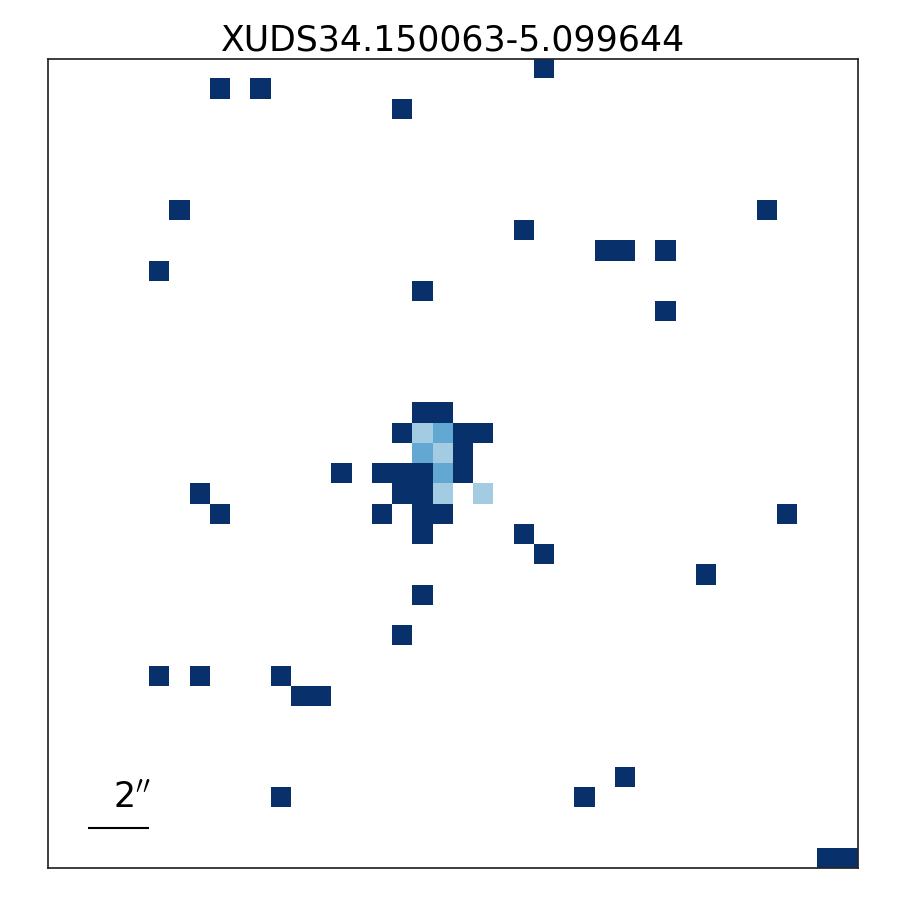}
    \end{subfigure}
    \hspace{-1cm}
    \hfill
    \begin{subfigure}
         \centering
         \includegraphics[width=0.33\textwidth]{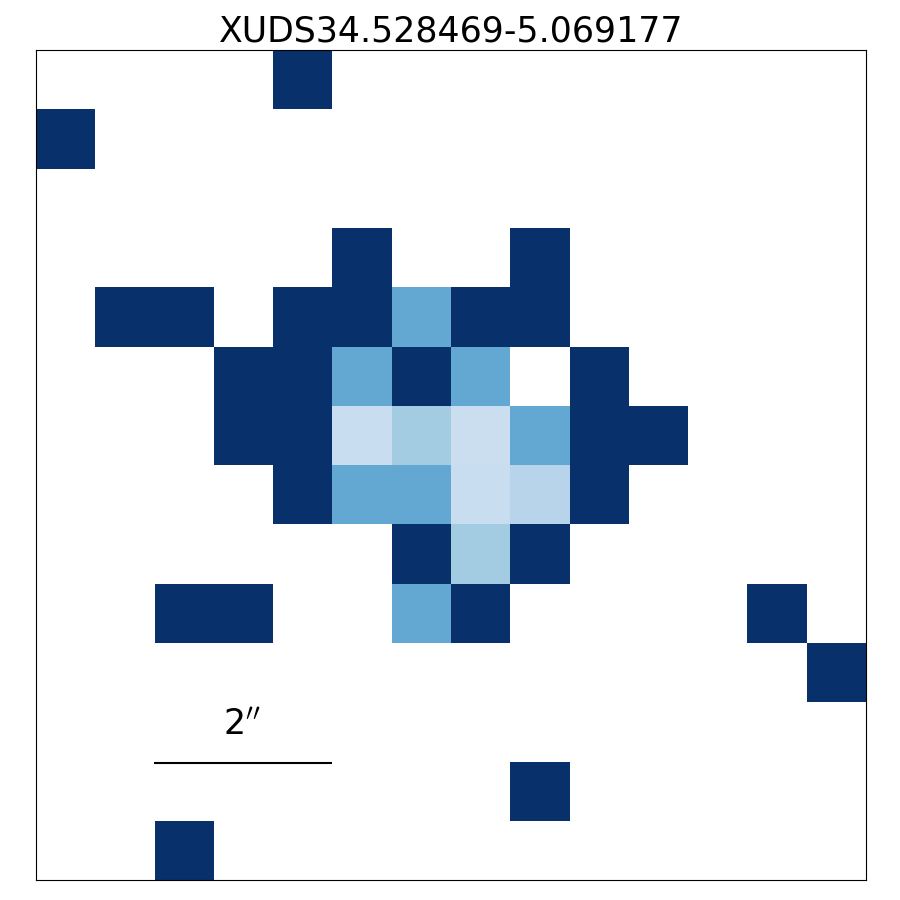}
    \end{subfigure}
    \hspace{5.7cm}
    \hfill

    \vspace{-0.3cm}
    \hspace{-5.3cm} (61) \hspace{5.3cm} (62) \\
    
\RaggedRight{\textbf{Figure 5.} (continued)} 
\label{}
\end{figure*}

\bibliography{dualagn}{}

\newcommand{\noop}[1]{}
\begin{thebibliography}{}
\expandafter\ifx\csname natexlab\endcsname\relax\def\natexlab#1{#1}\fi
\providecommand{\url}[1]{\href{#1}{#1}}
\providecommand{\dodoi}[1]{doi:~\href{http://doi.org/#1}{\nolinkurl{#1}}}
\providecommand{\doeprint}[1]{\href{http://ascl.net/#1}{\nolinkurl{http://ascl.net/#1}}}
\providecommand{\doarXiv}[1]{\href{https://arxiv.org/abs/#1}{\nolinkurl{https://arxiv.org/abs/#1}}}

\bibitem[{{Agazie} {et~al.}(2023{\natexlab{a}}){Agazie}, {Anumarlapudi},
  {Archibald}, {Arzoumanian}, {Baker}, {B{\'e}csy}, {Blecha}, {Brazier},
  {Brook}, {Burke-Spolaor}, {Burnette}, {Case}, {Charisi}, {Chatterjee},
  {Chatziioannou}, {Cheeseboro}, {Chen}, {Cohen}, {Cordes}, {Cornish},
  {Crawford}, {Cromartie}, {Crowter}, {Cutler}, {Decesar}, {Degan}, {Demorest},
  {Deng}, {Dolch}, {Drachler}, {Ellis}, {Ferrara}, {Fiore}, {Fonseca},
  {Freedman}, {Garver-Daniels}, {Gentile}, {Gersbach}, {Glaser}, {Good},
  {G{\"u}ltekin}, {Hazboun}, {Hourihane}, {Islo}, {Jennings}, {Johnson},
  {Jones}, {Kaiser}, {Kaplan}, {Kelley}, {Kerr}, {Key}, {Klein}, {Laal}, {Lam},
  {Lamb}, {Lazio}, {Lewandowska}, {Littenberg}, {Liu}, {Lommen}, {Lorimer},
  {Luo}, {Lynch}, {Ma}, {Madison}, {Mattson}, {McEwen}, {McKee}, {McLaughlin},
  {McMann}, {Meyers}, {Meyers}, {Mingarelli}, {Mitridate}, {Natarajan}, {Ng},
  {Nice}, {Ocker}, {Olum}, {Pennucci}, {Perera}, {Petrov}, {Pol}, {Radovan},
  {Ransom}, {Ray}, {Romano}, {Sardesai}, {Schmiedekamp}, {Schmiedekamp},
  {Schmitz}, {Schult}, {Shapiro-Albert}, {Siemens}, {Simon}, {Siwek}, {Stairs},
  {Stinebring}, {Stovall}, {Sun}, {Susobhanan}, {Swiggum}, {Taylor}, {Taylor},
  {Turner}, {Unal}, {Vallisneri}, {van Haasteren}, {Vigeland}, {Wahl}, {Wang},
  {Witt}, {Young}, \& {Nanograv Collaboration}}]{NG15yrGWB}
{Agazie}, G., {Anumarlapudi}, A., {Archibald}, A.~M., {et~al.}
  2023{\natexlab{a}}, \apjl, 951, L8, \dodoi{10.3847/2041-8213/acdac6}

\bibitem[{{Agazie} {et~al.}(2023{\natexlab{b}}){Agazie}, {Anumarlapudi},
  {Archibald}, {Baker}, {B{\'e}csy}, {Blecha}, {Bonilla}, {Brazier}, {Brook},
  {Burke-Spolaor}, {Burnette}, {Case}, {Casey-Clyde}, {Charisi}, {Chatterjee},
  {Chatziioannou}, {Cheeseboro}, {Chen}, {Cohen}, {Cordes}, {Cornish},
  {Crawford}, {Cromartie}, {Crowter}, {Cutler}, {D'Orazio}, {Decesar}, {Degan},
  {Demorest}, {Deng}, {Dolch}, {Drachler}, {Ferrara}, {Fiore}, {Fonseca},
  {Freedman}, {Gardiner}, {Garver-Daniels}, {Gentile}, {Gersbach}, {Glaser},
  {Good}, {G{\"u}ltekin}, {Hazboun}, {Hourihane}, {Islo}, {Jennings},
  {Johnson}, {Jones}, {Kaiser}, {Kaplan}, {Kelley}, {Kerr}, {Key}, {Laal},
  {Lam}, {Lamb}, {Lazio}, {Lewandowska}, {Littenberg}, {Liu}, {Luo}, {Lynch},
  {Ma}, {Madison}, {McEwen}, {McKee}, {McLaughlin}, {McMann}, {Meyers},
  {Meyers}, {Mingarelli}, {Mitridate}, {Natarajan}, {Ng}, {Nice}, {Ocker},
  {Olum}, {Pennucci}, {Perera}, {Petrov}, {Pol}, {Radovan}, {Ransom}, {Ray},
  {Romano}, {Runnoe}, {Sardesai}, {Schmiedekamp}, {Schmiedekamp}, {Schmitz},
  {Schult}, {Shapiro-Albert}, {Siemens}, {Simon}, {Siwek}, {Stairs},
  {Stinebring}, {Stovall}, {Sun}, {Susobhanan}, {Swiggum}, {Taylor}, {Taylor},
  {Turner}, {Unal}, {Vallisneri}, {Vigeland}, {Wachter}, {Wahl}, {Wang},
  {Witt}, {Wright}, {Young}, \& {Nanograv Collaboration}}]{NG15yrAstro}
---. 2023{\natexlab{b}}, \apjl, 952, L37, \dodoi{10.3847/2041-8213/ace18b}

\bibitem[{Amaro-Seoane {et~al.}(2012)Amaro-Seoane, Aoudia, Babak, Binétruy,
  Berti, Bohé, Caprini, Colpi, Cornish, Danzmann, Dufaux, Gair, Jennrich,
  Jetzer, Klein, Lang, Lobo, Littenberg, McWilliams, Nelemans, Petiteau,
  Porter, Schutz, Sesana, Stebbins, Sumner, Vallisneri, Vitale, Volonteri, \&
  Ward}]{lisa}
Amaro-Seoane, P., Aoudia, S., Babak, S., {et~al.} 2012, eLISA: Astrophysics and
  cosmology in the millihertz regime,  arXiv, \dodoi{10.48550/ARXIV.1201.3621}

\bibitem[{{Arnaud}(1996)}]{Arnaud1996}
{Arnaud}, K.~A. 1996, in Astronomical Society of the Pacific Conference Series,
  Vol. 101, Astronomical Data Analysis Software and Systems V, ed. G.~H.
  {Jacoby} \& J.~{Barnes}, 17

\bibitem[{{Barnes} \& {Hernquist}(1991)}]{agnfuel}
{Barnes}, J.~E., \& {Hernquist}, L.~E. 1991, \apjl, 370, L65,
  \dodoi{10.1086/185978}

\bibitem[{Begelman {et~al.}(1980)Begelman, Blandford, \&
  Rees}]{BegelmanDualAGNTimescale}
Begelman, M.~C., Blandford, R.~D., \& Rees, M.~J. 1980, Nature, 287, 307,
  \dodoi{10.1038/287307a0}

\bibitem[{{Blecha} {et~al.}(2018){Blecha}, {Snyder}, {Satyapal}, \&
  {Ellison}}]{Blecha2018}
{Blecha}, L., {Snyder}, G.~F., {Satyapal}, S., \& {Ellison}, S.~L. 2018,
  \mnras, 478, 3056, \dodoi{10.1093/mnras/sty1274}

\bibitem[{{Brusa} {et~al.}(2009){Brusa}, {Fiore}, {Santini}, {Grazian},
  {Comastri}, {Zamorani}, {Hasinger}, {Merloni}, {Civano}, {Fontana}, \&
  {Mainieri}}]{SMBHGalaxyMass}
{Brusa}, M., {Fiore}, F., {Santini}, P., {et~al.} 2009, \aap, 507, 1277,
  \dodoi{10.1051/0004-6361/200912261}

\bibitem[{{Chen} {et~al.}(2023{\natexlab{a}}){Chen}, {Di Matteo}, {Ni},
  {Tremmel}, {DeGraf}, {Shen}, {Holgado}, {Bird}, {Croft}, \&
  {Feng}}]{Chen2023}
{Chen}, N., {Di Matteo}, T., {Ni}, Y., {et~al.} 2023{\natexlab{a}}, \mnras,
  522, 1895, \dodoi{10.1093/mnras/stad834}

\bibitem[{{Chen} {et~al.}(2023{\natexlab{b}}){Chen}, {Liu}, {Foord}, {Shen},
  {Oguri}, {Chen}, {Di Matteo}, {Holgado}, {Hwang}, \& {Zakamska}}]{TChen2023}
{Chen}, Y.-C., {Liu}, X., {Foord}, A., {et~al.} 2023{\natexlab{b}}, \nat, 616,
  45, \dodoi{10.1038/s41586-023-05766-6}

\bibitem[{{Ciurlo} {et~al.}(2023){Ciurlo}, {Mannucci}, {Yeh}, {Amiri},
  {Carniani}, {Cicone}, {Cresci}, {Lusso}, {Marasco}, {Marconcini}, {Marconi},
  {Nardini}, {Pancino}, {Rosati}, {Rubinur}, {Severgnini}, {Scialpi}, {Tozzi},
  {Venturi}, {Vignali}, \& {Volonteri}}]{Ciurlo2023}
{Ciurlo}, A., {Mannucci}, F., {Yeh}, S., {et~al.} 2023, \aap, 671, L4,
  \dodoi{10.1051/0004-6361/202345853}

\bibitem[{Civano {et~al.}(2016)Civano, Marchesi, Comastri, Urry, Elvis,
  Cappelluti, Puccetti, Brusa, Zamorani, Hasinger, Aldcroft, Alexander,
  Allevato, Brunner, Capak, Finoguenov, Fiore, Fruscione, Gilli, Glotfelty,
  Griffiths, Hao, Harrison, Jahnke, Kartaltepe, Karim, LaMassa, Lanzuisi,
  Miyaji, Ranalli, Salvato, Sargent, Scoville, Schawinski, Schinnerer,
  Silverman, Smolcic, Stern, Toft, Trakhenbrot, Treister, \& Vignali}]{COSMOS}
Civano, F., Marchesi, S., Comastri, A., {et~al.} 2016, The Astrophysical
  Journal, 819, 62, \dodoi{10.3847/0004-637x/819/1/62}

\bibitem[{{Davis} {et~al.}(2012{\natexlab{a}}){Davis}, {Bautz}, {Dewey},
  {Heilmann}, {Houck}, {Huenemoerder}, {Marshall}, {Nowak}, {Schattenburg},
  {Schulz}, \& {Smith}}]{MARX}
{Davis}, J.~E., {Bautz}, M.~W., {Dewey}, D., {et~al.} 2012{\natexlab{a}}, in
  Society of Photo-Optical Instrumentation Engineers (SPIE) Conference Series,
  Vol. 8443, Space Telescopes and Instrumentation 2012: Ultraviolet to Gamma
  Ray, ed. T.~{Takahashi}, S.~S. {Murray}, \& J.-W.~A. {den Herder}, 84431A,
  \dodoi{10.1117/12.926937}

\bibitem[{{Davis} {et~al.}(2012{\natexlab{b}}){Davis}, {Bautz}, {Dewey},
  {Heilmann}, {Houck}, {Huenemoerder}, {Marshall}, {Nowak}, {Schattenburg},
  {Schulz}, \& {Smith}}]{Davis2012}
{Davis}, J.~E., {Bautz}, M.~W., {Dewey}, D., {et~al.} 2012{\natexlab{b}}, in
  \procspie, Vol. 8443, Space Telescopes and Instrumentation 2012: Ultraviolet
  to Gamma Ray, 84431A, \dodoi{10.1117/12.926937}

\bibitem[{{De Rosa} {et~al.}(2018){De Rosa}, {Vignali}, {Husemann}, {Bianchi},
  {Bogdanovi{\'c}}, {Guainazzi}, {Herrero-Illana}, {Komossa}, {Kun}, {Loiseau},
  {Paragi}, {Perez-Torres}, \& {Piconcelli}}]{DeRosa2018}
{De Rosa}, A., {Vignali}, C., {Husemann}, B., {et~al.} 2018, \mnras, 480, 1639,
  \dodoi{10.1093/mnras/sty1867}

\bibitem[{{Eftekharzadeh} {et~al.}(2017){Eftekharzadeh}, {Myers}, {Hennawi},
  {Djorgovski}, {Richards}, {Mahabal}, \& {Graham}}]{Eftekharzadeh2017}
{Eftekharzadeh}, S., {Myers}, A.~D., {Hennawi}, J.~F., {et~al.} 2017, \mnras,
  468, 77, \dodoi{10.1093/mnras/stx412}

\bibitem[{Engmann \& Cousineau(2011)}]{ADvsKS}
Engmann, S., \& Cousineau, D. 2011, Journal of Applied Quantitative Methods, 6,
  1

\bibitem[{{EPTA Collaboration} {et~al.}(2023){EPTA Collaboration}, {InPTA
  Collaboration}, {Antoniadis}, {Arumugam}, {Arumugam}, {Babak}, {Bagchi}, {Bak
  Nielsen}, {Bassa}, {Bathula}, {Berthereau}, {Bonetti}, {Bortolas}, {Brook},
  {Burgay}, {Caballero}, {Chalumeau}, {Champion}, {Chanlaridis}, {Chen},
  {Cognard}, {Dandapat}, {Deb}, {Desai}, {Desvignes}, {Dhanda-Batra},
  {Dwivedi}, {Falxa}, {Ferdman}, {Franchini}, {Gair}, {Goncharov}, {Gopakumar},
  {Graikou}, {Grie{\ss}meier}, {Guillemot}, {Guo}, {Gupta}, {Hisano}, {Hu},
  {Iraci}, {Izquierdo-Villalba}, {Jang}, {Jawor}, {Janssen}, {Jessner},
  {Joshi}, {Kareem}, {Karuppusamy}, {Keane}, {Keith}, {Kharbanda}, {Kikunaga},
  {Kolhe}, {Kramer}, {Krishnakumar}, {Lackeos}, {Lee}, {Liu}, {Liu}, {Lyne},
  {McKee}, {Maan}, {Main}, {Mickaliger}, {Ni{\c{t}}u}, {Nobleson}, {Paladi},
  {Parthasarathy}, {Perera}, {Perrodin}, {Petiteau}, {Porayko}, {Possenti},
  {Prabu}, {Quelquejay Leclere}, {Rana}, {Samajdar}, {Sanidas}, {Sesana},
  {Shaifullah}, {Singha}, {Speri}, {Spiewak}, {Srivastava}, {Stappers},
  {Surnis}, {Susarla}, {Susobhanan}, {Takahashi}, {Tarafdar}, {Theureau},
  {Tiburzi}, {van der Wateren}, {Vecchio}, {Venkatraman Krishnan}, {Verbiest},
  {Wang}, {Wang}, \& {Wu}}]{EPTA10yrGWB}
{EPTA Collaboration}, {InPTA Collaboration}, {Antoniadis}, J., {et~al.} 2023,
  \aap, 678, A50, \dodoi{10.1051/0004-6361/202346844}

\bibitem[{Foord {et~al.}(2020)Foord, Gültekin, Nevin, Comerford, Hodges-Kluck,
  Barrows, Goulding, \& Greene}]{BAYMAX1}
Foord, A., Gültekin, K., Nevin, R., {et~al.} 2020, The Astrophysical Journal,
  892, 29, \dodoi{10.3847/1538-4357/ab72fa}

\bibitem[{Foord {et~al.}(2021)Foord, Gültekin, Runnoe, \& Koss}]{BAYMAX2}
Foord, A., Gültekin, K., Runnoe, J.~C., \& Koss, M.~J. 2021, The Astrophysical
  Journal, 907, 71, \dodoi{10.3847/1538-4357/abce5d}

\bibitem[{Foord {et~al.}(2019)Foord, Gültekin, Reynolds, Hodges-Kluck,
  Cackett, Comerford, King, Miller, \& Runnoe}]{BAYMAX0}
Foord, A., Gültekin, K., Reynolds, M.~T., {et~al.} 2019, The Astrophysical
  Journal, 877, 17, \dodoi{10.3847/1538-4357/ab18a3}

\bibitem[{{Fruscione} {et~al.}(2006){Fruscione}, {McDowell}, {Allen},
  {Brickhouse}, {Burke}, {Davis}, {Durham}, {Elvis}, {Galle}, {Harris},
  {Huenemoerder}, {Houck}, {Ishibashi}, {Karovska}, {Nicastro}, {Noble},
  {Nowak}, {Primini}, {Siemiginowska}, {Smith}, \& {Wise}}]{Fruscione2006}
{Fruscione}, A., {McDowell}, J.~C., {Allen}, G.~E., {et~al.} 2006, in Society
  of Photo-Optical Instrumentation Engineers (SPIE) Conference Series, Vol.
  6270, Society of Photo-Optical Instrumentation Engineers (SPIE) Conference
  Series, 62701V, \dodoi{10.1117/12.671760}

\bibitem[{{Gaskin} {et~al.}(2019){Gaskin}, {Swartz}, {Vikhlinin}, {{\"O}zel},
  {Gelmis}, {Arenberg}, {Bandler}, {Bautz}, {Civitani}, {Dominguez}, {Eckart},
  {Falcone}, {Figueroa-Feliciano}, {Freeman}, {G{\"u}nther}, {Havey},
  {Heilmann}, {Kilaru}, {Kraft}, {McCarley}, {McEntaffer}, {Pareschi},
  {Purcell}, {Reid}, {Schattenburg}, {Schwartz}, {Schwartz}, {Tananbaum},
  {Tremblay}, {Zhang}, \& {Zuhone}}]{Lynx}
{Gaskin}, J.~A., {Swartz}, D.~A., {Vikhlinin}, A., {et~al.} 2019, Journal of
  Astronomical Telescopes, Instruments, and Systems, 5, 021001,
  \dodoi{10.1117/1.JATIS.5.2.021001}

\bibitem[{{Gehrels}(1986)}]{Gehrels1986}
{Gehrels}, N. 1986, \apj, 303, 336, \dodoi{10.1086/164079}

\bibitem[{Haehnelt(1994)}]{smbhwaves}
Haehnelt, M.~G. 1994, Monthly Notices of the Royal Astronomical Society, 269,
  199, \dodoi{10.1093/mnras/269.1.199}

\bibitem[{{Hennawi} {et~al.}(2006){Hennawi}, {Strauss}, {Oguri}, {Inada},
  {Richards}, {Pindor}, {Schneider}, {Becker}, {Gregg}, {Hall}, {Johnston},
  {Fan}, {Burles}, {Schlegel}, {Gunn}, {Lupton}, {Bahcall}, {Brunner}, \&
  {Brinkmann}}]{Hennawi2006}
{Hennawi}, J.~F., {Strauss}, M.~A., {Oguri}, M., {et~al.} 2006, \aj, 131, 1,
  \dodoi{10.1086/498235}

\bibitem[{{Hennawi} {et~al.}(2010){Hennawi}, {Myers}, {Shen}, {Strauss},
  {Djorgovski}, {Fan}, {Glikman}, {Mahabal}, {Martin}, {Richards}, {Schneider},
  \& {Shankar}}]{Hennawi2010}
{Hennawi}, J.~F., {Myers}, A.~D., {Shen}, Y., {et~al.} 2010, \apj, 719, 1672,
  \dodoi{10.1088/0004-637X/719/2/1672}

\bibitem[{{Hickox} \& {Alexander}(2018)}]{Hickox2018}
{Hickox}, R.~C., \& {Alexander}, D.~M. 2018, \araa, 56, 625,
  \dodoi{10.1146/annurev-astro-081817-051803}

\bibitem[{{Hopkins} {et~al.}(2008){Hopkins}, {Hernquist}, {Cox}, \&
  {Kere{\v{s}}}}]{HopkinsMergerTriggerAGN}
{Hopkins}, P.~F., {Hernquist}, L., {Cox}, T.~J., \& {Kere{\v{s}}}, D. 2008,
  \apjs, 175, 356, \dodoi{10.1086/524362}

\bibitem[{{Hwang} {et~al.}(2020){Hwang}, {Shen}, {Zakamska}, \&
  {Liu}}]{Hwang2020}
{Hwang}, H.-C., {Shen}, Y., {Zakamska}, N., \& {Liu}, X. 2020, \apj, 888, 73,
  \dodoi{10.3847/1538-4357/ab5c1a}

\bibitem[{Jeffreys(1935)}]{BFThresh}
Jeffreys, H. 1935, Mathematical Proceedings of the Cambridge Philosophical
  Society, 31, 203–222, \dodoi{10.1017/S030500410001330X}

\bibitem[{{Kayo} \& {Oguri}(2012)}]{Kayo2012}
{Kayo}, I., \& {Oguri}, M. 2012, \mnras, 424, 1363,
  \dodoi{10.1111/j.1365-2966.2012.21321.x}

\bibitem[{{Kelley} {et~al.}(2017){Kelley}, {Blecha}, \&
  {Hernquist}}]{Kelley2017}
{Kelley}, L.~Z., {Blecha}, L., \& {Hernquist}, L. 2017, \mnras, 464, 3131,
  \dodoi{10.1093/mnras/stw2452}

\bibitem[{{Kewley} {et~al.}(2006){Kewley}, {Groves}, {Kauffmann}, \&
  {Heckman}}]{Kewley2006}
{Kewley}, L.~J., {Groves}, B., {Kauffmann}, G., \& {Heckman}, T. 2006, \mnras,
  372, 961, \dodoi{10.1111/j.1365-2966.2006.10859.x}

\bibitem[{{Kocevski} {et~al.}(2015){Kocevski}, {Brightman}, {Nandra},
  {Koekemoer}, {Salvato}, {Aird}, {Bell}, {Hsu}, {Kartaltepe}, {Koo}, {Lotz},
  {McIntosh}, {Mozena}, {Rosario}, \& {Trump}}]{Kocevski2015}
{Kocevski}, D.~D., {Brightman}, M., {Nandra}, K., {et~al.} 2015, \apj, 814,
  104, \dodoi{10.1088/0004-637X/814/2/104}

\bibitem[{Kocevski {et~al.}(2018)Kocevski, Hasinger, Brightman, Nandra,
  Georgakakis, Cappelluti, Civano, Li, Li, Aird, Alexander, Almaini, Brusa,
  Buchner, Comastri, Conselice, Dickinson, Finoguenov, Gilli, Koekemoer,
  Miyaji, Mullaney, Papovich, Rosario, Salvato, Silverman, Somerville, \&
  Ueda}]{XUDS}
Kocevski, D.~D., Hasinger, G., Brightman, M., {et~al.} 2018, The Astrophysical
  Journal Supplement Series, 236, 48, \dodoi{10.3847/1538-4365/aab9b4}

\bibitem[{{Kormendy} \& {Richstone}(1995)}]{smbhcenter}
{Kormendy}, J., \& {Richstone}, D. 1995, \araa, 33, 581,
  \dodoi{10.1146/annurev.aa.33.090195.003053}

\bibitem[{Koss {et~al.}(2015)Koss, Romero-Ca{\~{n}}izales, Baronchelli, Teng,
  Balokovi{\'{c}}, Puccetti, Bauer, Ar{\'{e}}valo, Assef, Ballantyne, Brandt,
  Brightman, Comastri, Gandhi, Harrison, Luo, Schawinski, Stern, \&
  Treister}]{Koss}
Koss, M.~J., Romero-Ca{\~{n}}izales, C., Baronchelli, L., {et~al.} 2015, The
  Astrophysical Journal, 807, 149, \dodoi{10.1088/0004-637x/807/2/149}

\bibitem[{{Koss} {et~al.}(2016){Koss}, {Assef}, {Balokovi{\'c}}, {Stern},
  {Gandhi}, {Lamperti}, {Alexander}, {Ballantyne}, {Bauer}, {Berney}, {Brand
  t}, {Comastri}, {Gehrels}, {Harrison}, {Lansbury}, {Markwardt}, {Ricci},
  {Rivers}, {Schawinski}, {Trakhtenbrot}, {Treister}, \& {Urry}}]{Koss2016}
{Koss}, M.~J., {Assef}, R., {Balokovi{\'c}}, M., {et~al.} 2016, \apj, 825, 85,
  \dodoi{10.3847/0004-637X/825/2/85}

\bibitem[{{Koss} {et~al.}(2018){Koss}, {Blecha}, {Bernhard}, {Hung}, {Lu},
  {Trakhtenbrot}, {Treister}, {Weigel}, {Sartori}, {Mushotzky}, {Schawinski},
  {Ricci}, {Veilleux}, \& {Sanders}}]{Koss2018}
{Koss}, M.~J., {Blecha}, L., {Bernhard}, P., {et~al.} 2018, \nat, 563, 214,
  \dodoi{10.1038/s41586-018-0652-7}

\bibitem[{{Lanzuisi} {et~al.}(2018){Lanzuisi}, {Civano}, {Marchesi},
  {Comastri}, {Brusa}, {Gilli}, {Vignali}, {Zamorani}, {Brightman},
  {Griffiths}, \& {Koekemoer}}]{Lanzuisi2018}
{Lanzuisi}, G., {Civano}, F., {Marchesi}, S., {et~al.} 2018, \mnras, 480, 2578,
  \dodoi{10.1093/mnras/sty2025}

\bibitem[{{Lehmer} {et~al.}(2019){Lehmer}, {Eufrasio}, {Tzanavaris},
  {Basu-Zych}, {Fragos}, {Prestwich}, {Yukita}, {Zezas}, {Hornschemeier}, \&
  {Ptak}}]{agnlum}
{Lehmer}, B.~D., {Eufrasio}, R.~T., {Tzanavaris}, P., {et~al.} 2019, \apjs,
  243, 3, \dodoi{10.3847/1538-4365/ab22a8}

\bibitem[{{Liu} {et~al.}(2017){Liu}, {Tozzi}, {Wang}, {Brandt}, {Vignali},
  {Xue}, {Schneider}, {Comastri}, {Yang}, {Bauer}, {Paolillo}, {Luo}, {Gilli},
  {Wang}, {Giavalisco}, {Ji}, {Alexander}, {Mainieri}, {Shemmer}, {Koekemoer},
  \& {Risaliti}}]{Liu2017}
{Liu}, T., {Tozzi}, P., {Wang}, J.-X., {et~al.} 2017, \apjs, 232, 8,
  \dodoi{10.3847/1538-4365/aa7847}

\bibitem[{Luo {et~al.}(2016)Luo, Brandt, Xue, Lehmer, Alexander, Bauer, Vito,
  Yang, Basu-Zych, Comastri, Gilli, Gu, Hornschemeier, Koekemoer, Liu,
  Mainieri, Paolillo, Ranalli, Rosati, Schneider, Shemmer, Smail, Sun, Tozzi,
  Vignali, \& Wang}]{CDFS}
Luo, B., Brandt, W.~N., Xue, Y.~Q., {et~al.} 2016, The Astrophysical Journal
  Supplement Series, 228, 2, \dodoi{10.3847/1538-4365/228/1/2}

\bibitem[{{Manchester} {et~al.}(2013){Manchester}, {Hobbs}, {Bailes}, {Coles},
  {van Straten}, {Keith}, {Shannon}, {Bhat}, {Brown}, {Burke-Spolaor},
  {Champion}, {Chaudhary}, {Edwards}, {Hampson}, {Hotan}, {Jameson}, {Jenet},
  {Kesteven}, {Khoo}, {Kocz}, {Maciesiak}, {Oslowski}, {Ravi}, {Reynolds},
  {Sarkissian}, {Verbiest}, {Wen}, {Wilson}, {Yardley}, {Yan}, \& {You}}]{pta}
{Manchester}, R.~N., {Hobbs}, G., {Bailes}, M., {et~al.} 2013, \pasa, 30, e017,
  \dodoi{10.1017/pasa.2012.017}

\bibitem[{{Mannucci} {et~al.}(2022){Mannucci}, {Pancino}, {Belfiore}, {Cicone},
  {Ciurlo}, {Cresci}, {Lusso}, {Marasco}, {Marconi}, {Nardini}, {Pinna},
  {Severgnini}, {Saracco}, {Tozzi}, \& {Yeh}}]{Mannucci2022}
{Mannucci}, F., {Pancino}, E., {Belfiore}, F., {et~al.} 2022, Nature Astronomy,
  6, 1185, \dodoi{10.1038/s41550-022-01761-5}

\bibitem[{{Marchesi} {et~al.}(2016){Marchesi}, {Civano}, {Elvis}, {Salvato},
  {Brusa}, {Comastri}, {Gilli}, {Hasinger}, {Lanzuisi}, {Miyaji}, {Treister},
  {Urry}, {Vignali}, {Zamorani}, {Allevato}, {Cappelluti}, {Cardamone},
  {Finoguenov}, {Griffiths}, {Karim}, {Laigle}, {LaMassa}, {Jahnke}, {Ranalli},
  {Schawinski}, {Schinnerer}, {Silverman}, {Smolcic}, {Suh}, \&
  {Trakhtenbrot}}]{Marchesi2016}
{Marchesi}, S., {Civano}, F., {Elvis}, M., {et~al.} 2016, \apj, 817, 34,
  \dodoi{10.3847/0004-637X/817/1/34}

\bibitem[{Massey(1951)}]{KSGen}
Massey, F.~J. 1951, Journal of the American Statistical Association, 46, 68.
\newblock \url{http://www.jstor.org/stable/2280095}

\bibitem[{{Myers} {et~al.}(2008){Myers}, {Richards}, {Brunner}, {Schneider},
  {Strand}, {Hall}, {Blomquist}, \& {York}}]{Myers2008}
{Myers}, A.~D., {Richards}, G.~T., {Brunner}, R.~J., {et~al.} 2008, \apj, 678,
  635, \dodoi{10.1086/533491}

\bibitem[{Nandra {et~al.}(2015)Nandra, Laird, Aird, Salvato, Georgakakis,
  Barro, Perez-Gonzalez, Barmby, Chary, Coil, Cooper, Davis, Dickinson, Faber,
  Fazio, Guhathakurta, Gwyn, Hsu, Huang, Ivison, Koo, Newman, Rangel, Yamada,
  \& Willmer}]{AEGISXD}
Nandra, K., Laird, E.~S., Aird, J.~A., {et~al.} 2015, The Astrophysical Journal
  Supplement Series, 220, 10, \dodoi{10.1088/0067-0049/220/1/10}

\bibitem[{{Perna} {et~al.}(2023){Perna}, {Arribas}, {Lamperti}, {Circosta},
  {Bertola}, {P{\'e}rez-Gonz{\'a}lez}, {D'Eugenio}, {{\"U}bler}, {Cresci},
  {Maiolino}, {Rodr{\'\i}guez Del Pino}, {Bunker}, {Charlot}, {Willott},
  {Carniani}, {B{\"o}ker}, {Chevallard}, {Curti}, {Jones}, {Kumari},
  {Marshall}, {Saxena}, {Scholtz}, {Venturi}, \& {Witstok}}]{Perna2023}
{Perna}, M., {Arribas}, S., {Lamperti}, I., {et~al.} 2023, arXiv e-prints,
  arXiv:2310.03067, \dodoi{10.48550/arXiv.2310.03067}

\bibitem[{{Puerto-Sanchez}(in prep)}]{PuertoSanchez_inprep}
{Puerto-Sanchez}, C. in prep, \mnras

\bibitem[{Razali {et~al.}(2011)Razali, Wah, {et~al.}}]{AD&KSPower}
Razali, N.~M., Wah, Y.~B., {et~al.} 2011, Journal of statistical modeling and
  analytics, 2, 21

\bibitem[{{Reardon} {et~al.}(2023){Reardon}, {Zic}, {Shannon}, {Hobbs},
  {Bailes}, {Di Marco}, {Kapur}, {Rogers}, {Thrane}, {Askew}, {Bhat},
  {Cameron}, {Cury{\l}o}, {Coles}, {Dai}, {Goncharov}, {Kerr}, {Kulkarni},
  {Levin}, {Lower}, {Manchester}, {Mandow}, {Miles}, {Nathan}, {Os{\l}owski},
  {Russell}, {Spiewak}, {Zhang}, \& {Zhu}}]{PPTA18yrGWB}
{Reardon}, D.~J., {Zic}, A., {Shannon}, R.~M., {et~al.} 2023, \apjl, 951, L6,
  \dodoi{10.3847/2041-8213/acdd02}

\bibitem[{Reynolds {et~al.}(2023)Reynolds, Kara, Mushotzky, Ptak, Koss,
  Williams, Allen, Bauer, Bautz, Bogadhee, Burdge, Cappelluti, Cenko, Chartas,
  Chan, Corrales, Daylan, Falcone, Foord, Grant, Habouzit, Haggard, Herrmann,
  Hodges-Kluck, Kargaltsev, King, Kounkel, Lopez, Marchesi, McDonald, Meyer,
  Miller, Nynka, Okajima, Pacucci, Russell, Safi-Harb, Strassun, Falc{\~a}o,
  Walker, Wilms, Yukita, \& Zhang}]{AXIS}
Reynolds, C.~S., Kara, E.~A., Mushotzky, R.~F., {et~al.} 2023, in UV, X-Ray,
  and Gamma-Ray Space Instrumentation for Astronomy XXIII, ed. O.~H. Siegmund
  \& K.~Hoadley, Vol. 12678, International Society for Optics and Photonics
  (SPIE), 126781E, \dodoi{10.1117/12.2677468}

\bibitem[{{Ricci} {et~al.}(2017){Ricci}, {Bauer}, {Treister}, {Schawinski},
  {Privon}, {Blecha}, {Arevalo}, {Armus}, {Harrison}, {Ho}, {Iwasawa},
  {Sanders}, \& {Stern}}]{Ricci2017}
{Ricci}, C., {Bauer}, F.~E., {Treister}, E., {et~al.} 2017, \mnras, 468, 1273,
  \dodoi{10.1093/mnras/stx173}

\bibitem[{{Rosas-Guevara} {et~al.}(2019){Rosas-Guevara}, {Bower}, {McAlpine},
  {Bonoli}, \& {Tissera}}]{Rosas-Guevara2019}
{Rosas-Guevara}, Y.~M., {Bower}, R.~G., {McAlpine}, S., {Bonoli}, S., \&
  {Tissera}, P.~B. 2019, \mnras, 483, 2712, \dodoi{10.1093/mnras/sty3251}

\bibitem[{{Salvatier} {et~al.}(2016){Salvatier}, {Wiecki}, \&
  {Fonnesbeck}}]{Salvatier2016}
{Salvatier}, J., {Wiecki}, T., \& {Fonnesbeck}, C. 2016, PeerJ Computer
  Science, 2, \dodoi{10.7717/peerj-cs.55.}

\bibitem[{Scholz \& Stephens(1987)}]{ADGen}
Scholz, F.~W., \& Stephens, M.~A. 1987, Journal of the American Statistical
  Association, 82, 918.
\newblock \url{http://www.jstor.org/stable/2288805}

\bibitem[{{Shen} {et~al.}(2019){Shen}, {Hwang}, {Zakamska}, \&
  {Liu}}]{Shen2019}
{Shen}, Y., {Hwang}, H.-C., {Zakamska}, N., \& {Liu}, X. 2019, \apjl, 885, L4,
  \dodoi{10.3847/2041-8213/ab4b54}

\bibitem[{{Shen} {et~al.}(2021){Shen}, {Chen}, {Hwang}, {Liu}, {Zakamska},
  {Oguri}, {Li}, {Lazio}, \& {Breiding}}]{Shen2021}
{Shen}, Y., {Chen}, Y.-C., {Hwang}, H.-C., {et~al.} 2021, Nature Astronomy, 5,
  569, \dodoi{10.1038/s41550-021-01323-1}

\bibitem[{{Shen} {et~al.}(2023){Shen}, {Hwang}, {Oguri}, {Chen}, {Di Matteo},
  {Ni}, {Bird}, {Zakamska}, {Liu}, {Chen}, \& {Kratter}}]{Shen2023}
{Shen}, Y., {Hwang}, H.-C., {Oguri}, M., {et~al.} 2023, \apj, 943, 38,
  \dodoi{10.3847/1538-4357/aca662}

\bibitem[{{Sijacki} {et~al.}(2007){Sijacki}, {Springel}, {Di Matteo}, \&
  {Hernquist}}]{DeboraMergerTriggerAGN}
{Sijacki}, D., {Springel}, V., {Di Matteo}, T., \& {Hernquist}, L. 2007,
  \mnras, 380, 877, \dodoi{10.1111/j.1365-2966.2007.12153.x}

\bibitem[{{Silverman} {et~al.}(2020){Silverman}, {Tang}, {Lee}, {Hartwig},
  {Goulding}, {Strauss}, {Schramm}, {Ding}, {Riffel}, {Fujimoto}, {Hikage},
  {Imanishi}, {Iwasawa}, {Jahnke}, {Kayo}, {Kashikawa}, {Kawaguchi}, {Kohno},
  {Luo}, {Matsuoka}, {Matsuda}, {Nagao}, {Oguri}, {Ono}, {Onoue}, {Ouchi},
  {Shimasaku}, {Suh}, {Suzuki}, {Taniguchi}, {Toba}, {Ueda}, \&
  {Yasuda}}]{Silverman2020}
{Silverman}, J.~D., {Tang}, S., {Lee}, K.-G., {et~al.} 2020, \apj, 899, 154,
  \dodoi{10.3847/1538-4357/aba4a3}

\bibitem[{Skilling(2004)}]{nested}
Skilling, J. 2004in , American Institute of Physics, 395--405

\bibitem[{{Steinborn} {et~al.}(2016){Steinborn}, {Dolag}, {Comerford},
  {Hirschmann}, {Remus}, \& {Teklu}}]{Steinborn2016}
{Steinborn}, L.~K., {Dolag}, K., {Comerford}, J.~M., {et~al.} 2016, \mnras,
  458, 1013, \dodoi{10.1093/mnras/stw316}

\bibitem[{{Stemo} {et~al.}(2021){Stemo}, {Comerford}, {Barrows}, {Stern},
  {Assef}, {Griffith}, \& {Schechter}}]{Stemo2021}
{Stemo}, A., {Comerford}, J.~M., {Barrows}, R.~S., {et~al.} 2021, \apj, 923,
  36, \dodoi{10.3847/1538-4357/ac0bbf}

\bibitem[{{Torres-Alb{\`a}} {et~al.}(2018){Torres-Alb{\`a}}, {Iwasawa},
  {D{\'\i}az-Santos}, {Charmandaris}, {Ricci}, {Chu}, {Sanders}, {Armus},
  {Barcos-Mu{\~n}oz}, {Evans}, {Howell}, {Inami}, {Linden}, {Medling},
  {Privon}, {U}, \& {Yoon}}]{TorresAlba2018}
{Torres-Alb{\`a}}, N., {Iwasawa}, K., {D{\'\i}az-Santos}, T., {et~al.} 2018,
  \aap, 620, A140, \dodoi{10.1051/0004-6361/201834105}

\bibitem[{{Volonteri} {et~al.}(2022){Volonteri}, {Pfister}, {Beckmann},
  {Dotti}, {Dubois}, {Massonneau}, {Musoke}, \& {Tremmel}}]{Volonteri2022}
{Volonteri}, M., {Pfister}, H., {Beckmann}, R., {et~al.} 2022, \mnras, 514,
  640, \dodoi{10.1093/mnras/stac1217}

\bibitem[{{Weston} {et~al.}(2017){Weston}, {McIntosh}, {Brodwin}, {Mann},
  {Cooper}, {McConnell}, \& {Nielsen}}]{Weston2017}
{Weston}, M.~E., {McIntosh}, D.~H., {Brodwin}, M., {et~al.} 2017, \mnras, 464,
  3882, \dodoi{10.1093/mnras/stw2620}

\bibitem[{{White} \& {Rees}(1978)}]{hierarchical}
{White}, S.~D.~M., \& {Rees}, M.~J. 1978, \mnras, 183, 341,
  \dodoi{10.1093/mnras/183.3.341}

\bibitem[{{Xu} {et~al.}(2023){Xu}, {Chen}, {Guo}, {Jiang}, {Wang}, {Xu}, {Xue},
  {Nicolas Caballero}, {Yuan}, {Xu}, {Wang}, {Hao}, {Luo}, {Lee}, {Han},
  {Jiang}, {Shen}, {Wang}, {Wang}, {Xu}, {Wu}, {Manchester}, {Qian}, {Guan},
  {Huang}, {Sun}, \& {Zhu}}]{CPTAGWB}
{Xu}, H., {Chen}, S., {Guo}, Y., {et~al.} 2023, Research in Astronomy and
  Astrophysics, 23, 075024, \dodoi{10.1088/1674-4527/acdfa5}

\bibitem[{{Yue} {et~al.}(2021){Yue}, {Fan}, {Yang}, \& {Wang}}]{Yue2021}
{Yue}, M., {Fan}, X., {Yang}, J., \& {Wang}, F. 2021, \apjl, 921, L27,
  \dodoi{10.3847/2041-8213/ac31a9}

\bibitem[{{Yue} {et~al.}(2023){Yue}, {Fan}, {Yang}, \& {Wang}}]{Yue2023}
---. 2023, \aj, 165, 191, \dodoi{10.3847/1538-3881/acc2be}

\end{thebibliography}
\bibliographystyle{aasjournal}


\end{document}